\documentclass[%
 prl, twocolumn,
 amsmath,amssymb,
 aps, superscriptaddress]{revtex4-2}

\usepackage{graphicx}
\usepackage{dcolumn}
\usepackage{bm}
\usepackage{dsfont}
\usepackage{xcolor}
\usepackage{mathrsfs}
\usepackage{comment}
\usepackage{bbold}
\newcommand{\bs}{\boldsymbol}
\newcommand{\up}{\uparrow}
\newcommand{\dw}{\downarrow}
\usepackage[colorlinks=true,citecolor=blue]{hyperref}

\begin{document}

\title{Quantized spin pumping in topological ferromagnetic-superconducting  nanowires}

\author{V. Fern\'andez Becerra}
 \email{victor\_f\_becerra@outlook.com}
\affiliation{International Research Centre MagTop, Institute of Physics, Polish Academy of Sciences, Aleja Lotnikow 32/46, PL-02668 Warsaw, Poland}
\author{Mircea Trif}
\affiliation{International Research Centre MagTop, Institute of Physics, Polish Academy of Sciences, Aleja Lotnikow 32/46, PL-02668 Warsaw, Poland}
\author{Timo Hyart} 
\affiliation{International Research Centre MagTop, Institute of Physics, Polish Academy of Sciences, Aleja Lotnikow 32/46, PL-02668 Warsaw, Poland}
\affiliation{Department of Applied Physics, Aalto University, 00076 Aalto, Espoo, Finland}
\affiliation{Computational Physics Laboratory, Physics Unit, Faculty of Engineering and Natural Sciences, Tampere University, FI-33014 Tampere, Finland}

\date{\today}

\begin{abstract}
Semiconducting nanowires with strong spin-orbit coupling in the presence of induced superconductivity and ferromagnetism can support Majorana zero modes. We study the pumping due to the precession of the magnetization in single-subband nanowires and show that spin pumping is robustly quantized when the hybrid nanowire is in the topologically nontrivial phase, whereas charge pumping is not quantized. Moreover, there exists one-to-one correspondence between the quantized conductance, entropy change and spin pumping in long topologically nontrivial nanowires but these observables are uncorrelated in the case of accidental zero-energy Andreev bound states in the trivial phase. Thus, we conclude that observation of correlated and quantized peaks in the conductance, entropy change and spin pumping would provide strong evidence of Majorana zero modes, and we elaborate how  topological Majorana zero modes can be distinguished from quasi-Majorana modes potentially created by a smooth tunnel barrier at the lead-nanowire interface.  Finally, we discuss peculiar interference effects affecting the spin pumping in short nanowires at very low energies.
\end{abstract}

\maketitle

{\it Introduction.}$-$ One of the most challenging aims in the current condensed matter physics research is the demonstration of non-Abelian Majorana statistics -- the underlying fundamental property that would enable the realization of a topological quantum computer  
\cite{Nayak08, Hyart13, Aasen16, Karzig17, Beenakker20}. It is theoretically well established that Majorana zero modes (MZMs) can be realized in semiconducting nanowires with strong Rashba spin-orbit coupling
in the presence of induced superconductivity and external magnetic field
\cite{Lutchyn10, Oreg10}. One of the hallmark features of the MZMs is the resonant Andreev reflection, which gives rise to a quantized zero-bias peak in the conductance \cite{Law09, Wimmer11}. Although zero-bias conductance peaks have been observed in experiments \cite{Mourik12}, it is known that formation of unwanted quantum dots or unintentional inhomogeneities at the lead-nanowire interface can lead to non-Majorana zero-bias conductance peaks \cite{inh_Brouwer, QD_Liu,QD_Moore_one,QD_Moore_two, inh_workfunc, Chen19, qMajo_Vuik, RN14, Frolov21, ValentiniScience21}, and therefore the current attempts to demonstrate the existence of MZMs utilize multimodal experimental data with sophisticated protocols to reduce the likelihood of false positives \cite{Pikulin21, 2022arXiv220702472A}. Other techniques to detect MZMs based on noise \cite{LiuPRB15_1,LiuPRB15_2,SmirnovPRB19,SmirnovPRB22} and entropy change \cite{entro_smirnov_one, entro_Sela, entro_smirnov_two, entro_smirnov_three} are also being developed.

The advent of hybrid ferromagnetic insulator-superconducting (FI-SC) nanowire devices
\cite{YuNanoLett2020, Vaitieknas2020ZerofieldTS} opens a paths for novel approaches for probing the existence of MZMs. In this Letter, we study the charge and spin pumping in this system in the presence of precessing magnetization. The precessing magnetization is a well-established method for generating spin current in magnetic heterostructures and forms the basis of many contemporary spintronics applications 
\cite{Spump_theo, Spbattery, TserkovnyakRMP2005, spintroRev, SinovaRMP2015, Hoffman-review}. It is known that under some specific circumstances the precessing magnetization can also lead to quantized charge and spin pumping 
\cite{Qquanti_Zhang, MeidanPRB10,MeidanPRB11, qtizedspin,Becerra21} which can be understood as Thouless pumping in the Hamiltonian formalism \cite{Thouless83} and topological winding number in the scattering matrix formalism \cite{Becerra21}. Here, we apply the scattering matrix formalism to the case of FI-SC nanowire devices, and find an unprecedented case where the charge pumping is non-quantized, in agreement with the previous study \cite{Ojanen12}, but the spin-pumping is robustly quantized. We show that in long single-subband nanowires there exists one-to-one correspondence between the quantized conductance, entropy change and the quantized spin pumping in the topologically nontrivial nanowires but these observables are uncorrelated in the case of accidental zero-energy Andreev bound states in the trivial phase. Thus, we conclude that the observation of correlated and quantized peaks in the conductance, entropy change and spin pumping would provide strong evidence of MZMs, and we elaborate how  topological MZMs can be distinguished from quasi-Majorana modes potentially created by imperfections at the lead-nanowire interface. Furthermore, we identify a suitable regime of system parameters for observing the quantization and consider interference effects in short nanowires, which affect the conductance and spin pumping differently at very low energies. Finally, we show that the quantized spin pumping can be detected from the absorption linewidth of the precessing FI.   

\begin{figure}[t]
\centering
\includegraphics[width=
\linewidth]{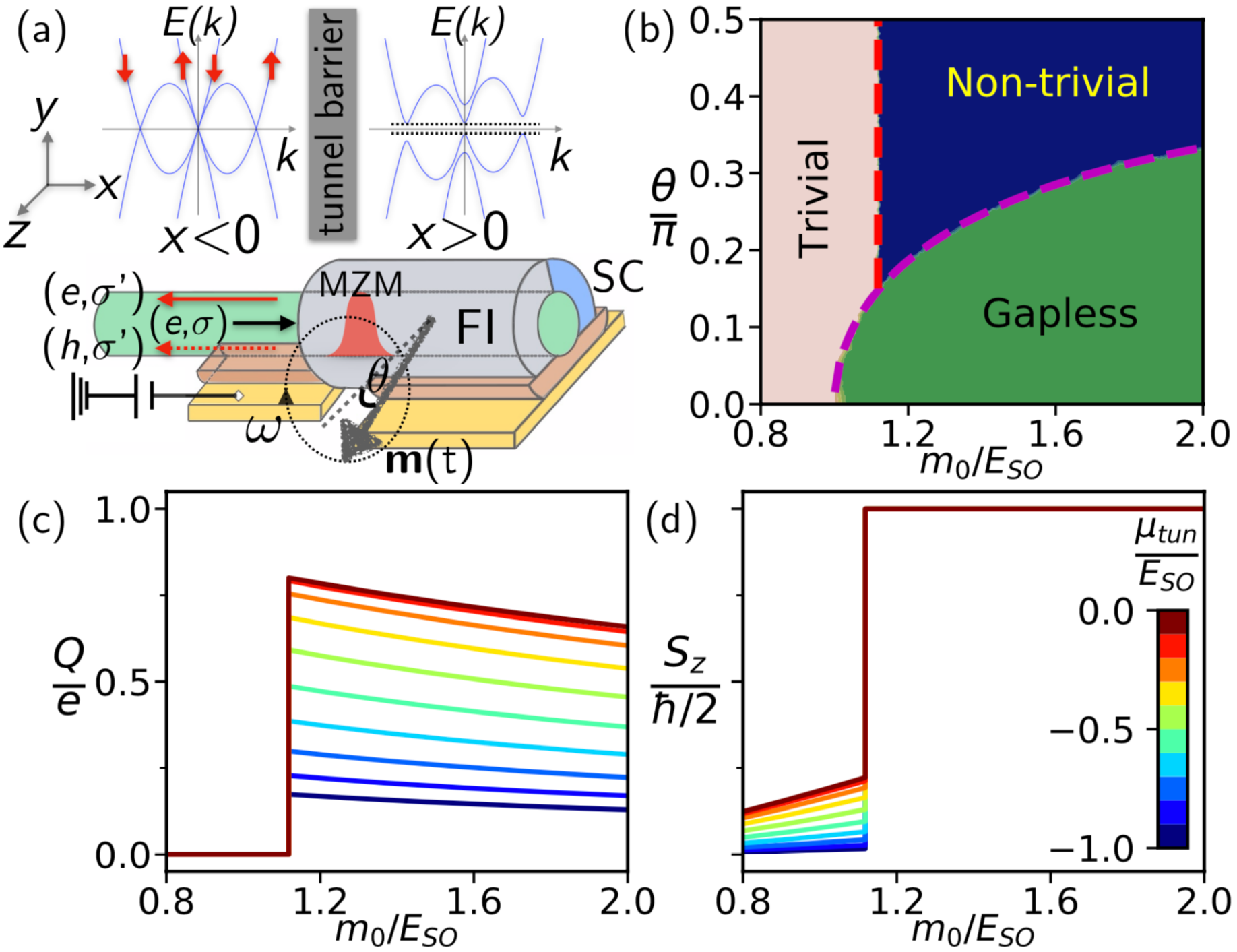}
\caption{(a) Rashba nanowire (green) with proximity induced superconductivity (blue) and magnetization (gray) supports MZM (red). The tunnel barrier $\mu_{tun}$ can be controlled with a gate voltage. The precessing magnetization ${\bs m}(t)$ pumps spin and charge into the  lead ($x<0$) due to the  normal and Andreev  reflection processes. 
(b) Topological phase diagram of the FI-SC nanowire as a function of   $m_0$ and  $\theta$. The phase transition line separating the gapped phases is given by $\Delta_0^2 + \mu^2 = m_0^2$ and the transition between gapped and gapless phase occurs at $\Delta_0 = m_0 \cos{\theta}$. (c),(d) The pumped charge $Q$ and spin $S_z$ as a function of $m_0$ for different $\mu_{tun}$. The pumped spin is quantized to $S_z=\hbar/2$ in the topologically nontrivial regime. The results have been calculated for $\Delta_0=E_{SO}$,  $\mu=0.5 E_{SO}$, and $\theta=2\pi/5$ in the limit $k_B T, \omega \to 0$.}
\label{fig1}
\end{figure}

{\it Spin and charge pumping in long nanowires.}$-$ We consider the system shown in Fig.~\ref{fig1}(a) described by a Bogoliubov-de Gennes (BdG) Hamiltonian
\begin{equation}
    \mathcal{H}(t)\!=\!\Bigl[\frac{p^2}{2m}- \alpha_R\,p\,\sigma_z-\mu(x)\Bigr]\tau_z  \!+\! \bs{m}(x,t)\cdot\bs{\sigma} \!+\! \Delta(x)\tau_x,
    \label{BdGen}
\end{equation}
where ${\bs \sigma}=(\sigma_x,\sigma_y,\sigma_z)$ and ${\bs \tau}=(\tau_x,\tau_y,\tau_z)$ are Pauli matrices that act in the spin and Nambu space respectively,  $p=-i\hbar\partial_x$ is the momentum operator along the wire ($x$-direction), $m$ is the effective electron mass, $\alpha_R$ is the Rashba spin-orbit coupling strength, ${\bs m}(x,t)=m_0\Theta(x)[\sin{\theta}\cos\phi(t),\sin{\theta} \sin{\phi(t)},\cos{\theta}]$ is the magnetic exchange field induced by the precessing magnetization of a ferromagnetic insulator, $\Delta(x)=\Delta_0\Theta(x)$ is the induced superconducting order parameter, $\Theta(x)$ is the Heaviside step function, and $\mu(x)$ is the chemical potential.  We denote the chemical potentials in the lead, tunnel barrier and FI-SC nanowire as $\mu_N$, $\mu_{tun}$ and $\mu$, respectively. We consider periodic driving ${\bs m}(x,t+\mathcal{T})={\bs m}(x,t)$ with period $\mathcal{T}=2\pi/\omega$ and frequency $\omega$, and  measure energies, momenta and lengths in units of $E_{SO} = m\alpha_R^2/2$, $p_{SO}= 2m\alpha_R$ and $\ell_{SO}=\hbar/p_{SO}$, respectively. The phase diagram of this system  is shown in Fig.~\ref{fig1}(b). The gapped phases of FI-SC nanowires are classified by the one-dimensional class D  $\mathbb{Z}_2$ topological invariant which determines the existence of topologically protected MZMs at the end of the wire \cite{Kitaev2001}.  The topologically nontrivial gapped phase emerges in the regime $\sqrt{\Delta_0^2 + \mu^2}<m_0<\Delta_0/\cos\theta$ and the system is gapless if $m_0>\Delta_0/\cos\theta$. In the following we stay in the parameter regime where the FI-SC nanowire is gapped. The width of the tunnel barrier is $4\ell_{SO}$ and $\mu_N=0$. For other barrier widths see Ref.~\cite{supplementary}.  

The precessing magnetization pumps charge and spin from the FI-SC nanowire into the lead. The pumped charge $Q$ and spin $S_z$ over one cycle in the adiabatic limit can be calculated from the expression 
\cite{MoskaletsBook, Blaauboer} 
\begin{equation}
{\cal O} = \int\!dE \Bigl(\frac{\partial f}{\partial E}\Bigr)\!
\int_0^{\mathcal{T}} \!dt\, \mathrm{Im}\Bigl\{\mathrm{Tr}\Bigr[ \hat{r}^{\dagger} \hat{{\cal O}} \frac{\partial \hat{r}}{\partial t} \Bigr]\Bigr\}\,,
\label{qqBPT}
\end{equation}
where $f(E)$ is the Fermi function, the operator
 $\hat{{\cal O}}$ for the pumped charge (spin) is $\hat{Q}=e \tau_z/4\pi$ ($\hat{S}_z=\hbar \sigma_z/8\pi$),  and $ \hat{r}(E, \phi(t))$ is the instantaneous scattering matrix 
\begin{equation*}
    \hat{r} = \begin{pmatrix}
    \hat{r}_{ee} & \hat{r}_{eh} \\
    \hat{r}_{he} & \hat{r}_{hh}
    \end{pmatrix}, \ \hat{r}_{ee}=\begin{pmatrix} r_{ee}^{\uparrow\uparrow} & r_{ee}^{\uparrow\downarrow} \\ r_{ee}^{\downarrow\uparrow} & r_{ee}^{\downarrow\downarrow} \end{pmatrix}, \ \hat{r}_{he}=\begin{pmatrix} r_{he}^{\downarrow\uparrow} & r_{he}^{\downarrow\downarrow} \\ r_{he}^{\uparrow\uparrow} & r_{he}^{\uparrow\downarrow} \end{pmatrix}
\end{equation*}
accounting for the normal $\hat{r}_{ee}$ and Andreev $\hat{r}_{he}$ processes.
The other blocks can be obtained via particle-hole symmetry
$\tau_y\sigma_y \hat{r}^*(-E)\tau_y\sigma_y = \hat{r}(E)$. In a continuous operation of the device the time averages of the pumped charge and spin currents can be written, respectively, as $\langle I_e\rangle= \omega\,Q/2\pi$ and $\langle I_s\rangle=\omega\,S_z/2\pi$.

In the case of uniform precession (or arbitrary adiabatic precession) the time-dependence of the Hamiltonian can be removed by switching to the rotating frame $\mathcal{H}_{\rm rot} = U^{\dagger}\mathcal{H}_{BdG}(t)U - i\hbar U^{\dagger}\partial_t U$ via unitary transformation $U=e^{-i\phi(t)\sigma_z/2}$. The coefficients $\tilde{r}_{ee}^{\sigma \sigma'}$ and  $\tilde{r}_{he}^{\sigma \sigma'}$ can then be obtained by finding a solution $\Psi_{\rm rot}(x)$ of the time-independent Hamiltonian $\mathcal{H}_{\rm rot}$, and the scattering states in the lab-frame are obtained using $\Psi(x,t)=U\Psi_{\rm rot}(x)$. This way we find that the only  $t$-dependent reflection coefficients are
%
\begin{equation*}
    \begin{array}{cc}
    r_{ee}^{\overline{\beta}\beta}(t) = \tilde{r}_{ee}^{\overline{\beta}\beta}e^{i\beta\phi(t)}, & r_{he}^{\beta\beta}(t) = \tilde{r}_{he}^{\beta\beta}e^{ i\beta\phi(t)},
        \end{array}
\end{equation*}
where $\bar{\beta}=-\beta$, and we use $\beta=+1,-1$ and $\uparrow, \downarrow$ interchangeably.
Using these expressions we obtain
\begin{eqnarray}
Q &=& \int\!dE \Bigl(-\frac{\partial f}{\partial E}\Bigr) {\cal Q}(E), \ S_z= \int\!dE \Bigl(-\frac{\partial f}{\partial E}\Bigr) {\cal S}_z(E),  \nonumber \\
&&\hspace{-0.5cm}{\cal Q}(E)=e\bigl( |r_{ee}^{\uparrow\downarrow}|^2-|r_{ee}^{\downarrow\uparrow}|^2 + |r_{he}^{\uparrow\uparrow}|^2
-|r_{he}^{\downarrow\downarrow}|^2  \bigr),\, \nonumber \\
&&\hspace{-0.6cm} {\cal S}_z(E) = 
\hbar\bigl(|r_{ee}^{\uparrow\downarrow}|^2+|r_{ee}^{\downarrow\uparrow}|^2+ |r_{he}^{\uparrow\uparrow}|^2   +|r_{he}^{\downarrow\downarrow}|^2  \bigr)/2. \label{eq-pumping}
\end{eqnarray}
In the limit of low temperature $k_B T \to 0$ and frequency $\omega \to 0$ the pumped charge (spin) is related to the corresponding  spectral density as $Q={\cal Q}(0)$ ($S_z={\cal S}_z(0)$).
These formulas can be generalized to  arbitrary $\omega$ by evaluating the contributions of the scattering paths  in the rotating frame taking into account the spin bias terms  in the distribution functions as described in Ref.~\cite{Becerra21}. The general expression is obtained from Eq.~(\ref{eq-pumping}) by replacing $\partial f/\partial E$ with a ``quantum derivative'' $[f(E+\hbar \omega/2)-f(E-\hbar \omega/2)]/\hbar \omega$ and computing the reflection coefficients in the presence of the term $- i\hbar U^{\dagger}\partial_t U=-\hbar \omega \sigma_z/2$. In the absence of a bias voltage the spectral densities are weighted with symmetric functions in the energy integrals. Hence, in some figures it is more illustrative to plot the symmetrized spectral density ${\cal S}^s_z(E)=\bigl({\cal S}_z(E)+{\cal S}_z(-E)\bigr)/2$.

Figs.~\ref{fig1}(c) and (d) show representative results for $Q$ and $S_z$ obtained by numerically calculating the reflection matrix for long FI-SC nanowires using Kwant software package \cite{Groth_2014}. Both quantities take arbitrary (typically small) values in the case of topologically trivial wires. By tuning the parameters of the system to the topologically nontrivial phase we observe a sharp transition in $Q$ and $S_z$. Within the topologically nontrivial phase $Q$ still takes arbitrary values depending on the $m_0$, $\theta$ and tunnel barrier $\mu_{tun}$ introduced between the lead and the system, similarly as obtained in the previous studies \cite{Ojanen12}. On the other hand, $S_z$ is now robustly quantized to $\hbar/2$ at low temperatures and frequencies. 

{\it Correspondence of spin pumping, conductance and entropy.}$-$ One of the hallmarks of the nontrivial topology is the robust quantization of the differential conductance 
\begin{eqnarray}
G&=&\int\!dE \Bigl(-\frac{\partial f}{\partial E}\Bigr) {\cal G}(E),  \ 
{\cal G}(E) = 2 G_0 \mathrm{Tr}[\hat{r}_{he}^{\dagger}\hat{r}_{he}] \label{conductance}
\end{eqnarray}
to a universal value $G=2G_0$ ($G_0=e^2/h$) at small bias voltages in the tunneling regime \cite{Law09}. Moreover, in ballistic point  contacts this zero-bias peak widens into a  plateau of quantized conductance  \cite{Wimmer11}. We show that in long nanowires there exists one-to-one correspondence between the quantization of the  conductance $G=2G_0$ and the spin pumping $S_z=\hbar/2$ in the nontrivial regime. For this purpose we consider the scattering states in the lead [$x<0$ in Fig.~\ref{fig1}(a)] at energy $E=0$ and $\omega \to 0$ 
\begin{eqnarray*}
    &\Psi_{\rm rot}^{\alpha, L}(x) = \left(\begin{array}{c}\chi_{_{\alpha}} \\ 0 \end{array}\right)e^{ixk_{in}^{\alpha}} + \nonumber \\
    &\sum_{\beta}\Bigl[
    \tilde{r}_{ee}^{\beta\alpha}\left(\begin{array}{c}\chi_{_{\beta}}\\ 0 \end{array}\right)e^{-ixk_o^{\beta}}
        +\tilde{r}_{he}^{\beta\alpha}\left(\begin{array}{c} 0 \\ \chi_{_{\overline{\beta}}} \end{array}\right)e^{ixk_o^{\beta}}
        \Bigr], 
\end{eqnarray*}
composed of an incoming electron with spin-$z$ eigenvalue $\alpha=\pm 1$ in eigenstate $\chi_\alpha=(1+\alpha, 1-\alpha)^T/2$ and  $k^{\alpha}_{in} \ell_{SO}=\bigl(\alpha  \!+\! \sqrt{1\!+\! \mu_N/E_{SO} \!}\,\bigr)\big/2$, and four outgoing states with $k_o^{\beta} \ell_{SO}=\bigl(-\beta \!+\! \sqrt{1\!+\! \mu_N/E_{SO} \!}\,\bigr)\big/2$. The reflection coefficients $\tilde{r}_{ee}^{\beta \alpha}$ and $\tilde{r}_{he}^{\beta \alpha}$ are obtained by matching the solutions $\Psi_{\rm rot}^{\alpha, L}(0)=\Psi_{\rm rot}^{\alpha, R}(0)$ and $\partial_x\Psi_{\rm rot}^{\alpha, L}(0)=\partial_x\Psi_{\rm rot}^{\alpha, R}(0)$, where $\Psi_{\rm rot}^{\alpha, R}(x)$ are the evanescent states in the FI-SC nanowire [$x>0$ in Fig.~\ref{fig1}(a)]. This way we arrive to the following expressions \cite{supplementary}
\begin{eqnarray}
|\tilde{r}_{ee}^{\beta\beta}| &=&   \frac{|k_o^{\overline{\beta}} - i z_4||k_o^{\beta} -iz_4 |}{\sum_\sigma (k_o^{\sigma})^2   + 2z_4^2 }, \ 
|\tilde{r}_{ee}^{\overline{\beta}\beta}| =   \frac{(k_o^{\beta})^2 +z_4^2 }{\sum_\sigma (k_o^{\sigma})^2  + 2z_4^2}, \nonumber\\
|\tilde{r}_{he}^{\overline{\beta}\beta}| &=& 
\frac{(k_o^{\overline{\beta}})^2  +  z_4^2 }{\sum_\sigma (k_o^{\sigma})^2  + 2z_4^2 }, \
|\tilde{r}_{he}^{\beta\beta}| =  \frac{|k_o^{\overline{\beta}} + i z_4 ||k_o^{\beta} -iz_4|}{\sum_\sigma (k_o^{\sigma})^2 + 2z_4^2}, \nonumber 
\end{eqnarray}
where $z_4$ is the only root of the quartic polynomial 
\begin{equation*}
\bigg(4z_4^2 \ell_{SO}^2 +\frac{\mu}{E_{SO}}\bigg)^2 +\bigg(4z_4 \ell_{SO} - \frac{\Delta_e}{E_{SO}}\bigg)^2 -\frac{m_{0}^2 \sin^2\theta}{E_{SO}^2}=0 \label{characteristic-eq} 
\end{equation*}
having $\Re[z_4]>0$ (evanescent mode). Thus, $z_4 \in \mathbb{R}$. Here, 
$\Delta_e = \sqrt{\Delta_0^2 -m_0^2 \cos^2\theta}>0$  in the gapped phase. 
\begin{figure}
\includegraphics[width=\linewidth]{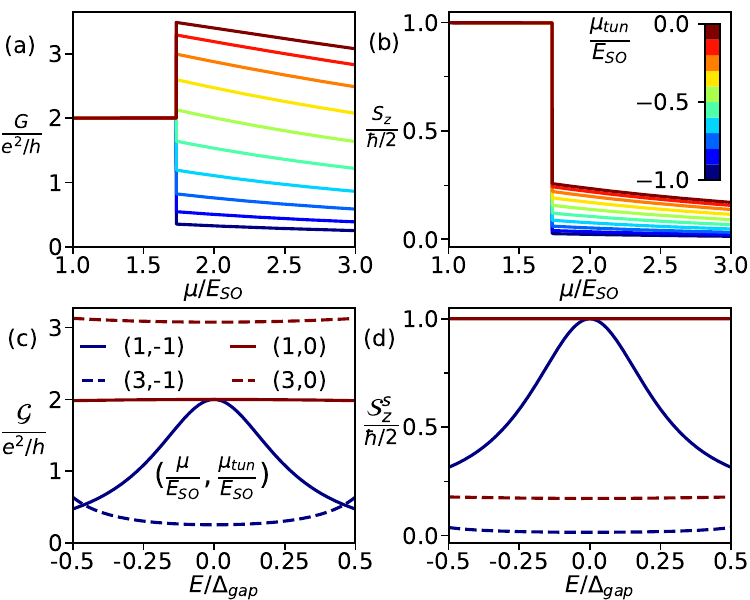}
\caption{(a) $G$ and (b) $S_z$ as a function of $\mu$ for different $\mu_{tun}$ in the limit $\omega, k_B T \to 0$. In the topologically nontrivial phase ($\mu< \sqrt{3} E_{SO}$) there is a perfect correspondence between the quantized conductance $G=2G_0$ and spin pumping $S_z=\hbar/2$, but these observables are unrelated in the trivial regime ($\mu > \sqrt{3} E_{SO}$). (c), (d) The spectral densities ${\cal G}(E)$ and ${\cal S}^s_z(E)$ in the nontrivial ($\mu=E_{SO}$) and trivial ($\mu=3 E_{SO}$) phases in the tunneling regime $\mu_{tun}=-E_{SO}$ and in the absence of tunnel barrier $\mu_{tun}=0$. The energy is normalized with the gap $\Delta_{gap}$ in the FI-SC nanowire. The model parameters are $\Delta_0 = E_{SO} $, $m_0 = 2 E_{SO}$ and  $\theta=2\pi/5$.} 
\label{cond_vs_spin}
\end{figure}
Because $|r_{ee}^{\beta\beta} | = |r_{he}^{\beta\beta}|$, it follows from Eqs.~(\ref{eq-pumping}) and (\ref{conductance}) that 
\begin{equation}
\frac{{\cal S}_z(0)}{\hbar/2}= \mathrm{Tr}[\hat{r}_{ee}^{\dagger}\hat{r}_{ee}]=2-\mathrm{Tr}[\hat{r}_{he}^{\dagger}\hat{r}_{he}]=2-\frac{{\cal G}(0)}{2G_0}.
\end{equation}
Therefore, in long nanowires there is a perfect correspondence between the quantized spin pumping ${\cal S}_z(0)=\hbar/2$ and the quantized conductance ${\cal G}(0)=2G_0$ in the topologically nontrivial regime. This correspondence is illustrated in Fig.~\ref{cond_vs_spin}. In the tunneling regime  there exists quantized peaks in ${\cal S}^s_z(E)$ and ${\cal G}(E)$ at $E=0$, which widen into plateaus upon decreasing the tunnel barrier to $\mu_{tun}=0$. Importantly,  ${\cal S}^s_z(E)$ and ${\cal G}(E)$ are unrelated to each other in the topologically trivial nanowires. We find that similar correlation exists also between quantized entropy change and spin pumping in nontrivial wires \cite{supplementary}.

\begin{figure}
\includegraphics[width=0.99\linewidth]{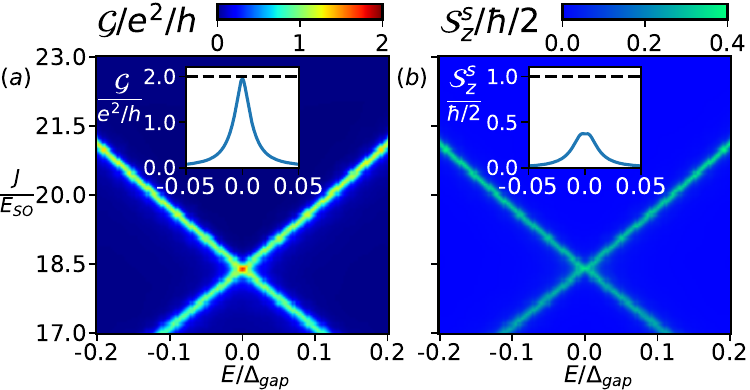}
\caption{The impurity induced Andreev bound states crossing zero energy give rise to peaks in (a) ${\cal G}(E)$ and (b) ${\cal S}^s_z(E)$ at $E=0$. The heights of the peaks are arbitrary but they can accidentally have a quantized value as shown in (a). However, the heights of the peaks in ${\cal G}(E)$ and ${\cal S}^s_z(E)$ are unrelated to each other. Therefore, accidental low-energy Andreev bound states can be distinguished from the MZMs, which give rise to correlated quantized peaks in ${\cal G}(E)$ and ${\cal S}^s_z(E)$ at $E=0$ (c.f. Fig.~\ref{cond_vs_spin}). Here the impurity states have been induced with a magnetic impurity $J \sigma_z$  located at the lattice site $x_i=6 \ell_{SO}$. The model parameters are $\Delta_0=E_{SO}$, $\mu=0$, $m_0=0.3 E_{SO}$, $\theta=\pi/2$, $\mu_{tun}=-1.5 E_{SO}$ and lattice constant $d=\ell_{SO}/4$.}
\label{Andreev-bound-states}
\end{figure}

{\it Distinguishing MZMs from Andreev bound states and quasi-MZMs.}$-$
The combined measurements of the spin pumping and conductance could also be helpful in  distinguishing non-Majorana zero-bias conductance peaks \cite{inh_Brouwer, QD_Liu,QD_Moore_one,QD_Moore_two, inh_workfunc, Chen19, qMajo_Vuik, RN14, Frolov21, ValentiniScience21} from the zero-bias peaks caused by the MZMs. To demonstrate this, we have computed the effect of non-Majorana Andreev bound states, induced by a magnetic impurity pointing along the spin $z$ direction, on the conductance and spin pumping in the trivial regime (see Fig.~\ref{Andreev-bound-states}). The impurity induced Andreev bound states can give rise to zero-energy peaks in ${\cal G}(E)$ and ${\cal S}^s_z(E)$. The heights of the peaks are arbitrary but they can accidentally have a quantized value. However, the heights of the peaks in ${\cal G}(E)$ and ${\cal S}^s_z(E)$ are unrelated to each other. Therefore accidental low-energy Andreev bound states can be distinguished from the MZMs, which give rise to correlated quantized peaks in ${\cal G}(E)$ and ${\cal S}^s_z(E)$ at $E=0$ (c.f.~Figs.~\ref{cond_vs_spin} and \ref{Andreev-bound-states}). We point out that a smooth tunnel barrier can induce two spatially separated MZMs at the lead-nanowire interface \cite{inh_Brouwer, QD_Liu, QD_Moore_one, QD_Moore_two, inh_workfunc, qMajo_Vuik}, and  in certain cases these quasi-MZMs are so weakly coupled to each other that they can mimic all properties of the MZMs, including quantized conductance, $4\pi$ Josephson effect and even the non-Abelian braiding statistics \cite{qMajo_Vuik, Fulga13}. However, the quantized spin pumping and conductance are independent of the strength of the tunnel barrier, and therefore the topological MZMs can be distinguished from quasi-MZMs by demonstrating the robustness of the quantized signatures upon lowering of the tunnel barrier \cite{supplementary}.

\begin{figure}
\includegraphics[width=0.493\linewidth]{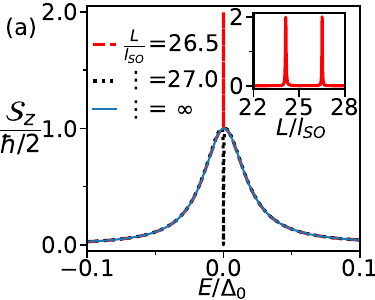}
\includegraphics[width=0.493\linewidth]{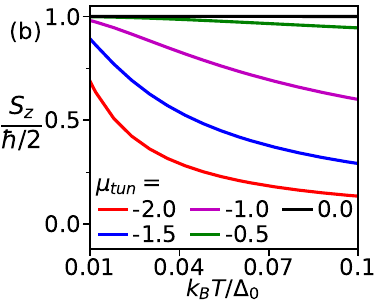}
\includegraphics[width=0.493\linewidth]{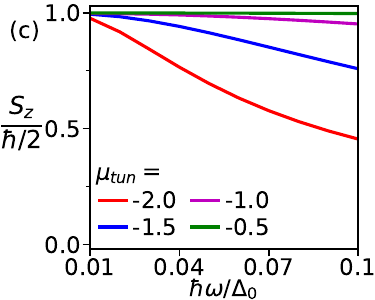}
\includegraphics[width=0.493\linewidth]{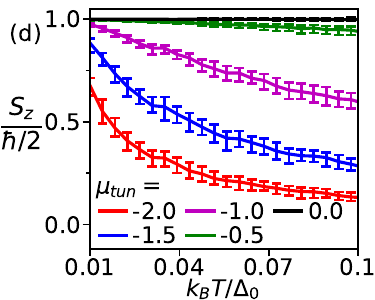}
\caption{(a) The dependence of ${\cal S}_z(E)$ on the length $L$ of the wire. The hybridization of the MZMs and the interference effects lead to appearance of a very sharp peak or a dip at very low energies. (b) $S_z$ as a function of $k_BT$ for $\omega \to 0$. (c) $S_z$ as a function of $\omega$ for $k_BT \to 0$. (d) $S_z$ as a function of $k_BT$ for $\omega \to 0$ in the presence of disorder  potential $V(x)\tau_z$, where the $V(x)$ at each lattice site $x$ are uncorrelated uniformly distributed random numbers between $[-8E_{SO}, 8E_{SO}]$ and we have used lattice constant $d=\ell_{SO}/50$. The error bars denote the $10$th and $90$th percentile values. The model parameters are $\Delta_0 = E_{SO} $, $\mu=0$, $m_0 = 2 E_{SO}$ and  $\theta=\pi/2$. In (a) $\mu_{tun} = -2E_{SO}$, in (b),(c) and (d) $L=26.5 \ell_{SO}$. }
\label{Sz-finiteL}
\end{figure}

{\it Parametric dependencies.}$-$ The dependence of ${\cal S}_z(E)$ on the length $L$ of the FI-SC wire is shown in Fig.~\ref{Sz-finiteL}(a), where $m(x)=m_0\Theta(L-x)\Theta(x)$. The overall shape  is similar to the case of $L \to \infty$, but the hybridization of the MZMs and the interference effects lead to  a very sharp dip or a peak at very low energies. Their widths decrease exponentially with $L$ so that robust quantization of $S_z$ exists at experimentally relevant temperatures and frequencies if $L$ is sufficiently large and the tunnel barrier is not too large [Figs.~\ref{Sz-finiteL}(b),(c)]. The quantization of $S_z$ is robust in the presence of disorder [Fig.~\ref{Sz-finiteL}(d)], but very strong disorder leads to large sample-to-sample fluctuations of $S_z$ at very small temperatures and frequencies due to the sharp peaks and dips in ${\cal S}_z(E)$ \cite{supplementary}. 

We can capture the essential physics behind the shape of the ${\cal S}_z(E)$ using the Mahaux-Weidenm\"{u}ller formula for the scattering matrix \cite{Mahaux69, Nilsson08, Schomerus2016}
\begin{eqnarray}
S&=&1-2 \pi i W^\dag (E-H_M+i \pi W W^\dag )^{-1} W, \nonumber\\
&& \hspace{-1.1cm}H_M=i \begin{pmatrix} 0 & E_M \\ - E_M & 0 \end{pmatrix}, W=\begin{pmatrix} w_{\uparrow}^L & w_{\downarrow}^L & w_{\uparrow}^{L*} & w_{\downarrow}^{L*} \\
                  w_{\uparrow}^R & w_{\downarrow}^R & w_{\uparrow}^{R*} & w_{\downarrow}^{R*} \end{pmatrix},
\end{eqnarray}
where $H_M$ describes the coupling $E_M$ between the left  and right  MZMs and $W$ the coupling of MZMs to the lead modes in the basis $(c^\dag_\uparrow, c^\dag_\downarrow,  -c_\uparrow, -c_\downarrow)$.
Without loss of generality we can choose $0< w^L_{\uparrow}, w^L_{\downarrow} \in \mathbb{R}$. Moreover, the couplings of the lead modes to the left MZM $w^L_{\sigma}$ are always much larger than the couplings to the right MZM $w^R_{\sigma}$. By neglecting the corrections caused by $w^R_{\sigma}$ we obtain  ${\cal G}(E)/(2 G_0)={\cal S}_z(E)/(\hbar/2)={\cal F}(E)$ \cite{supplementary}, where 
\begin{equation}
{\cal F}(E)= \frac{E^2}{E^2+(E_M^2-E^2)^2/\Gamma^2}, \ \frac{\Gamma}{2\pi}= \sum_\sigma (w^L_\sigma)^2.  \label{main:usual}
\end{equation}
This formula accurately describes the shape of the sharp dips typically occurring in the ${\cal S}_z(E)$ whereas the peaks originate from the interference effects when $E_M$ is small and the couplings $w^R_{\sigma}$ become relevant \cite{supplementary}. Our results agree with the earlierly reported dips in ${\cal G}(E)$ due to the coupling of MZMs \cite{PhysRevB.96.054520}, but interestingly the interference effects can turn the dips also into peaks in ${\cal S}_z(E)$. 

Finally, we note that the out-of-equilibrium electrons act back on the magnet, affecting its dynamics. In particular, the spin current ejected into the leads increases the Gilbert damping parameter $\tilde{\alpha}$ entering in the Landau-Lifshitz-Gilbert equation that describes the magnetization dynamics \cite{TserkovnyakRMP2005}. We find that in our setup the change in the Gilbert damping is $\Delta\tilde{\alpha}\propto S_z/\sin^2{\theta}$, which could be used to extract experimentally the pumped spin $S_z$ from ferromagnetic resonance measurements \cite{supplementary}. Alternatively, it could be possible to fabricate nanostructures where the pumped spin current could be measured by utilizing the inverse spin Hall effect \cite{Tinkham06, ZhangNP12}. 

{\it Conclusions.}$-$ We have shown that spin pumping is robustly quantized and there exists one-to-one correspondence to the quantized conductance and the entropy change in  topologically nontrivial nanowires, so that the observation of correlated and quantized peaks in the conductance, entropy change, and spin pumping would provide strong evidence of topological superconductivity. In the adiabatic limit our results  can be generalized to  arbitrary trajectory in $(\theta(t), \phi(t))$-space as long as the system stays gapped and topologically nontrivial, and the azimuthal angle $\phi(t)$ winds around the $z$-axis. We have neglected the non-equilibrium effects such as the dynamical spin accumulation at the interface of the FI-SC nanowire and the normal lead. Such effects can lead to nonlinear corrections to the calculated spin current as a function of $\omega$ but they do not affect the quantized response at low frequencies where the  spin current is proportional to $\omega$. The maximum frequency is limited by the topological energy gap, so we estimate that $\omega \lesssim 10$ GHz \cite{2022arXiv220702472A}.

\begin{acknowledgements}
{\em Acknowledgements.}
The work is supported by the Foundation for Polish Science through the IRA Programme
co-financed by EU within SG OP and the 
Academy of Finland Project No.~331094. 
We acknowledge
the computational resources provided by
the Aalto Science-IT project and the access to the computing facilities of  the Interdisciplinary Centre for Mathematical and Computational Modelling (ICM), University of Warsaw, under grant no G89-1264.
\end{acknowledgements}

\bibliography{amainrefbiblio}

\providecommand{\noopsort}[1]{}\providecommand{\singleletter}[1]{#1}%
\begin{thebibliography}{62}%
\makeatletter
\providecommand \@ifxundefined [1]{%
 \@ifx{#1\undefined}
}%
\providecommand \@ifnum [1]{%
 \ifnum #1\expandafter \@firstoftwo
 \else \expandafter \@secondoftwo
 \fi
}%
\providecommand \@ifx [1]{%
 \ifx #1\expandafter \@firstoftwo
 \else \expandafter \@secondoftwo
 \fi
}%
\providecommand \natexlab [1]{#1}%
\providecommand \enquote  [1]{``#1''}%
\providecommand \bibnamefont  [1]{#1}%
\providecommand \bibfnamefont [1]{#1}%
\providecommand \citenamefont [1]{#1}%
\providecommand \href@noop [0]{\@secondoftwo}%
\providecommand \href [0]{\begingroup \@sanitize@url \@href}%
\providecommand \@href[1]{\@@startlink{#1}\@@href}%
\providecommand \@@href[1]{\endgroup#1\@@endlink}%
\providecommand \@sanitize@url [0]{\catcode `\\12\catcode `\$12\catcode
  `\&12\catcode `\#12\catcode `\^12\catcode `\_12\catcode `\%12\relax}%
\providecommand \@@startlink[1]{}%
\providecommand \@@endlink[0]{}%
\providecommand \url  [0]{\begingroup\@sanitize@url \@url }%
\providecommand \@url [1]{\endgroup\@href {#1}{\urlprefix }}%
\providecommand \urlprefix  [0]{URL }%
\providecommand \Eprint [0]{\href }%
\providecommand \doibase [0]{https://doi.org/}%
\providecommand \selectlanguage [0]{\@gobble}%
\providecommand \bibinfo  [0]{\@secondoftwo}%
\providecommand \bibfield  [0]{\@secondoftwo}%
\providecommand \translation [1]{[#1]}%
\providecommand \BibitemOpen [0]{}%
\providecommand \bibitemStop [0]{}%
\providecommand \bibitemNoStop [0]{.\EOS\space}%
\providecommand \EOS [0]{\spacefactor3000\relax}%
\providecommand \BibitemShut  [1]{\csname bibitem#1\endcsname}%
\let\auto@bib@innerbib\@empty
\bibitem [{\citenamefont {Nayak}\ \emph {et~al.}(2008)\citenamefont {Nayak},
  \citenamefont {Simon}, \citenamefont {Stern}, \citenamefont {Freedman},\ and\
  \citenamefont {Das~Sarma}}]{Nayak08}%
  \BibitemOpen
  \bibfield  {author} {\bibinfo {author} {\bibfnamefont {C.}~\bibnamefont
  {Nayak}}, \bibinfo {author} {\bibfnamefont {S.~H.}\ \bibnamefont {Simon}},
  \bibinfo {author} {\bibfnamefont {A.}~\bibnamefont {Stern}}, \bibinfo
  {author} {\bibfnamefont {M.}~\bibnamefont {Freedman}},\ and\ \bibinfo
  {author} {\bibfnamefont {S.}~\bibnamefont {Das~Sarma}},\ }\bibfield  {title}
  {\bibinfo {title} {{Non-Abelian anyons and topological quantum
  computation}},\ }\href {https://doi.org/10.1103/RevModPhys.80.1083}
  {\bibfield  {journal} {\bibinfo  {journal} {Rev. Mod. Phys.}\ }\textbf
  {\bibinfo {volume} {80}},\ \bibinfo {pages} {1083} (\bibinfo {year}
  {2008})}\BibitemShut {NoStop}%
\bibitem [{\citenamefont {Hyart}\ \emph {et~al.}(2013)\citenamefont {Hyart},
  \citenamefont {van Heck}, \citenamefont {Fulga}, \citenamefont {Burrello},
  \citenamefont {Akhmerov},\ and\ \citenamefont {Beenakker}}]{Hyart13}%
  \BibitemOpen
  \bibfield  {author} {\bibinfo {author} {\bibfnamefont {T.}~\bibnamefont
  {Hyart}}, \bibinfo {author} {\bibfnamefont {B.}~\bibnamefont {van Heck}},
  \bibinfo {author} {\bibfnamefont {I.~C.}\ \bibnamefont {Fulga}}, \bibinfo
  {author} {\bibfnamefont {M.}~\bibnamefont {Burrello}}, \bibinfo {author}
  {\bibfnamefont {A.~R.}\ \bibnamefont {Akhmerov}},\ and\ \bibinfo {author}
  {\bibfnamefont {C.~W.~J.}\ \bibnamefont {Beenakker}},\ }\bibfield  {title}
  {\bibinfo {title} {{Flux-controlled quantum computation with Majorana
  fermions}},\ }\href {https://doi.org/10.1103/PhysRevB.88.035121} {\bibfield
  {journal} {\bibinfo  {journal} {Phys. Rev. B}\ }\textbf {\bibinfo {volume}
  {88}},\ \bibinfo {pages} {035121} (\bibinfo {year} {2013})}\BibitemShut
  {NoStop}%
\bibitem [{\citenamefont {Aasen}\ \emph {et~al.}(2016)\citenamefont {Aasen},
  \citenamefont {Hell}, \citenamefont {Mishmash}, \citenamefont {Higginbotham},
  \citenamefont {Danon}, \citenamefont {Leijnse}, \citenamefont {Jespersen},
  \citenamefont {Folk}, \citenamefont {Marcus}, \citenamefont {Flensberg},\
  and\ \citenamefont {Alicea}}]{Aasen16}%
  \BibitemOpen
  \bibfield  {author} {\bibinfo {author} {\bibfnamefont {D.}~\bibnamefont
  {Aasen}}, \bibinfo {author} {\bibfnamefont {M.}~\bibnamefont {Hell}},
  \bibinfo {author} {\bibfnamefont {R.~V.}\ \bibnamefont {Mishmash}}, \bibinfo
  {author} {\bibfnamefont {A.}~\bibnamefont {Higginbotham}}, \bibinfo {author}
  {\bibfnamefont {J.}~\bibnamefont {Danon}}, \bibinfo {author} {\bibfnamefont
  {M.}~\bibnamefont {Leijnse}}, \bibinfo {author} {\bibfnamefont {T.~S.}\
  \bibnamefont {Jespersen}}, \bibinfo {author} {\bibfnamefont {J.~A.}\
  \bibnamefont {Folk}}, \bibinfo {author} {\bibfnamefont {C.~M.}\ \bibnamefont
  {Marcus}}, \bibinfo {author} {\bibfnamefont {K.}~\bibnamefont {Flensberg}},\
  and\ \bibinfo {author} {\bibfnamefont {J.}~\bibnamefont {Alicea}},\
  }\bibfield  {title} {\bibinfo {title} {{Milestones Toward Majorana-Based
  Quantum Computing}},\ }\href {https://doi.org/10.1103/PhysRevX.6.031016}
  {\bibfield  {journal} {\bibinfo  {journal} {Phys. Rev. X}\ }\textbf {\bibinfo
  {volume} {6}},\ \bibinfo {pages} {031016} (\bibinfo {year}
  {2016})}\BibitemShut {NoStop}%
\bibitem [{\citenamefont {Karzig}\ \emph {et~al.}(2017)\citenamefont {Karzig},
  \citenamefont {Knapp}, \citenamefont {Lutchyn}, \citenamefont {Bonderson},
  \citenamefont {Hastings}, \citenamefont {Nayak}, \citenamefont {Alicea},
  \citenamefont {Flensberg}, \citenamefont {Plugge}, \citenamefont {Oreg},
  \citenamefont {Marcus},\ and\ \citenamefont {Freedman}}]{Karzig17}%
  \BibitemOpen
  \bibfield  {author} {\bibinfo {author} {\bibfnamefont {T.}~\bibnamefont
  {Karzig}}, \bibinfo {author} {\bibfnamefont {C.}~\bibnamefont {Knapp}},
  \bibinfo {author} {\bibfnamefont {R.~M.}\ \bibnamefont {Lutchyn}}, \bibinfo
  {author} {\bibfnamefont {P.}~\bibnamefont {Bonderson}}, \bibinfo {author}
  {\bibfnamefont {M.~B.}\ \bibnamefont {Hastings}}, \bibinfo {author}
  {\bibfnamefont {C.}~\bibnamefont {Nayak}}, \bibinfo {author} {\bibfnamefont
  {J.}~\bibnamefont {Alicea}}, \bibinfo {author} {\bibfnamefont
  {K.}~\bibnamefont {Flensberg}}, \bibinfo {author} {\bibfnamefont
  {S.}~\bibnamefont {Plugge}}, \bibinfo {author} {\bibfnamefont
  {Y.}~\bibnamefont {Oreg}}, \bibinfo {author} {\bibfnamefont {C.~M.}\
  \bibnamefont {Marcus}},\ and\ \bibinfo {author} {\bibfnamefont {M.~H.}\
  \bibnamefont {Freedman}},\ }\bibfield  {title} {\bibinfo {title} {{Scalable
  designs for quasiparticle-poisoning-protected topological quantum computation
  with Majorana zero modes}},\ }\href
  {https://doi.org/10.1103/PhysRevB.95.235305} {\bibfield  {journal} {\bibinfo
  {journal} {Phys. Rev. B}\ }\textbf {\bibinfo {volume} {95}},\ \bibinfo
  {pages} {235305} (\bibinfo {year} {2017})}\BibitemShut {NoStop}%
\bibitem [{\citenamefont {Beenakker}(2020)}]{Beenakker20}%
  \BibitemOpen
  \bibfield  {author} {\bibinfo {author} {\bibfnamefont {C.~W.~J.}\
  \bibnamefont {Beenakker}},\ }\bibfield  {title} {\bibinfo {title} {{Search
  for non-Abelian Majorana braiding statistics in superconductors}},\ }\href
  {https://doi.org/10.21468/SciPostPhysLectNotes.15} {\bibfield  {journal}
  {\bibinfo  {journal} {SciPost Phys. Lect. Notes}\ ,\ \bibinfo {pages} {15}}
  (\bibinfo {year} {2020})}\BibitemShut {NoStop}%
\bibitem [{\citenamefont {Lutchyn}\ \emph {et~al.}(2010)\citenamefont
  {Lutchyn}, \citenamefont {Sau},\ and\ \citenamefont {Das~Sarma}}]{Lutchyn10}%
  \BibitemOpen
  \bibfield  {author} {\bibinfo {author} {\bibfnamefont {R.~M.}\ \bibnamefont
  {Lutchyn}}, \bibinfo {author} {\bibfnamefont {J.~D.}\ \bibnamefont {Sau}},\
  and\ \bibinfo {author} {\bibfnamefont {S.}~\bibnamefont {Das~Sarma}},\
  }\bibfield  {title} {\bibinfo {title} {{Majorana Fermions and a Topological
  Phase Transition in Semiconductor-Superconductor Heterostructures}},\ }\href
  {https://doi.org/10.1103/PhysRevLett.105.077001} {\bibfield  {journal}
  {\bibinfo  {journal} {Phys. Rev. Lett.}\ }\textbf {\bibinfo {volume} {105}},\
  \bibinfo {pages} {077001} (\bibinfo {year} {2010})}\BibitemShut {NoStop}%
\bibitem [{\citenamefont {Oreg}\ \emph {et~al.}(2010)\citenamefont {Oreg},
  \citenamefont {Refael},\ and\ \citenamefont {von Oppen}}]{Oreg10}%
  \BibitemOpen
  \bibfield  {author} {\bibinfo {author} {\bibfnamefont {Y.}~\bibnamefont
  {Oreg}}, \bibinfo {author} {\bibfnamefont {G.}~\bibnamefont {Refael}},\ and\
  \bibinfo {author} {\bibfnamefont {F.}~\bibnamefont {von Oppen}},\ }\bibfield
  {title} {\bibinfo {title} {{Helical Liquids and Majorana Bound States in
  Quantum Wires}},\ }\href {https://doi.org/10.1103/PhysRevLett.105.177002}
  {\bibfield  {journal} {\bibinfo  {journal} {Phys. Rev. Lett.}\ }\textbf
  {\bibinfo {volume} {105}},\ \bibinfo {pages} {177002} (\bibinfo {year}
  {2010})}\BibitemShut {NoStop}%
\bibitem [{\citenamefont {Law}\ \emph {et~al.}(2009)\citenamefont {Law},
  \citenamefont {Lee},\ and\ \citenamefont {Ng}}]{Law09}%
  \BibitemOpen
  \bibfield  {author} {\bibinfo {author} {\bibfnamefont {K.~T.}\ \bibnamefont
  {Law}}, \bibinfo {author} {\bibfnamefont {P.~A.}\ \bibnamefont {Lee}},\ and\
  \bibinfo {author} {\bibfnamefont {T.~K.}\ \bibnamefont {Ng}},\ }\bibfield
  {title} {\bibinfo {title} {{Majorana Fermion Induced Resonant Andreev
  Reflection}},\ }\href {https://doi.org/10.1103/PhysRevLett.103.237001}
  {\bibfield  {journal} {\bibinfo  {journal} {Phys. Rev. Lett.}\ }\textbf
  {\bibinfo {volume} {103}},\ \bibinfo {pages} {237001} (\bibinfo {year}
  {2009})}\BibitemShut {NoStop}%
\bibitem [{\citenamefont {Wimmer}\ \emph {et~al.}(2011)\citenamefont {Wimmer},
  \citenamefont {Akhmerov}, \citenamefont {Dahlhaus},\ and\ \citenamefont
  {Beenakker}}]{Wimmer11}%
  \BibitemOpen
  \bibfield  {author} {\bibinfo {author} {\bibfnamefont {M.}~\bibnamefont
  {Wimmer}}, \bibinfo {author} {\bibfnamefont {A.~R.}\ \bibnamefont
  {Akhmerov}}, \bibinfo {author} {\bibfnamefont {J.~P.}\ \bibnamefont
  {Dahlhaus}},\ and\ \bibinfo {author} {\bibfnamefont {C.~W.~J.}\ \bibnamefont
  {Beenakker}},\ }\bibfield  {title} {\bibinfo {title} {{Quantum point contact
  as a probe of a topological superconductor}},\ }\href
  {https://doi.org/10.1088/1367-2630/13/5/053016} {\bibfield  {journal}
  {\bibinfo  {journal} {New Journal of Physics}\ }\textbf {\bibinfo {volume}
  {13}},\ \bibinfo {pages} {053016} (\bibinfo {year} {2011})}\BibitemShut
  {NoStop}%
\bibitem [{\citenamefont {Mourik}\ \emph {et~al.}(2012)\citenamefont {Mourik},
  \citenamefont {Zuo}, \citenamefont {Frolov}, \citenamefont {Plissard},
  \citenamefont {Bakkers},\ and\ \citenamefont {Kouwenhoven}}]{Mourik12}%
  \BibitemOpen
  \bibfield  {author} {\bibinfo {author} {\bibfnamefont {V.}~\bibnamefont
  {Mourik}}, \bibinfo {author} {\bibfnamefont {K.}~\bibnamefont {Zuo}},
  \bibinfo {author} {\bibfnamefont {S.~M.}\ \bibnamefont {Frolov}}, \bibinfo
  {author} {\bibfnamefont {S.~R.}\ \bibnamefont {Plissard}}, \bibinfo {author}
  {\bibfnamefont {E.~P. A.~M.}\ \bibnamefont {Bakkers}},\ and\ \bibinfo
  {author} {\bibfnamefont {L.~P.}\ \bibnamefont {Kouwenhoven}},\ }\bibfield
  {title} {\bibinfo {title} {{Signatures of Majorana Fermions in Hybrid
  Superconductor-Semiconductor Nanowire Devices}},\ }\href
  {https://doi.org/10.1126/science.1222360} {\bibfield  {journal} {\bibinfo
  {journal} {Science}\ }\textbf {\bibinfo {volume} {336}},\ \bibinfo {pages}
  {1003} (\bibinfo {year} {2012})}\BibitemShut {NoStop}%
\bibitem [{\citenamefont {Kells}\ \emph {et~al.}(2012)\citenamefont {Kells},
  \citenamefont {Meidan},\ and\ \citenamefont {Brouwer}}]{inh_Brouwer}%
  \BibitemOpen
  \bibfield  {author} {\bibinfo {author} {\bibfnamefont {G.}~\bibnamefont
  {Kells}}, \bibinfo {author} {\bibfnamefont {D.}~\bibnamefont {Meidan}},\ and\
  \bibinfo {author} {\bibfnamefont {P.~W.}\ \bibnamefont {Brouwer}},\
  }\bibfield  {title} {\bibinfo {title} {Near-zero-energy end states in
  topologically trivial spin-orbit coupled superconducting nanowires with a
  smooth confinement},\ }\href {https://doi.org/10.1103/PhysRevB.86.100503}
  {\bibfield  {journal} {\bibinfo  {journal} {Phys. Rev. B}\ }\textbf {\bibinfo
  {volume} {86}},\ \bibinfo {pages} {100503(R)} (\bibinfo {year}
  {2012})}\BibitemShut {NoStop}%
\bibitem [{\citenamefont {Liu}\ \emph {et~al.}(2017{\natexlab{a}})\citenamefont
  {Liu}, \citenamefont {Sau}, \citenamefont {Stanescu},\ and\ \citenamefont
  {Das~Sarma}}]{QD_Liu}%
  \BibitemOpen
  \bibfield  {author} {\bibinfo {author} {\bibfnamefont {C.-X.}\ \bibnamefont
  {Liu}}, \bibinfo {author} {\bibfnamefont {J.~D.}\ \bibnamefont {Sau}},
  \bibinfo {author} {\bibfnamefont {T.~D.}\ \bibnamefont {Stanescu}},\ and\
  \bibinfo {author} {\bibfnamefont {S.}~\bibnamefont {Das~Sarma}},\ }\bibfield
  {title} {\bibinfo {title} {{Andreev bound states versus Majorana bound states
  in quantum dot-nanowire-superconductor hybrid structures: Trivial versus
  topological zero-bias conductance peaks}},\ }\href
  {https://doi.org/10.1103/PhysRevB.96.075161} {\bibfield  {journal} {\bibinfo
  {journal} {Phys. Rev. B}\ }\textbf {\bibinfo {volume} {96}},\ \bibinfo
  {pages} {075161} (\bibinfo {year} {2017}{\natexlab{a}})}\BibitemShut
  {NoStop}%
\bibitem [{\citenamefont {Moore}\ \emph
  {et~al.}(2018{\natexlab{a}})\citenamefont {Moore}, \citenamefont {Stanescu},\
  and\ \citenamefont {Tewari}}]{QD_Moore_one}%
  \BibitemOpen
  \bibfield  {author} {\bibinfo {author} {\bibfnamefont {C.}~\bibnamefont
  {Moore}}, \bibinfo {author} {\bibfnamefont {T.~D.}\ \bibnamefont
  {Stanescu}},\ and\ \bibinfo {author} {\bibfnamefont {S.}~\bibnamefont
  {Tewari}},\ }\bibfield  {title} {\bibinfo {title} {{Two-terminal charge
  tunneling: Disentangling Majorana zero modes from partially separated Andreev
  bound states in semiconductor-superconductor heterostructures}},\ }\href
  {https://doi.org/10.1103/PhysRevB.97.165302} {\bibfield  {journal} {\bibinfo
  {journal} {Phys. Rev. B}\ }\textbf {\bibinfo {volume} {97}},\ \bibinfo
  {pages} {165302} (\bibinfo {year} {2018}{\natexlab{a}})}\BibitemShut
  {NoStop}%
\bibitem [{\citenamefont {Moore}\ \emph
  {et~al.}(2018{\natexlab{b}})\citenamefont {Moore}, \citenamefont {Zeng},
  \citenamefont {Stanescu},\ and\ \citenamefont {Tewari}}]{QD_Moore_two}%
  \BibitemOpen
  \bibfield  {author} {\bibinfo {author} {\bibfnamefont {C.}~\bibnamefont
  {Moore}}, \bibinfo {author} {\bibfnamefont {C.}~\bibnamefont {Zeng}},
  \bibinfo {author} {\bibfnamefont {T.~D.}\ \bibnamefont {Stanescu}},\ and\
  \bibinfo {author} {\bibfnamefont {S.}~\bibnamefont {Tewari}},\ }\bibfield
  {title} {\bibinfo {title} {{Quantized zero-bias conductance plateau in
  semiconductor-superconductor heterostructures without topological Majorana
  zero modes}},\ }\href {https://doi.org/10.1103/PhysRevB.98.155314} {\bibfield
   {journal} {\bibinfo  {journal} {Phys. Rev. B}\ }\textbf {\bibinfo {volume}
  {98}},\ \bibinfo {pages} {155314} (\bibinfo {year}
  {2018}{\natexlab{b}})}\BibitemShut {NoStop}%
\bibitem [{\citenamefont {Woods}\ \emph {et~al.}(2018)\citenamefont {Woods},
  \citenamefont {Stanescu},\ and\ \citenamefont {Das~Sarma}}]{inh_workfunc}%
  \BibitemOpen
  \bibfield  {author} {\bibinfo {author} {\bibfnamefont {B.~D.}\ \bibnamefont
  {Woods}}, \bibinfo {author} {\bibfnamefont {T.~D.}\ \bibnamefont
  {Stanescu}},\ and\ \bibinfo {author} {\bibfnamefont {S.}~\bibnamefont
  {Das~Sarma}},\ }\bibfield  {title} {\bibinfo {title} {{Effective theory
  approach to the Schr\"odinger-Poisson problem in semiconductor Majorana
  devices}},\ }\href {https://doi.org/10.1103/PhysRevB.98.035428} {\bibfield
  {journal} {\bibinfo  {journal} {Phys. Rev. B}\ }\textbf {\bibinfo {volume}
  {98}},\ \bibinfo {pages} {035428} (\bibinfo {year} {2018})}\BibitemShut
  {NoStop}%
\bibitem [{\citenamefont {Chen}\ \emph {et~al.}(2019)\citenamefont {Chen},
  \citenamefont {Woods}, \citenamefont {Yu}, \citenamefont {Hocevar},
  \citenamefont {Car}, \citenamefont {Plissard}, \citenamefont {Bakkers},
  \citenamefont {Stanescu},\ and\ \citenamefont {Frolov}}]{Chen19}%
  \BibitemOpen
  \bibfield  {author} {\bibinfo {author} {\bibfnamefont {J.}~\bibnamefont
  {Chen}}, \bibinfo {author} {\bibfnamefont {B.~D.}\ \bibnamefont {Woods}},
  \bibinfo {author} {\bibfnamefont {P.}~\bibnamefont {Yu}}, \bibinfo {author}
  {\bibfnamefont {M.}~\bibnamefont {Hocevar}}, \bibinfo {author} {\bibfnamefont
  {D.}~\bibnamefont {Car}}, \bibinfo {author} {\bibfnamefont {S.~R.}\
  \bibnamefont {Plissard}}, \bibinfo {author} {\bibfnamefont {E.~P. A.~M.}\
  \bibnamefont {Bakkers}}, \bibinfo {author} {\bibfnamefont {T.~D.}\
  \bibnamefont {Stanescu}},\ and\ \bibinfo {author} {\bibfnamefont {S.~M.}\
  \bibnamefont {Frolov}},\ }\bibfield  {title} {\bibinfo {title} {{Ubiquitous
  Non-Majorana Zero-Bias Conductance Peaks in Nanowire Devices}},\ }\href
  {https://doi.org/10.1103/PhysRevLett.123.107703} {\bibfield  {journal}
  {\bibinfo  {journal} {Phys. Rev. Lett.}\ }\textbf {\bibinfo {volume} {123}},\
  \bibinfo {pages} {107703} (\bibinfo {year} {2019})}\BibitemShut {NoStop}%
\bibitem [{\citenamefont {Vuik}\ \emph {et~al.}(2019)\citenamefont {Vuik},
  \citenamefont {Nijholt}, \citenamefont {Akhmerov},\ and\ \citenamefont
  {Wimmer}}]{qMajo_Vuik}%
  \BibitemOpen
  \bibfield  {author} {\bibinfo {author} {\bibfnamefont {A.}~\bibnamefont
  {Vuik}}, \bibinfo {author} {\bibfnamefont {B.}~\bibnamefont {Nijholt}},
  \bibinfo {author} {\bibfnamefont {A.~R.}\ \bibnamefont {Akhmerov}},\ and\
  \bibinfo {author} {\bibfnamefont {M.}~\bibnamefont {Wimmer}},\ }\bibfield
  {title} {\bibinfo {title} {{Reproducing topological properties with
  quasi-Majorana states}},\ }\href
  {https://doi.org/10.21468/SciPostPhys.7.5.061} {\bibfield  {journal}
  {\bibinfo  {journal} {SciPost Phys.}\ }\textbf {\bibinfo {volume} {7}},\
  \bibinfo {pages} {061} (\bibinfo {year} {2019})}\BibitemShut {NoStop}%
\bibitem [{\citenamefont {Prada}\ \emph {et~al.}(2020)\citenamefont {Prada},
  \citenamefont {San-Jose}, \citenamefont {de~Moor}, \citenamefont {Geresdi},
  \citenamefont {Lee}, \citenamefont {Klinovaja}, \citenamefont {Loss},
  \citenamefont {Nygård}, \citenamefont {Aguado},\ and\ \citenamefont
  {Kouwenhoven}}]{RN14}%
  \BibitemOpen
  \bibfield  {author} {\bibinfo {author} {\bibfnamefont {E.}~\bibnamefont
  {Prada}}, \bibinfo {author} {\bibfnamefont {P.}~\bibnamefont {San-Jose}},
  \bibinfo {author} {\bibfnamefont {M.~W.~A.}\ \bibnamefont {de~Moor}},
  \bibinfo {author} {\bibfnamefont {A.}~\bibnamefont {Geresdi}}, \bibinfo
  {author} {\bibfnamefont {E.~J.~H.}\ \bibnamefont {Lee}}, \bibinfo {author}
  {\bibfnamefont {J.}~\bibnamefont {Klinovaja}}, \bibinfo {author}
  {\bibfnamefont {D.}~\bibnamefont {Loss}}, \bibinfo {author} {\bibfnamefont
  {J.}~\bibnamefont {Nygård}}, \bibinfo {author} {\bibfnamefont
  {R.}~\bibnamefont {Aguado}},\ and\ \bibinfo {author} {\bibfnamefont {L.~P.}\
  \bibnamefont {Kouwenhoven}},\ }\bibfield  {title} {\bibinfo {title} {{From
  Andreev to Majorana bound states in hybrid superconductor–semiconductor
  nanowires}},\ }\href {https://doi.org/10.1038/s42254-020-0228-y} {\bibfield
  {journal} {\bibinfo  {journal} {Nature Reviews Physics}\ }\textbf {\bibinfo
  {volume} {2}},\ \bibinfo {pages} {575} (\bibinfo {year} {2020})}\BibitemShut
  {NoStop}%
\bibitem [{\citenamefont {Yu}\ \emph {et~al.}(2021)\citenamefont {Yu},
  \citenamefont {Chen}, \citenamefont {Gomanko}, \citenamefont {Badawy},
  \citenamefont {Bakkers}, \citenamefont {Zuo}, \citenamefont {Mourik},\ and\
  \citenamefont {Frolov}}]{Frolov21}%
  \BibitemOpen
  \bibfield  {author} {\bibinfo {author} {\bibfnamefont {P.}~\bibnamefont
  {Yu}}, \bibinfo {author} {\bibfnamefont {J.}~\bibnamefont {Chen}}, \bibinfo
  {author} {\bibfnamefont {M.}~\bibnamefont {Gomanko}}, \bibinfo {author}
  {\bibfnamefont {G.}~\bibnamefont {Badawy}}, \bibinfo {author} {\bibfnamefont
  {E.~P. A.~M.}\ \bibnamefont {Bakkers}}, \bibinfo {author} {\bibfnamefont
  {K.}~\bibnamefont {Zuo}}, \bibinfo {author} {\bibfnamefont {V.}~\bibnamefont
  {Mourik}},\ and\ \bibinfo {author} {\bibfnamefont {S.~M.}\ \bibnamefont
  {Frolov}},\ }\bibfield  {title} {\bibinfo {title} {{Non-Majorana states yield
  nearly quantized conductance in proximatized nanowires}},\ }\href
  {https://doi.org/10.1038/s41567-020-01107-w} {\bibfield  {journal} {\bibinfo
  {journal} {Nature Physics}\ }\textbf {\bibinfo {volume} {17}},\ \bibinfo
  {pages} {482} (\bibinfo {year} {2021})}\BibitemShut {NoStop}%
\bibitem [{\citenamefont {Valentini}\ \emph {et~al.}(2021)\citenamefont
  {Valentini}, \citenamefont {Peñaranda}, \citenamefont {Hofmann},
  \citenamefont {Brauns}, \citenamefont {Hauschild}, \citenamefont {Krogstrup},
  \citenamefont {San-Jose}, \citenamefont {Prada}, \citenamefont {Aguado},\
  and\ \citenamefont {Katsaros}}]{ValentiniScience21}%
  \BibitemOpen
  \bibfield  {author} {\bibinfo {author} {\bibfnamefont {M.}~\bibnamefont
  {Valentini}}, \bibinfo {author} {\bibfnamefont {F.}~\bibnamefont
  {Peñaranda}}, \bibinfo {author} {\bibfnamefont {A.}~\bibnamefont {Hofmann}},
  \bibinfo {author} {\bibfnamefont {M.}~\bibnamefont {Brauns}}, \bibinfo
  {author} {\bibfnamefont {R.}~\bibnamefont {Hauschild}}, \bibinfo {author}
  {\bibfnamefont {P.}~\bibnamefont {Krogstrup}}, \bibinfo {author}
  {\bibfnamefont {P.}~\bibnamefont {San-Jose}}, \bibinfo {author}
  {\bibfnamefont {E.}~\bibnamefont {Prada}}, \bibinfo {author} {\bibfnamefont
  {R.}~\bibnamefont {Aguado}},\ and\ \bibinfo {author} {\bibfnamefont
  {G.}~\bibnamefont {Katsaros}},\ }\bibfield  {title} {\bibinfo {title}
  {{Nontopological zero-bias peaks in full-shell nanowires induced by
  flux-tunable Andreev states}},\ }\href
  {https://doi.org/10.1126/science.abf1513} {\bibfield  {journal} {\bibinfo
  {journal} {Science}\ }\textbf {\bibinfo {volume} {373}},\ \bibinfo {pages}
  {82} (\bibinfo {year} {2021})}\BibitemShut {NoStop}%
\bibitem [{\citenamefont {{Pikulin}}\ \emph {et~al.}()\citenamefont
  {{Pikulin}}, \citenamefont {{van Heck}}, \citenamefont {{Karzig}},
  \citenamefont {{Martinez}}, \citenamefont {{Nijholt}}, \citenamefont
  {{Laeven}}, \citenamefont {{Winkler}}, \citenamefont {{Watson}},
  \citenamefont {{Heedt}}, \citenamefont {{Temurhan}}, \citenamefont
  {{Svidenko}}, \citenamefont {{Lutchyn}}, \citenamefont {{Thomas}},
  \citenamefont {{de Lange}}, \citenamefont {{Casparis}},\ and\ \citenamefont
  {{Nayak}}}]{Pikulin21}%
  \BibitemOpen
  \bibfield  {author} {\bibinfo {author} {\bibfnamefont {D.~I.}\ \bibnamefont
  {{Pikulin}}}, \bibinfo {author} {\bibfnamefont {B.}~\bibnamefont {{van
  Heck}}}, \bibinfo {author} {\bibfnamefont {T.}~\bibnamefont {{Karzig}}},
  \bibinfo {author} {\bibfnamefont {E.~A.}\ \bibnamefont {{Martinez}}},
  \bibinfo {author} {\bibfnamefont {B.}~\bibnamefont {{Nijholt}}}, \bibinfo
  {author} {\bibfnamefont {T.}~\bibnamefont {{Laeven}}}, \bibinfo {author}
  {\bibfnamefont {G.~W.}\ \bibnamefont {{Winkler}}}, \bibinfo {author}
  {\bibfnamefont {J.~D.}\ \bibnamefont {{Watson}}}, \bibinfo {author}
  {\bibfnamefont {S.}~\bibnamefont {{Heedt}}}, \bibinfo {author} {\bibfnamefont
  {M.}~\bibnamefont {{Temurhan}}}, \bibinfo {author} {\bibfnamefont
  {V.}~\bibnamefont {{Svidenko}}}, \bibinfo {author} {\bibfnamefont {R.~M.}\
  \bibnamefont {{Lutchyn}}}, \bibinfo {author} {\bibfnamefont {M.}~\bibnamefont
  {{Thomas}}}, \bibinfo {author} {\bibfnamefont {G.}~\bibnamefont {{de
  Lange}}}, \bibinfo {author} {\bibfnamefont {L.}~\bibnamefont {{Casparis}}},\
  and\ \bibinfo {author} {\bibfnamefont {C.}~\bibnamefont {{Nayak}}},\
  }\bibfield  {title} {\bibinfo {title} {{Protocol to identify a topological
  superconducting phase in a three-terminal device}},\ }\href@noop {} {\
  }\Eprint {https://arxiv.org/abs/2103.12217} {arXiv:2103.12217
  [cond-mat.mes-hall]} \BibitemShut {NoStop}%
\bibitem [{\citenamefont {{Aghaee}}\ \emph {et~al.}()\citenamefont {{Aghaee}},
  \citenamefont {{Akkala}}, \citenamefont {{Alam}}, \citenamefont {{Ali}},
  \citenamefont {{Alcaraz Ramirez}}, \citenamefont {{Andrzejczuk}},
  \citenamefont {{E Antipov}}, \citenamefont {{Aseev}}, \citenamefont
  {{Astafev}}, \citenamefont {{Bauer}}, \citenamefont {{Becker}}, \citenamefont
  {{Boddapati}}, \citenamefont {{Boekhout}}, \citenamefont {{Bommer}},
  \citenamefont {{Bork Hansen}}, \citenamefont {{Bosma}}, \citenamefont
  {{Bourdet}}, \citenamefont {{Boutin}}, \citenamefont {{Caroff}},
  \citenamefont {{Casparis}}, \citenamefont {{Cassidy}}, \citenamefont {{Wulf
  Christensen}}, \citenamefont {{Clay}}, \citenamefont {{Cole}}, \citenamefont
  {{Corsetti}}, \citenamefont {{Cui}}, \citenamefont {{Dalampiras}},
  \citenamefont {{Dokania}}, \citenamefont {{de Lange}}, \citenamefont {{de
  Moor}}, \citenamefont {{Estrada Salda{\~n}a}}, \citenamefont {{Fallahi}},
  \citenamefont {{Heidarnia Fathabad}}, \citenamefont {{Gamble}}, \citenamefont
  {{Gardner}}, \citenamefont {{Govender}}, \citenamefont {{Griggio}},
  \citenamefont {{Grigoryan}}, \citenamefont {{Gronin}}, \citenamefont
  {{Gukelberger}}, \citenamefont {{Heedt}}, \citenamefont {{Herranz Zamorano}},
  \citenamefont {{Ho}}, \citenamefont {{Laurens Holgaard}}, \citenamefont
  {{Hvidtfelt Padk{\ae}r Nielsen}}, \citenamefont {{Ingerslev}}, \citenamefont
  {{Jeppesen Krogstrup}}, \citenamefont {{Johansson}}, \citenamefont {{Jones}},
  \citenamefont {{Kallaher}}, \citenamefont {{Karimi}}, \citenamefont
  {{Karzig}}, \citenamefont {{King}}, \citenamefont {{Kloster}}, \citenamefont
  {{Knapp}}, \citenamefont {{Kocon}}, \citenamefont {{Koski}}, \citenamefont
  {{Kostamo}}, \citenamefont {{Kumar}}, \citenamefont {{Laeven}}, \citenamefont
  {{Larsen}}, \citenamefont {{Li}}, \citenamefont {{Lindemann}}, \citenamefont
  {{Love}}, \citenamefont {{Lutchyn}}, \citenamefont {{Manfra}}, \citenamefont
  {{Memisevic}}, \citenamefont {{Nayak}}, \citenamefont {{Nijholt}},
  \citenamefont {{Hannibal Madsen}}, \citenamefont {{Markussen}}, \citenamefont
  {{Martinez}}, \citenamefont {{McNeil}}, \citenamefont {{Mullally}},
  \citenamefont {{Nielsen}}, \citenamefont {{Nurmohamed}}, \citenamefont
  {{O'Farrell}}, \citenamefont {{Otani}}, \citenamefont {{Pauka}},
  \citenamefont {{Petersson}}, \citenamefont {{Petit}}, \citenamefont
  {{Pikulin}}, \citenamefont {{Preiss}}, \citenamefont {{Quintero Perez}},
  \citenamefont {{Rasmussen}}, \citenamefont {{Rajpalke}}, \citenamefont
  {{Razmadze}}, \citenamefont {{Reentila}}, \citenamefont {{Reilly}},
  \citenamefont {{Rouse}}, \citenamefont {{Sadovskyy}}, \citenamefont
  {{Sainiemi}}, \citenamefont {{Schreppler}}, \citenamefont {{Sidorkin}},
  \citenamefont {{Singh}}, \citenamefont {{Singh}}, \citenamefont {{Sinha}},
  \citenamefont {{Sohr}}, \citenamefont {{Stankevi{\v{c}}}}, \citenamefont
  {{Stek}}, \citenamefont {{Suominen}}, \citenamefont {{Suter}}, \citenamefont
  {{Svidenko}}, \citenamefont {{Teicher}}, \citenamefont {{Temuerhan}},
  \citenamefont {{Thiyagarajah}}, \citenamefont {{Tholapi}}, \citenamefont
  {{Thomas}}, \citenamefont {{Toomey}}, \citenamefont {{Upadhyay}},
  \citenamefont {{Urban}}, \citenamefont {{Vaitiek{\.{e}}nas}}, \citenamefont
  {{Van Hoogdalem}}, \citenamefont {{Viazmitinov}}, \citenamefont {{Waddy}},
  \citenamefont {{Van Woerkom}}, \citenamefont {{Vogel}}, \citenamefont
  {{Watson}}, \citenamefont {{Weston}}, \citenamefont {{Winkler}},
  \citenamefont {{Yang}}, \citenamefont {{Yau}}, \citenamefont {{Yi}},
  \citenamefont {{Yucelen}}, \citenamefont {{Webster}}, \citenamefont
  {{Zeisel}},\ and\ \citenamefont {{Zhao}}}]{2022arXiv220702472A}%
  \BibitemOpen
  \bibfield  {author} {\bibinfo {author} {\bibfnamefont {M.}~\bibnamefont
  {{Aghaee}}}, \bibinfo {author} {\bibfnamefont {A.}~\bibnamefont {{Akkala}}},
  \bibinfo {author} {\bibfnamefont {Z.}~\bibnamefont {{Alam}}}, \bibinfo
  {author} {\bibfnamefont {R.}~\bibnamefont {{Ali}}}, \bibinfo {author}
  {\bibfnamefont {A.}~\bibnamefont {{Alcaraz Ramirez}}}, \bibinfo {author}
  {\bibfnamefont {M.}~\bibnamefont {{Andrzejczuk}}}, \bibinfo {author}
  {\bibfnamefont {A.}~\bibnamefont {{E Antipov}}}, \bibinfo {author}
  {\bibfnamefont {P.}~\bibnamefont {{Aseev}}}, \bibinfo {author} {\bibfnamefont
  {M.}~\bibnamefont {{Astafev}}}, \bibinfo {author} {\bibfnamefont
  {B.}~\bibnamefont {{Bauer}}}, \bibinfo {author} {\bibfnamefont
  {J.}~\bibnamefont {{Becker}}}, \bibinfo {author} {\bibfnamefont
  {S.}~\bibnamefont {{Boddapati}}}, \bibinfo {author} {\bibfnamefont
  {F.}~\bibnamefont {{Boekhout}}}, \bibinfo {author} {\bibfnamefont
  {J.}~\bibnamefont {{Bommer}}}, \bibinfo {author} {\bibfnamefont
  {E.}~\bibnamefont {{Bork Hansen}}}, \bibinfo {author} {\bibfnamefont
  {T.}~\bibnamefont {{Bosma}}}, \bibinfo {author} {\bibfnamefont
  {L.}~\bibnamefont {{Bourdet}}}, \bibinfo {author} {\bibfnamefont
  {S.}~\bibnamefont {{Boutin}}}, \bibinfo {author} {\bibfnamefont
  {P.}~\bibnamefont {{Caroff}}}, \bibinfo {author} {\bibfnamefont
  {L.}~\bibnamefont {{Casparis}}}, \bibinfo {author} {\bibfnamefont
  {M.}~\bibnamefont {{Cassidy}}}, \bibinfo {author} {\bibfnamefont
  {A.}~\bibnamefont {{Wulf Christensen}}}, \bibinfo {author} {\bibfnamefont
  {N.}~\bibnamefont {{Clay}}}, \bibinfo {author} {\bibfnamefont {W.~S.}\
  \bibnamefont {{Cole}}}, \bibinfo {author} {\bibfnamefont {F.}~\bibnamefont
  {{Corsetti}}}, \bibinfo {author} {\bibfnamefont {A.}~\bibnamefont {{Cui}}},
  \bibinfo {author} {\bibfnamefont {P.}~\bibnamefont {{Dalampiras}}}, \bibinfo
  {author} {\bibfnamefont {A.}~\bibnamefont {{Dokania}}}, \bibinfo {author}
  {\bibfnamefont {G.}~\bibnamefont {{de Lange}}}, \bibinfo {author}
  {\bibfnamefont {M.}~\bibnamefont {{de Moor}}}, \bibinfo {author}
  {\bibfnamefont {J.~C.}\ \bibnamefont {{Estrada Salda{\~n}a}}}, \bibinfo
  {author} {\bibfnamefont {S.}~\bibnamefont {{Fallahi}}}, \bibinfo {author}
  {\bibfnamefont {Z.}~\bibnamefont {{Heidarnia Fathabad}}}, \bibinfo {author}
  {\bibfnamefont {J.}~\bibnamefont {{Gamble}}}, \bibinfo {author}
  {\bibfnamefont {G.}~\bibnamefont {{Gardner}}}, \bibinfo {author}
  {\bibfnamefont {D.}~\bibnamefont {{Govender}}}, \bibinfo {author}
  {\bibfnamefont {F.}~\bibnamefont {{Griggio}}}, \bibinfo {author}
  {\bibfnamefont {R.}~\bibnamefont {{Grigoryan}}}, \bibinfo {author}
  {\bibfnamefont {S.}~\bibnamefont {{Gronin}}}, \bibinfo {author}
  {\bibfnamefont {J.}~\bibnamefont {{Gukelberger}}}, \bibinfo {author}
  {\bibfnamefont {S.}~\bibnamefont {{Heedt}}}, \bibinfo {author} {\bibfnamefont
  {J.}~\bibnamefont {{Herranz Zamorano}}}, \bibinfo {author} {\bibfnamefont
  {S.}~\bibnamefont {{Ho}}}, \bibinfo {author} {\bibfnamefont {U.}~\bibnamefont
  {{Laurens Holgaard}}}, \bibinfo {author} {\bibfnamefont {W.}~\bibnamefont
  {{Hvidtfelt Padk{\ae}r Nielsen}}}, \bibinfo {author} {\bibfnamefont
  {H.}~\bibnamefont {{Ingerslev}}}, \bibinfo {author} {\bibfnamefont
  {P.}~\bibnamefont {{Jeppesen Krogstrup}}}, \bibinfo {author} {\bibfnamefont
  {L.}~\bibnamefont {{Johansson}}}, \bibinfo {author} {\bibfnamefont
  {J.}~\bibnamefont {{Jones}}}, \bibinfo {author} {\bibfnamefont
  {R.}~\bibnamefont {{Kallaher}}}, \bibinfo {author} {\bibfnamefont
  {F.}~\bibnamefont {{Karimi}}}, \bibinfo {author} {\bibfnamefont
  {T.}~\bibnamefont {{Karzig}}}, \bibinfo {author} {\bibfnamefont
  {C.}~\bibnamefont {{King}}}, \bibinfo {author} {\bibfnamefont {M.~E.}\
  \bibnamefont {{Kloster}}}, \bibinfo {author} {\bibfnamefont {C.}~\bibnamefont
  {{Knapp}}}, \bibinfo {author} {\bibfnamefont {D.}~\bibnamefont {{Kocon}}},
  \bibinfo {author} {\bibfnamefont {J.}~\bibnamefont {{Koski}}}, \bibinfo
  {author} {\bibfnamefont {P.}~\bibnamefont {{Kostamo}}}, \bibinfo {author}
  {\bibfnamefont {M.}~\bibnamefont {{Kumar}}}, \bibinfo {author} {\bibfnamefont
  {T.}~\bibnamefont {{Laeven}}}, \bibinfo {author} {\bibfnamefont
  {T.}~\bibnamefont {{Larsen}}}, \bibinfo {author} {\bibfnamefont
  {K.}~\bibnamefont {{Li}}}, \bibinfo {author} {\bibfnamefont {T.}~\bibnamefont
  {{Lindemann}}}, \bibinfo {author} {\bibfnamefont {J.}~\bibnamefont {{Love}}},
  \bibinfo {author} {\bibfnamefont {R.}~\bibnamefont {{Lutchyn}}}, \bibinfo
  {author} {\bibfnamefont {M.}~\bibnamefont {{Manfra}}}, \bibinfo {author}
  {\bibfnamefont {E.}~\bibnamefont {{Memisevic}}}, \bibinfo {author}
  {\bibfnamefont {C.}~\bibnamefont {{Nayak}}}, \bibinfo {author} {\bibfnamefont
  {B.}~\bibnamefont {{Nijholt}}}, \bibinfo {author} {\bibfnamefont
  {M.}~\bibnamefont {{Hannibal Madsen}}}, \bibinfo {author} {\bibfnamefont
  {S.}~\bibnamefont {{Markussen}}}, \bibinfo {author} {\bibfnamefont
  {E.}~\bibnamefont {{Martinez}}}, \bibinfo {author} {\bibfnamefont
  {R.}~\bibnamefont {{McNeil}}}, \bibinfo {author} {\bibfnamefont
  {A.}~\bibnamefont {{Mullally}}}, \bibinfo {author} {\bibfnamefont
  {J.}~\bibnamefont {{Nielsen}}}, \bibinfo {author} {\bibfnamefont
  {A.}~\bibnamefont {{Nurmohamed}}}, \bibinfo {author} {\bibfnamefont
  {E.}~\bibnamefont {{O'Farrell}}}, \bibinfo {author} {\bibfnamefont
  {K.}~\bibnamefont {{Otani}}}, \bibinfo {author} {\bibfnamefont
  {S.}~\bibnamefont {{Pauka}}}, \bibinfo {author} {\bibfnamefont
  {K.}~\bibnamefont {{Petersson}}}, \bibinfo {author} {\bibfnamefont
  {L.}~\bibnamefont {{Petit}}}, \bibinfo {author} {\bibfnamefont
  {D.}~\bibnamefont {{Pikulin}}}, \bibinfo {author} {\bibfnamefont
  {F.}~\bibnamefont {{Preiss}}}, \bibinfo {author} {\bibfnamefont
  {M.}~\bibnamefont {{Quintero Perez}}}, \bibinfo {author} {\bibfnamefont
  {K.}~\bibnamefont {{Rasmussen}}}, \bibinfo {author} {\bibfnamefont
  {M.}~\bibnamefont {{Rajpalke}}}, \bibinfo {author} {\bibfnamefont
  {D.}~\bibnamefont {{Razmadze}}}, \bibinfo {author} {\bibfnamefont
  {O.}~\bibnamefont {{Reentila}}}, \bibinfo {author} {\bibfnamefont
  {D.}~\bibnamefont {{Reilly}}}, \bibinfo {author} {\bibfnamefont
  {R.}~\bibnamefont {{Rouse}}}, \bibinfo {author} {\bibfnamefont
  {I.}~\bibnamefont {{Sadovskyy}}}, \bibinfo {author} {\bibfnamefont
  {L.}~\bibnamefont {{Sainiemi}}}, \bibinfo {author} {\bibfnamefont
  {S.}~\bibnamefont {{Schreppler}}}, \bibinfo {author} {\bibfnamefont
  {V.}~\bibnamefont {{Sidorkin}}}, \bibinfo {author} {\bibfnamefont
  {A.}~\bibnamefont {{Singh}}}, \bibinfo {author} {\bibfnamefont
  {S.}~\bibnamefont {{Singh}}}, \bibinfo {author} {\bibfnamefont
  {S.}~\bibnamefont {{Sinha}}}, \bibinfo {author} {\bibfnamefont
  {P.}~\bibnamefont {{Sohr}}}, \bibinfo {author} {\bibfnamefont
  {T.}~\bibnamefont {{Stankevi{\v{c}}}}}, \bibinfo {author} {\bibfnamefont
  {L.}~\bibnamefont {{Stek}}}, \bibinfo {author} {\bibfnamefont
  {H.}~\bibnamefont {{Suominen}}}, \bibinfo {author} {\bibfnamefont
  {J.}~\bibnamefont {{Suter}}}, \bibinfo {author} {\bibfnamefont
  {V.}~\bibnamefont {{Svidenko}}}, \bibinfo {author} {\bibfnamefont
  {S.}~\bibnamefont {{Teicher}}}, \bibinfo {author} {\bibfnamefont
  {M.}~\bibnamefont {{Temuerhan}}}, \bibinfo {author} {\bibfnamefont
  {N.}~\bibnamefont {{Thiyagarajah}}}, \bibinfo {author} {\bibfnamefont
  {R.}~\bibnamefont {{Tholapi}}}, \bibinfo {author} {\bibfnamefont
  {M.}~\bibnamefont {{Thomas}}}, \bibinfo {author} {\bibfnamefont
  {E.}~\bibnamefont {{Toomey}}}, \bibinfo {author} {\bibfnamefont
  {S.}~\bibnamefont {{Upadhyay}}}, \bibinfo {author} {\bibfnamefont
  {I.}~\bibnamefont {{Urban}}}, \bibinfo {author} {\bibfnamefont
  {S.}~\bibnamefont {{Vaitiek{\.{e}}nas}}}, \bibinfo {author} {\bibfnamefont
  {K.}~\bibnamefont {{Van Hoogdalem}}}, \bibinfo {author} {\bibfnamefont
  {D.~V.}\ \bibnamefont {{Viazmitinov}}}, \bibinfo {author} {\bibfnamefont
  {S.}~\bibnamefont {{Waddy}}}, \bibinfo {author} {\bibfnamefont
  {D.}~\bibnamefont {{Van Woerkom}}}, \bibinfo {author} {\bibfnamefont
  {D.}~\bibnamefont {{Vogel}}}, \bibinfo {author} {\bibfnamefont
  {J.}~\bibnamefont {{Watson}}}, \bibinfo {author} {\bibfnamefont
  {J.}~\bibnamefont {{Weston}}}, \bibinfo {author} {\bibfnamefont {G.~W.}\
  \bibnamefont {{Winkler}}}, \bibinfo {author} {\bibfnamefont {C.~K.}\
  \bibnamefont {{Yang}}}, \bibinfo {author} {\bibfnamefont {S.}~\bibnamefont
  {{Yau}}}, \bibinfo {author} {\bibfnamefont {D.}~\bibnamefont {{Yi}}},
  \bibinfo {author} {\bibfnamefont {E.}~\bibnamefont {{Yucelen}}}, \bibinfo
  {author} {\bibfnamefont {A.}~\bibnamefont {{Webster}}}, \bibinfo {author}
  {\bibfnamefont {R.}~\bibnamefont {{Zeisel}}},\ and\ \bibinfo {author}
  {\bibfnamefont {R.}~\bibnamefont {{Zhao}}},\ }\bibfield  {title} {\bibinfo
  {title} {{InAs-Al Hybrid Devices Passing the Topological Gap Protocol}},\
  }\href@noop {} {\ }\Eprint {https://arxiv.org/abs/2207.02472}
  {arXiv:2207.02472 [cond-mat.mes-hall]} \BibitemShut {NoStop}%
\bibitem [{\citenamefont {Liu}\ \emph {et~al.}(2015{\natexlab{a}})\citenamefont
  {Liu}, \citenamefont {Cheng},\ and\ \citenamefont {Lutchyn}}]{LiuPRB15_1}%
  \BibitemOpen
  \bibfield  {author} {\bibinfo {author} {\bibfnamefont {D.~E.}\ \bibnamefont
  {Liu}}, \bibinfo {author} {\bibfnamefont {M.}~\bibnamefont {Cheng}},\ and\
  \bibinfo {author} {\bibfnamefont {R.~M.}\ \bibnamefont {Lutchyn}},\
  }\bibfield  {title} {\bibinfo {title} {{Probing Majorana physics in
  quantum-dot shot-noise experiments}},\ }\href
  {https://doi.org/10.1103/PhysRevB.91.081405} {\bibfield  {journal} {\bibinfo
  {journal} {Phys. Rev. B}\ }\textbf {\bibinfo {volume} {91}},\ \bibinfo
  {pages} {081405(R)} (\bibinfo {year} {2015}{\natexlab{a}})}\BibitemShut
  {NoStop}%
\bibitem [{\citenamefont {Liu}\ \emph {et~al.}(2015{\natexlab{b}})\citenamefont
  {Liu}, \citenamefont {Levchenko},\ and\ \citenamefont
  {Lutchyn}}]{LiuPRB15_2}%
  \BibitemOpen
  \bibfield  {author} {\bibinfo {author} {\bibfnamefont {D.~E.}\ \bibnamefont
  {Liu}}, \bibinfo {author} {\bibfnamefont {A.}~\bibnamefont {Levchenko}},\
  and\ \bibinfo {author} {\bibfnamefont {R.~M.}\ \bibnamefont {Lutchyn}},\
  }\bibfield  {title} {\bibinfo {title} {{Majorana zero modes choose Euler
  numbers as revealed by full counting statistics}},\ }\href
  {https://doi.org/10.1103/PhysRevB.92.205422} {\bibfield  {journal} {\bibinfo
  {journal} {Phys. Rev. B}\ }\textbf {\bibinfo {volume} {92}},\ \bibinfo
  {pages} {205422} (\bibinfo {year} {2015}{\natexlab{b}})}\BibitemShut
  {NoStop}%
\bibitem [{\citenamefont {Smirnov}(2019)}]{SmirnovPRB19}%
  \BibitemOpen
  \bibfield  {author} {\bibinfo {author} {\bibfnamefont {S.}~\bibnamefont
  {Smirnov}},\ }\bibfield  {title} {\bibinfo {title} {Majorana finite-frequency
  nonequilibrium quantum noise},\ }\href
  {https://doi.org/10.1103/PhysRevB.99.165427} {\bibfield  {journal} {\bibinfo
  {journal} {Phys. Rev. B}\ }\textbf {\bibinfo {volume} {99}},\ \bibinfo
  {pages} {165427} (\bibinfo {year} {2019})}\BibitemShut {NoStop}%
\bibitem [{\citenamefont {Smirnov}(2022)}]{SmirnovPRB22}%
  \BibitemOpen
  \bibfield  {author} {\bibinfo {author} {\bibfnamefont {S.}~\bibnamefont
  {Smirnov}},\ }\bibfield  {title} {\bibinfo {title} {{Revealing universal
  Majorana fractionalization using differential shot noise and conductance in
  nonequilibrium states controlled by tunneling phases}},\ }\href
  {https://doi.org/10.1103/PhysRevB.105.205430} {\bibfield  {journal} {\bibinfo
   {journal} {Phys. Rev. B}\ }\textbf {\bibinfo {volume} {105}},\ \bibinfo
  {pages} {205430} (\bibinfo {year} {2022})}\BibitemShut {NoStop}%
\bibitem [{\citenamefont {Smirnov}(2015)}]{entro_smirnov_one}%
  \BibitemOpen
  \bibfield  {author} {\bibinfo {author} {\bibfnamefont {S.}~\bibnamefont
  {Smirnov}},\ }\bibfield  {title} {\bibinfo {title} {Majorana tunneling
  entropy},\ }\href {https://doi.org/10.1103/PhysRevB.92.195312} {\bibfield
  {journal} {\bibinfo  {journal} {Phys. Rev. B}\ }\textbf {\bibinfo {volume}
  {92}},\ \bibinfo {pages} {195312} (\bibinfo {year} {2015})}\BibitemShut
  {NoStop}%
\bibitem [{\citenamefont {Sela}\ \emph {et~al.}(2019)\citenamefont {Sela},
  \citenamefont {Oreg}, \citenamefont {Plugge}, \citenamefont {Hartman},
  \citenamefont {L\"uscher},\ and\ \citenamefont {Folk}}]{entro_Sela}%
  \BibitemOpen
  \bibfield  {author} {\bibinfo {author} {\bibfnamefont {E.}~\bibnamefont
  {Sela}}, \bibinfo {author} {\bibfnamefont {Y.}~\bibnamefont {Oreg}}, \bibinfo
  {author} {\bibfnamefont {S.}~\bibnamefont {Plugge}}, \bibinfo {author}
  {\bibfnamefont {N.}~\bibnamefont {Hartman}}, \bibinfo {author} {\bibfnamefont
  {S.}~\bibnamefont {L\"uscher}},\ and\ \bibinfo {author} {\bibfnamefont
  {J.}~\bibnamefont {Folk}},\ }\bibfield  {title} {\bibinfo {title} {{Detecting
  the Universal Fractional Entropy of Majorana Zero Modes}},\ }\href
  {https://doi.org/10.1103/PhysRevLett.123.147702} {\bibfield  {journal}
  {\bibinfo  {journal} {Phys. Rev. Lett.}\ }\textbf {\bibinfo {volume} {123}},\
  \bibinfo {pages} {147702} (\bibinfo {year} {2019})}\BibitemShut {NoStop}%
\bibitem [{\citenamefont {Smirnov}(2021{\natexlab{a}})}]{entro_smirnov_two}%
  \BibitemOpen
  \bibfield  {author} {\bibinfo {author} {\bibfnamefont {S.}~\bibnamefont
  {Smirnov}},\ }\bibfield  {title} {\bibinfo {title} {Majorana entropy revival
  via tunneling phases},\ }\href {https://doi.org/10.1103/PhysRevB.103.075440}
  {\bibfield  {journal} {\bibinfo  {journal} {Phys. Rev. B}\ }\textbf {\bibinfo
  {volume} {103}},\ \bibinfo {pages} {075440} (\bibinfo {year}
  {2021}{\natexlab{a}})}\BibitemShut {NoStop}%
\bibitem [{\citenamefont {Smirnov}(2021{\natexlab{b}})}]{entro_smirnov_three}%
  \BibitemOpen
  \bibfield  {author} {\bibinfo {author} {\bibfnamefont {S.}~\bibnamefont
  {Smirnov}},\ }\bibfield  {title} {\bibinfo {title} {Majorana ensembles with
  fractional entropy and conductance in nanoscopic systems},\ }\href
  {https://doi.org/10.1103/PhysRevB.104.205406} {\bibfield  {journal} {\bibinfo
   {journal} {Phys. Rev. B}\ }\textbf {\bibinfo {volume} {104}},\ \bibinfo
  {pages} {205406} (\bibinfo {year} {2021}{\natexlab{b}})}\BibitemShut
  {NoStop}%
\bibitem [{\citenamefont {Liu}\ \emph {et~al.}(2020)\citenamefont {Liu},
  \citenamefont {Vaitiek{\.e}nas}, \citenamefont {Mart{\'\i}-S{\'a}nchez},
  \citenamefont {Koch}, \citenamefont {Hart}, \citenamefont {Cui},
  \citenamefont {Kanne}, \citenamefont {Khan}, \citenamefont {Tanta},
  \citenamefont {Upadhyay}, \citenamefont {Cachaza}, \citenamefont {Marcus},
  \citenamefont {Arbiol}, \citenamefont {Moler},\ and\ \citenamefont
  {Krogstrup}}]{YuNanoLett2020}%
  \BibitemOpen
  \bibfield  {author} {\bibinfo {author} {\bibfnamefont {Y.}~\bibnamefont
  {Liu}}, \bibinfo {author} {\bibfnamefont {S.}~\bibnamefont
  {Vaitiek{\.e}nas}}, \bibinfo {author} {\bibfnamefont {S.}~\bibnamefont
  {Mart{\'\i}-S{\'a}nchez}}, \bibinfo {author} {\bibfnamefont {C.}~\bibnamefont
  {Koch}}, \bibinfo {author} {\bibfnamefont {S.}~\bibnamefont {Hart}}, \bibinfo
  {author} {\bibfnamefont {Z.}~\bibnamefont {Cui}}, \bibinfo {author}
  {\bibfnamefont {T.}~\bibnamefont {Kanne}}, \bibinfo {author} {\bibfnamefont
  {S.~A.}\ \bibnamefont {Khan}}, \bibinfo {author} {\bibfnamefont
  {R.}~\bibnamefont {Tanta}}, \bibinfo {author} {\bibfnamefont
  {S.}~\bibnamefont {Upadhyay}}, \bibinfo {author} {\bibfnamefont {M.~E.}\
  \bibnamefont {Cachaza}}, \bibinfo {author} {\bibfnamefont {C.~M.}\
  \bibnamefont {Marcus}}, \bibinfo {author} {\bibfnamefont {J.}~\bibnamefont
  {Arbiol}}, \bibinfo {author} {\bibfnamefont {K.~A.}\ \bibnamefont {Moler}},\
  and\ \bibinfo {author} {\bibfnamefont {P.}~\bibnamefont {Krogstrup}},\
  }\bibfield  {title} {\bibinfo {title} {Semiconductor--ferromagnetic
  insulator--superconductor nanowires: Stray field and exchange field},\ }\href
  {https://doi.org/10.1021/acs.nanolett.9b04187} {\bibfield  {journal}
  {\bibinfo  {journal} {Nano Letters}\ }\textbf {\bibinfo {volume} {20}},\
  \bibinfo {pages} {456} (\bibinfo {year} {2020})}\BibitemShut {NoStop}%
\bibitem [{\citenamefont {Vaitiek{\.e}nas}\ \emph {et~al.}(2021)\citenamefont
  {Vaitiek{\.e}nas}, \citenamefont {Liu}, \citenamefont {Krogstrup},\ and\
  \citenamefont {Marcus}}]{Vaitieknas2020ZerofieldTS}%
  \BibitemOpen
  \bibfield  {author} {\bibinfo {author} {\bibfnamefont {S.}~\bibnamefont
  {Vaitiek{\.e}nas}}, \bibinfo {author} {\bibfnamefont {Y.}~\bibnamefont
  {Liu}}, \bibinfo {author} {\bibfnamefont {P.}~\bibnamefont {Krogstrup}},\
  and\ \bibinfo {author} {\bibfnamefont {C.~M.}\ \bibnamefont {Marcus}},\
  }\bibfield  {title} {\bibinfo {title} {Zero-bias peaks at zero magnetic field
  in ferromagnetic hybrid nanowires},\ }\href
  {https://doi.org/10.1038/s41567-020-1017-3} {\bibfield  {journal} {\bibinfo
  {journal} {Nature Physics}\ }\textbf {\bibinfo {volume} {17}},\ \bibinfo
  {pages} {43} (\bibinfo {year} {2021})}\BibitemShut {NoStop}%
\bibitem [{\citenamefont {Tserkovnyak}\ \emph {et~al.}(2002)\citenamefont
  {Tserkovnyak}, \citenamefont {Brataas},\ and\ \citenamefont
  {Bauer}}]{Spump_theo}%
  \BibitemOpen
  \bibfield  {author} {\bibinfo {author} {\bibfnamefont {Y.}~\bibnamefont
  {Tserkovnyak}}, \bibinfo {author} {\bibfnamefont {A.}~\bibnamefont
  {Brataas}},\ and\ \bibinfo {author} {\bibfnamefont {G.~E.~W.}\ \bibnamefont
  {Bauer}},\ }\bibfield  {title} {\bibinfo {title} {{Enhanced Gilbert Damping
  in Thin Ferromagnetic Films}},\ }\href
  {https://doi.org/10.1103/PhysRevLett.88.117601} {\bibfield  {journal}
  {\bibinfo  {journal} {Phys. Rev. Lett.}\ }\textbf {\bibinfo {volume} {88}},\
  \bibinfo {pages} {117601} (\bibinfo {year} {2002})}\BibitemShut {NoStop}%
\bibitem [{\citenamefont {Brataas}\ \emph {et~al.}(2002)\citenamefont
  {Brataas}, \citenamefont {Tserkovnyak}, \citenamefont {Bauer},\ and\
  \citenamefont {Halperin}}]{Spbattery}%
  \BibitemOpen
  \bibfield  {author} {\bibinfo {author} {\bibfnamefont {A.}~\bibnamefont
  {Brataas}}, \bibinfo {author} {\bibfnamefont {Y.}~\bibnamefont
  {Tserkovnyak}}, \bibinfo {author} {\bibfnamefont {G.~E.~W.}\ \bibnamefont
  {Bauer}},\ and\ \bibinfo {author} {\bibfnamefont {B.~I.}\ \bibnamefont
  {Halperin}},\ }\bibfield  {title} {\bibinfo {title} {{Spin battery operated
  by ferromagnetic resonance}},\ }\href
  {https://doi.org/10.1103/PhysRevB.66.060404} {\bibfield  {journal} {\bibinfo
  {journal} {Phys. Rev. B}\ }\textbf {\bibinfo {volume} {66}},\ \bibinfo
  {pages} {060404(R)} (\bibinfo {year} {2002})}\BibitemShut {NoStop}%
\bibitem [{\citenamefont {Tserkovnyak}\ \emph {et~al.}(2005)\citenamefont
  {Tserkovnyak}, \citenamefont {Brataas}, \citenamefont {Bauer},\ and\
  \citenamefont {Halperin}}]{TserkovnyakRMP2005}%
  \BibitemOpen
  \bibfield  {author} {\bibinfo {author} {\bibfnamefont {Y.}~\bibnamefont
  {Tserkovnyak}}, \bibinfo {author} {\bibfnamefont {A.}~\bibnamefont
  {Brataas}}, \bibinfo {author} {\bibfnamefont {G.~E.~W.}\ \bibnamefont
  {Bauer}},\ and\ \bibinfo {author} {\bibfnamefont {B.~I.}\ \bibnamefont
  {Halperin}},\ }\bibfield  {title} {\bibinfo {title} {Nonlocal magnetization
  dynamics in ferromagnetic heterostructures},\ }\href
  {https://doi.org/10.1103/RevModPhys.77.1375} {\bibfield  {journal} {\bibinfo
  {journal} {Rev. Mod. Phys.}\ }\textbf {\bibinfo {volume} {77}},\ \bibinfo
  {pages} {1375} (\bibinfo {year} {2005})}\BibitemShut {NoStop}%
\bibitem [{\citenamefont {Hirohata}\ \emph {et~al.}(2020)\citenamefont
  {Hirohata}, \citenamefont {Yamada}, \citenamefont {Nakatani}, \citenamefont
  {Prejbeanu}, \citenamefont {Diény}, \citenamefont {Pirro},\ and\
  \citenamefont {Hillebrands}}]{spintroRev}%
  \BibitemOpen
  \bibfield  {author} {\bibinfo {author} {\bibfnamefont {A.}~\bibnamefont
  {Hirohata}}, \bibinfo {author} {\bibfnamefont {K.}~\bibnamefont {Yamada}},
  \bibinfo {author} {\bibfnamefont {Y.}~\bibnamefont {Nakatani}}, \bibinfo
  {author} {\bibfnamefont {I.-L.}\ \bibnamefont {Prejbeanu}}, \bibinfo {author}
  {\bibfnamefont {B.}~\bibnamefont {Diény}}, \bibinfo {author} {\bibfnamefont
  {P.}~\bibnamefont {Pirro}},\ and\ \bibinfo {author} {\bibfnamefont
  {B.}~\bibnamefont {Hillebrands}},\ }\bibfield  {title} {\bibinfo {title}
  {Review on spintronics: Principles and device applications},\ }\href
  {https://doi.org/https://doi.org/10.1016/j.jmmm.2020.166711} {\bibfield
  {journal} {\bibinfo  {journal} {Journal of Magnetism and Magnetic Materials}\
  }\textbf {\bibinfo {volume} {509}},\ \bibinfo {pages} {166711} (\bibinfo
  {year} {2020})}\BibitemShut {NoStop}%
\bibitem [{\citenamefont {Sinova}\ \emph {et~al.}(2015)\citenamefont {Sinova},
  \citenamefont {Valenzuela}, \citenamefont {Wunderlich}, \citenamefont
  {Back},\ and\ \citenamefont {Jungwirth}}]{SinovaRMP2015}%
  \BibitemOpen
  \bibfield  {author} {\bibinfo {author} {\bibfnamefont {J.}~\bibnamefont
  {Sinova}}, \bibinfo {author} {\bibfnamefont {S.~O.}\ \bibnamefont
  {Valenzuela}}, \bibinfo {author} {\bibfnamefont {J.}~\bibnamefont
  {Wunderlich}}, \bibinfo {author} {\bibfnamefont {C.~H.}\ \bibnamefont
  {Back}},\ and\ \bibinfo {author} {\bibfnamefont {T.}~\bibnamefont
  {Jungwirth}},\ }\bibfield  {title} {\bibinfo {title} {{Spin Hall effects}},\
  }\href {https://doi.org/10.1103/RevModPhys.87.1213} {\bibfield  {journal}
  {\bibinfo  {journal} {Rev. Mod. Phys.}\ }\textbf {\bibinfo {volume} {87}},\
  \bibinfo {pages} {1213} (\bibinfo {year} {2015})}\BibitemShut {NoStop}%
\bibitem [{\citenamefont {Hoffmann}\ and\ \citenamefont
  {Bader}(2015)}]{Hoffman-review}%
  \BibitemOpen
  \bibfield  {author} {\bibinfo {author} {\bibfnamefont {A.}~\bibnamefont
  {Hoffmann}}\ and\ \bibinfo {author} {\bibfnamefont {S.~D.}\ \bibnamefont
  {Bader}},\ }\bibfield  {title} {\bibinfo {title} {Opportunities at the
  frontiers of spintronics},\ }\href
  {https://doi.org/10.1103/PhysRevApplied.4.047001} {\bibfield  {journal}
  {\bibinfo  {journal} {Phys. Rev. Applied}\ }\textbf {\bibinfo {volume} {4}},\
  \bibinfo {pages} {047001} (\bibinfo {year} {2015})}\BibitemShut {NoStop}%
\bibitem [{\citenamefont {Qi}\ \emph {et~al.}(2008)\citenamefont {Qi},
  \citenamefont {Hughes},\ and\ \citenamefont {Zhang}}]{Qquanti_Zhang}%
  \BibitemOpen
  \bibfield  {author} {\bibinfo {author} {\bibfnamefont {X.-L.}\ \bibnamefont
  {Qi}}, \bibinfo {author} {\bibfnamefont {T.~L.}\ \bibnamefont {Hughes}},\
  and\ \bibinfo {author} {\bibfnamefont {S.-C.}\ \bibnamefont {Zhang}},\
  }\bibfield  {title} {\bibinfo {title} {{Fractional charge and quantized
  current in the quantum spin Hall state}},\ }\href
  {https://doi.org/10.1038/nphys913} {\bibfield  {journal} {\bibinfo  {journal}
  {Nature Physics}\ }\textbf {\bibinfo {volume} {4}},\ \bibinfo {pages} {273}
  (\bibinfo {year} {2008})}\BibitemShut {NoStop}%
\bibitem [{\citenamefont {Meidan}\ \emph {et~al.}(2010)\citenamefont {Meidan},
  \citenamefont {Micklitz},\ and\ \citenamefont {Brouwer}}]{MeidanPRB10}%
  \BibitemOpen
  \bibfield  {author} {\bibinfo {author} {\bibfnamefont {D.}~\bibnamefont
  {Meidan}}, \bibinfo {author} {\bibfnamefont {T.}~\bibnamefont {Micklitz}},\
  and\ \bibinfo {author} {\bibfnamefont {P.~W.}\ \bibnamefont {Brouwer}},\
  }\bibfield  {title} {\bibinfo {title} {Optimal topological spin pump},\
  }\href {https://doi.org/10.1103/PhysRevB.82.161303} {\bibfield  {journal}
  {\bibinfo  {journal} {Phys. Rev. B}\ }\textbf {\bibinfo {volume} {82}},\
  \bibinfo {pages} {161303(R)} (\bibinfo {year} {2010})}\BibitemShut {NoStop}%
\bibitem [{\citenamefont {Meidan}\ \emph {et~al.}(2011)\citenamefont {Meidan},
  \citenamefont {Micklitz},\ and\ \citenamefont {Brouwer}}]{MeidanPRB11}%
  \BibitemOpen
  \bibfield  {author} {\bibinfo {author} {\bibfnamefont {D.}~\bibnamefont
  {Meidan}}, \bibinfo {author} {\bibfnamefont {T.}~\bibnamefont {Micklitz}},\
  and\ \bibinfo {author} {\bibfnamefont {P.~W.}\ \bibnamefont {Brouwer}},\
  }\bibfield  {title} {\bibinfo {title} {Topological classification of
  interaction-driven spin pumps},\ }\href
  {https://doi.org/10.1103/PhysRevB.84.075325} {\bibfield  {journal} {\bibinfo
  {journal} {Phys. Rev. B}\ }\textbf {\bibinfo {volume} {84}},\ \bibinfo
  {pages} {075325} (\bibinfo {year} {2011})}\BibitemShut {NoStop}%
\bibitem [{\citenamefont {Mahfouzi}\ \emph {et~al.}(2010)\citenamefont
  {Mahfouzi}, \citenamefont {Nikoli\ifmmode~\acute{c}\else \'{c}\fi{}},
  \citenamefont {Chen},\ and\ \citenamefont {Chang}}]{qtizedspin}%
  \BibitemOpen
  \bibfield  {author} {\bibinfo {author} {\bibfnamefont {F.}~\bibnamefont
  {Mahfouzi}}, \bibinfo {author} {\bibfnamefont {B.~K.}\ \bibnamefont
  {Nikoli\ifmmode~\acute{c}\else \'{c}\fi{}}}, \bibinfo {author} {\bibfnamefont
  {S.-H.}\ \bibnamefont {Chen}},\ and\ \bibinfo {author} {\bibfnamefont
  {C.-R.}\ \bibnamefont {Chang}},\ }\bibfield  {title} {\bibinfo {title}
  {Microwave-driven ferromagnet--topological-insulator heterostructures: The
  prospect for giant spin battery effect and quantized charge pump devices},\
  }\href {https://doi.org/10.1103/PhysRevB.82.195440} {\bibfield  {journal}
  {\bibinfo  {journal} {Phys. Rev. B}\ }\textbf {\bibinfo {volume} {82}},\
  \bibinfo {pages} {195440} (\bibinfo {year} {2010})}\BibitemShut {NoStop}%
\bibitem [{\citenamefont {Becerra}\ \emph {et~al.}(2021)\citenamefont
  {Becerra}, \citenamefont {Trif},\ and\ \citenamefont {Hyart}}]{Becerra21}%
  \BibitemOpen
  \bibfield  {author} {\bibinfo {author} {\bibfnamefont {V.~F.}\ \bibnamefont
  {Becerra}}, \bibinfo {author} {\bibfnamefont {M.}~\bibnamefont {Trif}},\ and\
  \bibinfo {author} {\bibfnamefont {T.}~\bibnamefont {Hyart}},\ }\bibfield
  {title} {\bibinfo {title} {{Topological charge, spin, and heat transistor}},\
  }\href {https://doi.org/10.1103/PhysRevB.103.205410} {\bibfield  {journal}
  {\bibinfo  {journal} {Phys. Rev. B}\ }\textbf {\bibinfo {volume} {103}},\
  \bibinfo {pages} {205410} (\bibinfo {year} {2021})}\BibitemShut {NoStop}%
\bibitem [{\citenamefont {Thouless}(1983)}]{Thouless83}%
  \BibitemOpen
  \bibfield  {author} {\bibinfo {author} {\bibfnamefont {D.~J.}\ \bibnamefont
  {Thouless}},\ }\bibfield  {title} {\bibinfo {title} {Quantization of particle
  transport},\ }\href {https://doi.org/10.1103/PhysRevB.27.6083} {\bibfield
  {journal} {\bibinfo  {journal} {Phys. Rev. B}\ }\textbf {\bibinfo {volume}
  {27}},\ \bibinfo {pages} {6083} (\bibinfo {year} {1983})}\BibitemShut
  {NoStop}%
\bibitem [{\citenamefont {Ojanen}(2012)}]{Ojanen12}%
  \BibitemOpen
  \bibfield  {author} {\bibinfo {author} {\bibfnamefont {T.}~\bibnamefont
  {Ojanen}},\ }\bibfield  {title} {\bibinfo {title} {{Magnetoelectric Effects
  in Superconducting Nanowires with Rashba Spin-Orbit Coupling}},\ }\href
  {https://doi.org/10.1103/PhysRevLett.109.226804} {\bibfield  {journal}
  {\bibinfo  {journal} {Phys. Rev. Lett.}\ }\textbf {\bibinfo {volume} {109}},\
  \bibinfo {pages} {226804} (\bibinfo {year} {2012})}\BibitemShut {NoStop}%
\bibitem [{\citenamefont {Kitaev}(2001)}]{Kitaev2001}%
  \BibitemOpen
  \bibfield  {author} {\bibinfo {author} {\bibfnamefont {A.~Y.}\ \bibnamefont
  {Kitaev}},\ }\bibfield  {title} {\bibinfo {title} {{Unpaired Majorana
  fermions in quantum wires}},\ }\href
  {https://doi.org/10.1070/1063-7869/44/10s/s29} {\bibfield  {journal}
  {\bibinfo  {journal} {Physics-Uspekhi}\ }\textbf {\bibinfo {volume} {44}},\
  \bibinfo {pages} {131} (\bibinfo {year} {2001})}\BibitemShut {NoStop}%
\bibitem [{sup()}]{supplementary}%
  \BibitemOpen
  \bibinfo {note} {See Supplementary Material for more details on the
  dependence of the conductance and spin pumping with other barrier widths, the
  reflection coefficients, the fractional entropy, quasi-MZMs, robustness
  against disorder of the main results, and the magnetization dynamics, which
  includes Refs.
  \cite{entro_expone,entro_exptwo,entro_expthree,entro_Kondo,entro_setup}.}\BibitemShut
  {Stop}%
\bibitem [{\citenamefont {Moskalets}(2012)}]{MoskaletsBook}%
  \BibitemOpen
  \bibfield  {author} {\bibinfo {author} {\bibfnamefont {M.~V.}\ \bibnamefont
  {Moskalets}},\ }\href@noop {} {\emph {\bibinfo {title} {Scattering Matrix
  Approach to Non-Stationary Quantum Transport}}}\ (\bibinfo  {publisher}
  {World Scientific, London},\ \bibinfo {year} {2012})\BibitemShut {NoStop}%
\bibitem [{\citenamefont {Blaauboer}(2002)}]{Blaauboer}%
  \BibitemOpen
  \bibfield  {author} {\bibinfo {author} {\bibfnamefont {M.}~\bibnamefont
  {Blaauboer}},\ }\bibfield  {title} {\bibinfo {title} {{Charge pumping in
  mesoscopic systems coupled to a superconducting lead}},\ }\href
  {https://doi.org/10.1103/PhysRevB.65.235318} {\bibfield  {journal} {\bibinfo
  {journal} {Phys. Rev. B}\ }\textbf {\bibinfo {volume} {65}},\ \bibinfo
  {pages} {235318} (\bibinfo {year} {2002})}\BibitemShut {NoStop}%
\bibitem [{\citenamefont {Groth}\ \emph {et~al.}(2014)\citenamefont {Groth},
  \citenamefont {Wimmer}, \citenamefont {Akhmerov},\ and\ \citenamefont
  {Waintal}}]{Groth_2014}%
  \BibitemOpen
  \bibfield  {author} {\bibinfo {author} {\bibfnamefont {C.~W.}\ \bibnamefont
  {Groth}}, \bibinfo {author} {\bibfnamefont {M.}~\bibnamefont {Wimmer}},
  \bibinfo {author} {\bibfnamefont {A.~R.}\ \bibnamefont {Akhmerov}},\ and\
  \bibinfo {author} {\bibfnamefont {X.}~\bibnamefont {Waintal}},\ }\bibfield
  {title} {\bibinfo {title} {Kwant: a software package for quantum transport},\
  }\href {https://doi.org/10.1088/1367-2630/16/6/063065} {\bibfield  {journal}
  {\bibinfo  {journal} {New Journal of Physics}\ }\textbf {\bibinfo {volume}
  {16}},\ \bibinfo {pages} {063065} (\bibinfo {year} {2014})}\BibitemShut
  {NoStop}%
\bibitem [{\citenamefont {Fulga}\ \emph {et~al.}(2013)\citenamefont {Fulga},
  \citenamefont {van Heck}, \citenamefont {Burrello},\ and\ \citenamefont
  {Hyart}}]{Fulga13}%
  \BibitemOpen
  \bibfield  {author} {\bibinfo {author} {\bibfnamefont {I.~C.}\ \bibnamefont
  {Fulga}}, \bibinfo {author} {\bibfnamefont {B.}~\bibnamefont {van Heck}},
  \bibinfo {author} {\bibfnamefont {M.}~\bibnamefont {Burrello}},\ and\
  \bibinfo {author} {\bibfnamefont {T.}~\bibnamefont {Hyart}},\ }\bibfield
  {title} {\bibinfo {title} {{Effects of disorder on Coulomb-assisted braiding
  of Majorana zero modes}},\ }\href
  {https://doi.org/10.1103/PhysRevB.88.155435} {\bibfield  {journal} {\bibinfo
  {journal} {Phys. Rev. B}\ }\textbf {\bibinfo {volume} {88}},\ \bibinfo
  {pages} {155435} (\bibinfo {year} {2013})}\BibitemShut {NoStop}%
\bibitem [{\citenamefont {{Mahaux}}\ and\ \citenamefont
  {{Weidenm{\"u}ller}}(1969)}]{Mahaux69}%
  \BibitemOpen
  \bibfield  {author} {\bibinfo {author} {\bibfnamefont {C.}~\bibnamefont
  {{Mahaux}}}\ and\ \bibinfo {author} {\bibfnamefont {H.~A.}\ \bibnamefont
  {{Weidenm{\"u}ller}}},\ }\href@noop {} {\emph {\bibinfo {title} {{Shell-model
  approach to nuclear reactions}}}}\ (\bibinfo  {publisher} {{{North-Holland,
  Amsterdam}}},\ \bibinfo {year} {1969})\BibitemShut {NoStop}%
\bibitem [{\citenamefont {Nilsson}\ \emph {et~al.}(2008)\citenamefont
  {Nilsson}, \citenamefont {Akhmerov},\ and\ \citenamefont
  {Beenakker}}]{Nilsson08}%
  \BibitemOpen
  \bibfield  {author} {\bibinfo {author} {\bibfnamefont {J.}~\bibnamefont
  {Nilsson}}, \bibinfo {author} {\bibfnamefont {A.~R.}\ \bibnamefont
  {Akhmerov}},\ and\ \bibinfo {author} {\bibfnamefont {C.~W.~J.}\ \bibnamefont
  {Beenakker}},\ }\bibfield  {title} {\bibinfo {title} {{Splitting of a Cooper
  Pair by a Pair of Majorana Bound States}},\ }\href
  {https://doi.org/10.1103/PhysRevLett.101.120403} {\bibfield  {journal}
  {\bibinfo  {journal} {Phys. Rev. Lett.}\ }\textbf {\bibinfo {volume} {101}},\
  \bibinfo {pages} {120403} (\bibinfo {year} {2008})}\BibitemShut {NoStop}%
\bibitem [{\citenamefont {{Schomerus}}()}]{Schomerus2016}%
  \BibitemOpen
  \bibfield  {author} {\bibinfo {author} {\bibfnamefont {H.}~\bibnamefont
  {{Schomerus}}},\ }\bibfield  {title} {\bibinfo {title} {{Random matrix
  approaches to open quantum systems}},\ }\href@noop {} {\ }\Eprint
  {https://arxiv.org/abs/1610.05816} {arXiv:1610.05816 [cond-mat.dis-nn]}
  \BibitemShut {NoStop}%
\bibitem [{\citenamefont {Liu}\ \emph {et~al.}(2017{\natexlab{b}})\citenamefont
  {Liu}, \citenamefont {Setiawan}, \citenamefont {Sau},\ and\ \citenamefont
  {Das~Sarma}}]{PhysRevB.96.054520}%
  \BibitemOpen
  \bibfield  {author} {\bibinfo {author} {\bibfnamefont {C.-X.}\ \bibnamefont
  {Liu}}, \bibinfo {author} {\bibfnamefont {F.}~\bibnamefont {Setiawan}},
  \bibinfo {author} {\bibfnamefont {J.~D.}\ \bibnamefont {Sau}},\ and\ \bibinfo
  {author} {\bibfnamefont {S.}~\bibnamefont {Das~Sarma}},\ }\bibfield  {title}
  {\bibinfo {title} {{Phenomenology of the soft gap, zero-bias peak, and
  zero-mode splitting in ideal Majorana nanowires}},\ }\href
  {https://doi.org/10.1103/PhysRevB.96.054520} {\bibfield  {journal} {\bibinfo
  {journal} {Phys. Rev. B}\ }\textbf {\bibinfo {volume} {96}},\ \bibinfo
  {pages} {054520} (\bibinfo {year} {2017}{\natexlab{b}})}\BibitemShut
  {NoStop}%
\bibitem [{\citenamefont {Valenzuela}\ and\ \citenamefont
  {Tinkham}(2006)}]{Tinkham06}%
  \BibitemOpen
  \bibfield  {author} {\bibinfo {author} {\bibfnamefont {S.~O.}\ \bibnamefont
  {Valenzuela}}\ and\ \bibinfo {author} {\bibfnamefont {M.}~\bibnamefont
  {Tinkham}},\ }\bibfield  {title} {\bibinfo {title} {{Direct electronic
  measurement of the spin Hall effect}},\ }\href
  {https://doi.org/10.1038/nature04937} {\bibfield  {journal} {\bibinfo
  {journal} {Nature}\ }\textbf {\bibinfo {volume} {442}},\ \bibinfo {pages}
  {176} (\bibinfo {year} {2006})}\BibitemShut {NoStop}%
\bibitem [{\citenamefont {Br{\"u}ne}\ \emph {et~al.}(2012)\citenamefont
  {Br{\"u}ne}, \citenamefont {Roth}, \citenamefont {Buhmann}, \citenamefont
  {Hankiewicz}, \citenamefont {Molenkamp}, \citenamefont {Maciejko},
  \citenamefont {Qi},\ and\ \citenamefont {Zhang}}]{ZhangNP12}%
  \BibitemOpen
  \bibfield  {author} {\bibinfo {author} {\bibfnamefont {C.}~\bibnamefont
  {Br{\"u}ne}}, \bibinfo {author} {\bibfnamefont {A.}~\bibnamefont {Roth}},
  \bibinfo {author} {\bibfnamefont {H.}~\bibnamefont {Buhmann}}, \bibinfo
  {author} {\bibfnamefont {E.~M.}\ \bibnamefont {Hankiewicz}}, \bibinfo
  {author} {\bibfnamefont {L.~W.}\ \bibnamefont {Molenkamp}}, \bibinfo {author}
  {\bibfnamefont {J.}~\bibnamefont {Maciejko}}, \bibinfo {author}
  {\bibfnamefont {X.-L.}\ \bibnamefont {Qi}},\ and\ \bibinfo {author}
  {\bibfnamefont {S.-C.}\ \bibnamefont {Zhang}},\ }\bibfield  {title} {\bibinfo
  {title} {{Spin polarization of the quantum spin Hall edge states}},\ }\href
  {https://doi.org/10.1038/nphys2322} {\bibfield  {journal} {\bibinfo
  {journal} {Nature Physics}\ }\textbf {\bibinfo {volume} {8}},\ \bibinfo
  {pages} {485} (\bibinfo {year} {2012})}\BibitemShut {NoStop}%
\bibitem [{\citenamefont {Hartman}\ \emph {et~al.}(2018)\citenamefont
  {Hartman}, \citenamefont {Olsen}, \citenamefont
  {L{\ifmmode\ddot{u}\else\"{u}\fi}scher}, \citenamefont {Samani},
  \citenamefont {Fallahi}, \citenamefont {Gardner}, \citenamefont {Manfra},\
  and\ \citenamefont {Folk}}]{entro_expone}%
  \BibitemOpen
  \bibfield  {author} {\bibinfo {author} {\bibfnamefont {N.}~\bibnamefont
  {Hartman}}, \bibinfo {author} {\bibfnamefont {C.}~\bibnamefont {Olsen}},
  \bibinfo {author} {\bibfnamefont {S.}~\bibnamefont
  {L{\ifmmode\ddot{u}\else\"{u}\fi}scher}}, \bibinfo {author} {\bibfnamefont
  {M.}~\bibnamefont {Samani}}, \bibinfo {author} {\bibfnamefont
  {S.}~\bibnamefont {Fallahi}}, \bibinfo {author} {\bibfnamefont {G.~C.}\
  \bibnamefont {Gardner}}, \bibinfo {author} {\bibfnamefont {M.}~\bibnamefont
  {Manfra}},\ and\ \bibinfo {author} {\bibfnamefont {J.}~\bibnamefont {Folk}},\
  }\bibfield  {title} {\bibinfo {title} {{Direct entropy measurement in a
  mesoscopic quantum system}},\ }\href
  {https://doi.org/10.1038/s41567-018-0250-5} {\bibfield  {journal} {\bibinfo
  {journal} {Nat. Phys.}\ }\textbf {\bibinfo {volume} {14}},\ \bibinfo {pages}
  {1083} (\bibinfo {year} {2018})}\BibitemShut {NoStop}%
\bibitem [{\citenamefont {Child}\ \emph
  {et~al.}(2022{\natexlab{a}})\citenamefont {Child}, \citenamefont {Sheekey},
  \citenamefont {L\"uscher}, \citenamefont {Fallahi}, \citenamefont {Gardner},
  \citenamefont {Manfra}, \citenamefont {Mitchell}, \citenamefont {Sela},
  \citenamefont {Kleeorin}, \citenamefont {Meir},\ and\ \citenamefont
  {Folk}}]{entro_exptwo}%
  \BibitemOpen
  \bibfield  {author} {\bibinfo {author} {\bibfnamefont {T.}~\bibnamefont
  {Child}}, \bibinfo {author} {\bibfnamefont {O.}~\bibnamefont {Sheekey}},
  \bibinfo {author} {\bibfnamefont {S.}~\bibnamefont {L\"uscher}}, \bibinfo
  {author} {\bibfnamefont {S.}~\bibnamefont {Fallahi}}, \bibinfo {author}
  {\bibfnamefont {G.~C.}\ \bibnamefont {Gardner}}, \bibinfo {author}
  {\bibfnamefont {M.}~\bibnamefont {Manfra}}, \bibinfo {author} {\bibfnamefont
  {A.}~\bibnamefont {Mitchell}}, \bibinfo {author} {\bibfnamefont
  {E.}~\bibnamefont {Sela}}, \bibinfo {author} {\bibfnamefont {Y.}~\bibnamefont
  {Kleeorin}}, \bibinfo {author} {\bibfnamefont {Y.}~\bibnamefont {Meir}},\
  and\ \bibinfo {author} {\bibfnamefont {J.}~\bibnamefont {Folk}},\ }\bibfield
  {title} {\bibinfo {title} {{Entropy Measurement of a Strongly Coupled Quantum
  Dot}},\ }\href {https://doi.org/10.1103/PhysRevLett.129.227702} {\bibfield
  {journal} {\bibinfo  {journal} {Phys. Rev. Lett.}\ }\textbf {\bibinfo
  {volume} {129}},\ \bibinfo {pages} {227702} (\bibinfo {year}
  {2022}{\natexlab{a}})}\BibitemShut {NoStop}%
\bibitem [{\citenamefont {Kleeorin}\ \emph {et~al.}(2019)\citenamefont
  {Kleeorin}, \citenamefont {Thierschmann}, \citenamefont {Buhmann},
  \citenamefont {Georges}, \citenamefont {Molenkamp},\ and\ \citenamefont
  {Meir}}]{entro_expthree}%
  \BibitemOpen
  \bibfield  {author} {\bibinfo {author} {\bibfnamefont {Y.}~\bibnamefont
  {Kleeorin}}, \bibinfo {author} {\bibfnamefont {H.}~\bibnamefont
  {Thierschmann}}, \bibinfo {author} {\bibfnamefont {H.}~\bibnamefont
  {Buhmann}}, \bibinfo {author} {\bibfnamefont {A.}~\bibnamefont {Georges}},
  \bibinfo {author} {\bibfnamefont {L.~W.}\ \bibnamefont {Molenkamp}},\ and\
  \bibinfo {author} {\bibfnamefont {Y.}~\bibnamefont {Meir}},\ }\bibfield
  {title} {\bibinfo {title} {{How to measure the entropy of a mesoscopic system
  via thermoelectric transport}},\ }\href
  {https://doi.org/10.1038/s41467-019-13630-3} {\bibfield  {journal} {\bibinfo
  {journal} {Nat. Commun.}\ }\textbf {\bibinfo {volume} {10}},\ \bibinfo
  {pages} {5801} (\bibinfo {year} {2019})}\BibitemShut {NoStop}%
\bibitem [{\citenamefont {Han}\ \emph {et~al.}(2022)\citenamefont {Han},
  \citenamefont {Iftikhar}, \citenamefont {Kleeorin}, \citenamefont {Anthore},
  \citenamefont {Pierre}, \citenamefont {Meir}, \citenamefont {Mitchell},\ and\
  \citenamefont {Sela}}]{entro_Kondo}%
  \BibitemOpen
  \bibfield  {author} {\bibinfo {author} {\bibfnamefont {C.}~\bibnamefont
  {Han}}, \bibinfo {author} {\bibfnamefont {Z.}~\bibnamefont {Iftikhar}},
  \bibinfo {author} {\bibfnamefont {Y.}~\bibnamefont {Kleeorin}}, \bibinfo
  {author} {\bibfnamefont {A.}~\bibnamefont {Anthore}}, \bibinfo {author}
  {\bibfnamefont {F.}~\bibnamefont {Pierre}}, \bibinfo {author} {\bibfnamefont
  {Y.}~\bibnamefont {Meir}}, \bibinfo {author} {\bibfnamefont {A.~K.}\
  \bibnamefont {Mitchell}},\ and\ \bibinfo {author} {\bibfnamefont
  {E.}~\bibnamefont {Sela}},\ }\bibfield  {title} {\bibinfo {title}
  {{Fractional Entropy of Multichannel Kondo Systems from Conductance-Charge
  Relations}},\ }\href {https://doi.org/10.1103/PhysRevLett.128.146803}
  {\bibfield  {journal} {\bibinfo  {journal} {Phys. Rev. Lett.}\ }\textbf
  {\bibinfo {volume} {128}},\ \bibinfo {pages} {146803} (\bibinfo {year}
  {2022})}\BibitemShut {NoStop}%
\bibitem [{\citenamefont {Child}\ \emph
  {et~al.}(2022{\natexlab{b}})\citenamefont {Child}, \citenamefont {Sheekey},
  \citenamefont {Lüscher}, \citenamefont {Fallahi}, \citenamefont {Gardner},
  \citenamefont {Manfra},\ and\ \citenamefont {Folk}}]{entro_setup}%
  \BibitemOpen
  \bibfield  {author} {\bibinfo {author} {\bibfnamefont {T.}~\bibnamefont
  {Child}}, \bibinfo {author} {\bibfnamefont {O.}~\bibnamefont {Sheekey}},
  \bibinfo {author} {\bibfnamefont {S.}~\bibnamefont {Lüscher}}, \bibinfo
  {author} {\bibfnamefont {S.}~\bibnamefont {Fallahi}}, \bibinfo {author}
  {\bibfnamefont {G.~C.}\ \bibnamefont {Gardner}}, \bibinfo {author}
  {\bibfnamefont {M.}~\bibnamefont {Manfra}},\ and\ \bibinfo {author}
  {\bibfnamefont {J.}~\bibnamefont {Folk}},\ }\bibfield  {title} {\bibinfo
  {title} {{A Robust Protocol for Entropy Measurement in Mesoscopic
  Circuits}},\ }\bibfield  {journal} {\bibinfo  {journal} {Entropy}\ }\textbf
  {\bibinfo {volume} {24}},\ \href {https://doi.org/10.3390/e24030417}
  {10.3390/e24030417} (\bibinfo {year} {2022}{\natexlab{b}})\BibitemShut
  {NoStop}%
\end{thebibliography}%


\providecommand{\noopsort}[1]{}\providecommand{\singleletter}[1]{#1}%
\begin{thebibliography}{16}%
\makeatletter
\providecommand \@ifxundefined [1]{%
 \@ifx{#1\undefined}
}%
\providecommand \@ifnum [1]{%
 \ifnum #1\expandafter \@firstoftwo
 \else \expandafter \@secondoftwo
 \fi
}%
\providecommand \@ifx [1]{%
 \ifx #1\expandafter \@firstoftwo
 \else \expandafter \@secondoftwo
 \fi
}%
\providecommand \natexlab [1]{#1}%
\providecommand \enquote  [1]{``#1''}%
\providecommand \bibnamefont  [1]{#1}%
\providecommand \bibfnamefont [1]{#1}%
\providecommand \citenamefont [1]{#1}%
\providecommand \href@noop [0]{\@secondoftwo}%
\providecommand \href [0]{\begingroup \@sanitize@url \@href}%
\providecommand \@href[1]{\@@startlink{#1}\@@href}%
\providecommand \@@href[1]{\endgroup#1\@@endlink}%
\providecommand \@sanitize@url [0]{\catcode `\\12\catcode `\$12\catcode
  `\&12\catcode `\#12\catcode `\^12\catcode `\_12\catcode `\%12\relax}%
\providecommand \@@startlink[1]{}%
\providecommand \@@endlink[0]{}%
\providecommand \url  [0]{\begingroup\@sanitize@url \@url }%
\providecommand \@url [1]{\endgroup\@href {#1}{\urlprefix }}%
\providecommand \urlprefix  [0]{URL }%
\providecommand \Eprint [0]{\href }%
\providecommand \doibase [0]{https://doi.org/}%
\providecommand \selectlanguage [0]{\@gobble}%
\providecommand \bibinfo  [0]{\@secondoftwo}%
\providecommand \bibfield  [0]{\@secondoftwo}%
\providecommand \translation [1]{[#1]}%
\providecommand \BibitemOpen [0]{}%
\providecommand \bibitemStop [0]{}%
\providecommand \bibitemNoStop [0]{.\EOS\space}%
\providecommand \EOS [0]{\spacefactor3000\relax}%
\providecommand \BibitemShut  [1]{\csname bibitem#1\endcsname}%
\let\auto@bib@innerbib\@empty
\bibitem [{\citenamefont {Tserkovnyak}\ \emph {et~al.}(2005)\citenamefont
  {Tserkovnyak}, \citenamefont {Brataas}, \citenamefont {Bauer},\ and\
  \citenamefont {Halperin}}]{TserkovnyakRMP2005}%
  \BibitemOpen
  \bibfield  {author} {\bibinfo {author} {\bibfnamefont {Y.}~\bibnamefont
  {Tserkovnyak}}, \bibinfo {author} {\bibfnamefont {A.}~\bibnamefont
  {Brataas}}, \bibinfo {author} {\bibfnamefont {G.~E.~W.}\ \bibnamefont
  {Bauer}},\ and\ \bibinfo {author} {\bibfnamefont {B.~I.}\ \bibnamefont
  {Halperin}},\ }\bibfield  {title} {\bibinfo {title} {Nonlocal magnetization
  dynamics in ferromagnetic heterostructures},\ }\href
  {https://doi.org/10.1103/RevModPhys.77.1375} {\bibfield  {journal} {\bibinfo
  {journal} {Rev. Mod. Phys.}\ }\textbf {\bibinfo {volume} {77}},\ \bibinfo
  {pages} {1375} (\bibinfo {year} {2005})}\BibitemShut {NoStop}%
\bibitem [{\citenamefont {Kells}\ \emph {et~al.}(2012)\citenamefont {Kells},
  \citenamefont {Meidan},\ and\ \citenamefont {Brouwer}}]{inh_Brouwer}%
  \BibitemOpen
  \bibfield  {author} {\bibinfo {author} {\bibfnamefont {G.}~\bibnamefont
  {Kells}}, \bibinfo {author} {\bibfnamefont {D.}~\bibnamefont {Meidan}},\ and\
  \bibinfo {author} {\bibfnamefont {P.~W.}\ \bibnamefont {Brouwer}},\
  }\bibfield  {title} {\bibinfo {title} {Near-zero-energy end states in
  topologically trivial spin-orbit coupled superconducting nanowires with a
  smooth confinement},\ }\href {https://doi.org/10.1103/PhysRevB.86.100503}
  {\bibfield  {journal} {\bibinfo  {journal} {Phys. Rev. B}\ }\textbf {\bibinfo
  {volume} {86}},\ \bibinfo {pages} {100503(R)} (\bibinfo {year}
  {2012})}\BibitemShut {NoStop}%
\bibitem [{\citenamefont {Liu}\ \emph {et~al.}(2017)\citenamefont {Liu},
  \citenamefont {Sau}, \citenamefont {Stanescu},\ and\ \citenamefont
  {Das~Sarma}}]{QD_Liu}%
  \BibitemOpen
  \bibfield  {author} {\bibinfo {author} {\bibfnamefont {C.-X.}\ \bibnamefont
  {Liu}}, \bibinfo {author} {\bibfnamefont {J.~D.}\ \bibnamefont {Sau}},
  \bibinfo {author} {\bibfnamefont {T.~D.}\ \bibnamefont {Stanescu}},\ and\
  \bibinfo {author} {\bibfnamefont {S.}~\bibnamefont {Das~Sarma}},\ }\bibfield
  {title} {\bibinfo {title} {{Andreev bound states versus Majorana bound states
  in quantum dot-nanowire-superconductor hybrid structures: Trivial versus
  topological zero-bias conductance peaks}},\ }\href
  {https://doi.org/10.1103/PhysRevB.96.075161} {\bibfield  {journal} {\bibinfo
  {journal} {Phys. Rev. B}\ }\textbf {\bibinfo {volume} {96}},\ \bibinfo
  {pages} {075161} (\bibinfo {year} {2017})}\BibitemShut {NoStop}%
\bibitem [{\citenamefont {Moore}\ \emph
  {et~al.}(2018{\natexlab{a}})\citenamefont {Moore}, \citenamefont {Stanescu},\
  and\ \citenamefont {Tewari}}]{QD_Moore_one}%
  \BibitemOpen
  \bibfield  {author} {\bibinfo {author} {\bibfnamefont {C.}~\bibnamefont
  {Moore}}, \bibinfo {author} {\bibfnamefont {T.~D.}\ \bibnamefont
  {Stanescu}},\ and\ \bibinfo {author} {\bibfnamefont {S.}~\bibnamefont
  {Tewari}},\ }\bibfield  {title} {\bibinfo {title} {{Two-terminal charge
  tunneling: Disentangling Majorana zero modes from partially separated Andreev
  bound states in semiconductor-superconductor heterostructures}},\ }\href
  {https://doi.org/10.1103/PhysRevB.97.165302} {\bibfield  {journal} {\bibinfo
  {journal} {Phys. Rev. B}\ }\textbf {\bibinfo {volume} {97}},\ \bibinfo
  {pages} {165302} (\bibinfo {year} {2018}{\natexlab{a}})}\BibitemShut
  {NoStop}%
\bibitem [{\citenamefont {Moore}\ \emph
  {et~al.}(2018{\natexlab{b}})\citenamefont {Moore}, \citenamefont {Zeng},
  \citenamefont {Stanescu},\ and\ \citenamefont {Tewari}}]{QD_Moore_two}%
  \BibitemOpen
  \bibfield  {author} {\bibinfo {author} {\bibfnamefont {C.}~\bibnamefont
  {Moore}}, \bibinfo {author} {\bibfnamefont {C.}~\bibnamefont {Zeng}},
  \bibinfo {author} {\bibfnamefont {T.~D.}\ \bibnamefont {Stanescu}},\ and\
  \bibinfo {author} {\bibfnamefont {S.}~\bibnamefont {Tewari}},\ }\bibfield
  {title} {\bibinfo {title} {{Quantized zero-bias conductance plateau in
  semiconductor-superconductor heterostructures without topological Majorana
  zero modes}},\ }\href {https://doi.org/10.1103/PhysRevB.98.155314} {\bibfield
   {journal} {\bibinfo  {journal} {Phys. Rev. B}\ }\textbf {\bibinfo {volume}
  {98}},\ \bibinfo {pages} {155314} (\bibinfo {year}
  {2018}{\natexlab{b}})}\BibitemShut {NoStop}%
\bibitem [{\citenamefont {Woods}\ \emph {et~al.}(2018)\citenamefont {Woods},
  \citenamefont {Stanescu},\ and\ \citenamefont {Das~Sarma}}]{inh_workfunc}%
  \BibitemOpen
  \bibfield  {author} {\bibinfo {author} {\bibfnamefont {B.~D.}\ \bibnamefont
  {Woods}}, \bibinfo {author} {\bibfnamefont {T.~D.}\ \bibnamefont
  {Stanescu}},\ and\ \bibinfo {author} {\bibfnamefont {S.}~\bibnamefont
  {Das~Sarma}},\ }\bibfield  {title} {\bibinfo {title} {{Effective theory
  approach to the Schr\"odinger-Poisson problem in semiconductor Majorana
  devices}},\ }\href {https://doi.org/10.1103/PhysRevB.98.035428} {\bibfield
  {journal} {\bibinfo  {journal} {Phys. Rev. B}\ }\textbf {\bibinfo {volume}
  {98}},\ \bibinfo {pages} {035428} (\bibinfo {year} {2018})}\BibitemShut
  {NoStop}%
\bibitem [{\citenamefont {Vuik}\ \emph {et~al.}(2019)\citenamefont {Vuik},
  \citenamefont {Nijholt}, \citenamefont {Akhmerov},\ and\ \citenamefont
  {Wimmer}}]{qMajo_Vuik}%
  \BibitemOpen
  \bibfield  {author} {\bibinfo {author} {\bibfnamefont {A.}~\bibnamefont
  {Vuik}}, \bibinfo {author} {\bibfnamefont {B.}~\bibnamefont {Nijholt}},
  \bibinfo {author} {\bibfnamefont {A.~R.}\ \bibnamefont {Akhmerov}},\ and\
  \bibinfo {author} {\bibfnamefont {M.}~\bibnamefont {Wimmer}},\ }\bibfield
  {title} {\bibinfo {title} {{Reproducing topological properties with
  quasi-Majorana states}},\ }\href
  {https://doi.org/10.21468/SciPostPhys.7.5.061} {\bibfield  {journal}
  {\bibinfo  {journal} {SciPost Phys.}\ }\textbf {\bibinfo {volume} {7}},\
  \bibinfo {pages} {061} (\bibinfo {year} {2019})}\BibitemShut {NoStop}%
\bibitem [{\citenamefont {Smirnov}(2015)}]{entro_smirnov_one}%
  \BibitemOpen
  \bibfield  {author} {\bibinfo {author} {\bibfnamefont {S.}~\bibnamefont
  {Smirnov}},\ }\bibfield  {title} {\bibinfo {title} {Majorana tunneling
  entropy},\ }\href {https://doi.org/10.1103/PhysRevB.92.195312} {\bibfield
  {journal} {\bibinfo  {journal} {Phys. Rev. B}\ }\textbf {\bibinfo {volume}
  {92}},\ \bibinfo {pages} {195312} (\bibinfo {year} {2015})}\BibitemShut
  {NoStop}%
\bibitem [{\citenamefont {Sela}\ \emph {et~al.}(2019)\citenamefont {Sela},
  \citenamefont {Oreg}, \citenamefont {Plugge}, \citenamefont {Hartman},
  \citenamefont {L\"uscher},\ and\ \citenamefont {Folk}}]{entro_Sela}%
  \BibitemOpen
  \bibfield  {author} {\bibinfo {author} {\bibfnamefont {E.}~\bibnamefont
  {Sela}}, \bibinfo {author} {\bibfnamefont {Y.}~\bibnamefont {Oreg}}, \bibinfo
  {author} {\bibfnamefont {S.}~\bibnamefont {Plugge}}, \bibinfo {author}
  {\bibfnamefont {N.}~\bibnamefont {Hartman}}, \bibinfo {author} {\bibfnamefont
  {S.}~\bibnamefont {L\"uscher}},\ and\ \bibinfo {author} {\bibfnamefont
  {J.}~\bibnamefont {Folk}},\ }\bibfield  {title} {\bibinfo {title} {{Detecting
  the Universal Fractional Entropy of Majorana Zero Modes}},\ }\href
  {https://doi.org/10.1103/PhysRevLett.123.147702} {\bibfield  {journal}
  {\bibinfo  {journal} {Phys. Rev. Lett.}\ }\textbf {\bibinfo {volume} {123}},\
  \bibinfo {pages} {147702} (\bibinfo {year} {2019})}\BibitemShut {NoStop}%
\bibitem [{\citenamefont {Hartman}\ \emph {et~al.}(2018)\citenamefont
  {Hartman}, \citenamefont {Olsen}, \citenamefont
  {L{\ifmmode\ddot{u}\else\"{u}\fi}scher}, \citenamefont {Samani},
  \citenamefont {Fallahi}, \citenamefont {Gardner}, \citenamefont {Manfra},\
  and\ \citenamefont {Folk}}]{entro_expone}%
  \BibitemOpen
  \bibfield  {author} {\bibinfo {author} {\bibfnamefont {N.}~\bibnamefont
  {Hartman}}, \bibinfo {author} {\bibfnamefont {C.}~\bibnamefont {Olsen}},
  \bibinfo {author} {\bibfnamefont {S.}~\bibnamefont
  {L{\ifmmode\ddot{u}\else\"{u}\fi}scher}}, \bibinfo {author} {\bibfnamefont
  {M.}~\bibnamefont {Samani}}, \bibinfo {author} {\bibfnamefont
  {S.}~\bibnamefont {Fallahi}}, \bibinfo {author} {\bibfnamefont {G.~C.}\
  \bibnamefont {Gardner}}, \bibinfo {author} {\bibfnamefont {M.}~\bibnamefont
  {Manfra}},\ and\ \bibinfo {author} {\bibfnamefont {J.}~\bibnamefont {Folk}},\
  }\bibfield  {title} {\bibinfo {title} {{Direct entropy measurement in a
  mesoscopic quantum system}},\ }\href
  {https://doi.org/10.1038/s41567-018-0250-5} {\bibfield  {journal} {\bibinfo
  {journal} {Nat. Phys.}\ }\textbf {\bibinfo {volume} {14}},\ \bibinfo {pages}
  {1083} (\bibinfo {year} {2018})}\BibitemShut {NoStop}%
\bibitem [{\citenamefont {Child}\ \emph
  {et~al.}(2022{\natexlab{a}})\citenamefont {Child}, \citenamefont {Sheekey},
  \citenamefont {L\"uscher}, \citenamefont {Fallahi}, \citenamefont {Gardner},
  \citenamefont {Manfra}, \citenamefont {Mitchell}, \citenamefont {Sela},
  \citenamefont {Kleeorin}, \citenamefont {Meir},\ and\ \citenamefont
  {Folk}}]{entro_exptwo}%
  \BibitemOpen
  \bibfield  {author} {\bibinfo {author} {\bibfnamefont {T.}~\bibnamefont
  {Child}}, \bibinfo {author} {\bibfnamefont {O.}~\bibnamefont {Sheekey}},
  \bibinfo {author} {\bibfnamefont {S.}~\bibnamefont {L\"uscher}}, \bibinfo
  {author} {\bibfnamefont {S.}~\bibnamefont {Fallahi}}, \bibinfo {author}
  {\bibfnamefont {G.~C.}\ \bibnamefont {Gardner}}, \bibinfo {author}
  {\bibfnamefont {M.}~\bibnamefont {Manfra}}, \bibinfo {author} {\bibfnamefont
  {A.}~\bibnamefont {Mitchell}}, \bibinfo {author} {\bibfnamefont
  {E.}~\bibnamefont {Sela}}, \bibinfo {author} {\bibfnamefont {Y.}~\bibnamefont
  {Kleeorin}}, \bibinfo {author} {\bibfnamefont {Y.}~\bibnamefont {Meir}},\
  and\ \bibinfo {author} {\bibfnamefont {J.}~\bibnamefont {Folk}},\ }\bibfield
  {title} {\bibinfo {title} {{Entropy Measurement of a Strongly Coupled Quantum
  Dot}},\ }\href {https://doi.org/10.1103/PhysRevLett.129.227702} {\bibfield
  {journal} {\bibinfo  {journal} {Phys. Rev. Lett.}\ }\textbf {\bibinfo
  {volume} {129}},\ \bibinfo {pages} {227702} (\bibinfo {year}
  {2022}{\natexlab{a}})}\BibitemShut {NoStop}%
\bibitem [{\citenamefont {Kleeorin}\ \emph {et~al.}(2019)\citenamefont
  {Kleeorin}, \citenamefont {Thierschmann}, \citenamefont {Buhmann},
  \citenamefont {Georges}, \citenamefont {Molenkamp},\ and\ \citenamefont
  {Meir}}]{entro_expthree}%
  \BibitemOpen
  \bibfield  {author} {\bibinfo {author} {\bibfnamefont {Y.}~\bibnamefont
  {Kleeorin}}, \bibinfo {author} {\bibfnamefont {H.}~\bibnamefont
  {Thierschmann}}, \bibinfo {author} {\bibfnamefont {H.}~\bibnamefont
  {Buhmann}}, \bibinfo {author} {\bibfnamefont {A.}~\bibnamefont {Georges}},
  \bibinfo {author} {\bibfnamefont {L.~W.}\ \bibnamefont {Molenkamp}},\ and\
  \bibinfo {author} {\bibfnamefont {Y.}~\bibnamefont {Meir}},\ }\bibfield
  {title} {\bibinfo {title} {{How to measure the entropy of a mesoscopic system
  via thermoelectric transport}},\ }\href
  {https://doi.org/10.1038/s41467-019-13630-3} {\bibfield  {journal} {\bibinfo
  {journal} {Nat. Commun.}\ }\textbf {\bibinfo {volume} {10}},\ \bibinfo
  {pages} {5801} (\bibinfo {year} {2019})}\BibitemShut {NoStop}%
\bibitem [{\citenamefont {Han}\ \emph {et~al.}(2022)\citenamefont {Han},
  \citenamefont {Iftikhar}, \citenamefont {Kleeorin}, \citenamefont {Anthore},
  \citenamefont {Pierre}, \citenamefont {Meir}, \citenamefont {Mitchell},\ and\
  \citenamefont {Sela}}]{entro_Kondo}%
  \BibitemOpen
  \bibfield  {author} {\bibinfo {author} {\bibfnamefont {C.}~\bibnamefont
  {Han}}, \bibinfo {author} {\bibfnamefont {Z.}~\bibnamefont {Iftikhar}},
  \bibinfo {author} {\bibfnamefont {Y.}~\bibnamefont {Kleeorin}}, \bibinfo
  {author} {\bibfnamefont {A.}~\bibnamefont {Anthore}}, \bibinfo {author}
  {\bibfnamefont {F.}~\bibnamefont {Pierre}}, \bibinfo {author} {\bibfnamefont
  {Y.}~\bibnamefont {Meir}}, \bibinfo {author} {\bibfnamefont {A.~K.}\
  \bibnamefont {Mitchell}},\ and\ \bibinfo {author} {\bibfnamefont
  {E.}~\bibnamefont {Sela}},\ }\bibfield  {title} {\bibinfo {title}
  {{Fractional Entropy of Multichannel Kondo Systems from Conductance-Charge
  Relations}},\ }\href {https://doi.org/10.1103/PhysRevLett.128.146803}
  {\bibfield  {journal} {\bibinfo  {journal} {Phys. Rev. Lett.}\ }\textbf
  {\bibinfo {volume} {128}},\ \bibinfo {pages} {146803} (\bibinfo {year}
  {2022})}\BibitemShut {NoStop}%
\bibitem [{\citenamefont {Child}\ \emph
  {et~al.}(2022{\natexlab{b}})\citenamefont {Child}, \citenamefont {Sheekey},
  \citenamefont {Lüscher}, \citenamefont {Fallahi}, \citenamefont {Gardner},
  \citenamefont {Manfra},\ and\ \citenamefont {Folk}}]{entro_setup}%
  \BibitemOpen
  \bibfield  {author} {\bibinfo {author} {\bibfnamefont {T.}~\bibnamefont
  {Child}}, \bibinfo {author} {\bibfnamefont {O.}~\bibnamefont {Sheekey}},
  \bibinfo {author} {\bibfnamefont {S.}~\bibnamefont {Lüscher}}, \bibinfo
  {author} {\bibfnamefont {S.}~\bibnamefont {Fallahi}}, \bibinfo {author}
  {\bibfnamefont {G.~C.}\ \bibnamefont {Gardner}}, \bibinfo {author}
  {\bibfnamefont {M.}~\bibnamefont {Manfra}},\ and\ \bibinfo {author}
  {\bibfnamefont {J.}~\bibnamefont {Folk}},\ }\bibfield  {title} {\bibinfo
  {title} {{A Robust Protocol for Entropy Measurement in Mesoscopic
  Circuits}},\ }\bibfield  {journal} {\bibinfo  {journal} {Entropy}\ }\textbf
  {\bibinfo {volume} {24}},\ \href {https://doi.org/10.3390/e24030417}
  {10.3390/e24030417} (\bibinfo {year} {2022}{\natexlab{b}})\BibitemShut
  {NoStop}%
\bibitem [{\citenamefont {Smirnov}(2021{\natexlab{a}})}]{entro_smirnov_two}%
  \BibitemOpen
  \bibfield  {author} {\bibinfo {author} {\bibfnamefont {S.}~\bibnamefont
  {Smirnov}},\ }\bibfield  {title} {\bibinfo {title} {Majorana entropy revival
  via tunneling phases},\ }\href {https://doi.org/10.1103/PhysRevB.103.075440}
  {\bibfield  {journal} {\bibinfo  {journal} {Phys. Rev. B}\ }\textbf {\bibinfo
  {volume} {103}},\ \bibinfo {pages} {075440} (\bibinfo {year}
  {2021}{\natexlab{a}})}\BibitemShut {NoStop}%
\bibitem [{\citenamefont {Smirnov}(2021{\natexlab{b}})}]{entro_smirnov_three}%
  \BibitemOpen
  \bibfield  {author} {\bibinfo {author} {\bibfnamefont {S.}~\bibnamefont
  {Smirnov}},\ }\bibfield  {title} {\bibinfo {title} {Majorana ensembles with
  fractional entropy and conductance in nanoscopic systems},\ }\href
  {https://doi.org/10.1103/PhysRevB.104.205406} {\bibfield  {journal} {\bibinfo
   {journal} {Phys. Rev. B}\ }\textbf {\bibinfo {volume} {104}},\ \bibinfo
  {pages} {205406} (\bibinfo {year} {2021}{\natexlab{b}})}\BibitemShut
  {NoStop}%
\end{thebibliography}%

\end{document}


\title{Supplementary Material for ``Quantized spin pumping in topological ferromagnetic-superconducting  nanowires"}

\author{V. Fern\'andez Becerra}
\affiliation{International Research Centre MagTop, Institute of Physics, Polish Academy of Sciences, Aleja Lotnikow 32/46, PL-02668 Warsaw, Poland}
\author{Mircea Trif}
\affiliation{International Research Centre MagTop, Institute of Physics, Polish Academy of Sciences, Aleja Lotnikow 32/46, PL-02668 Warsaw, Poland}
\author{Timo Hyart} 
\affiliation{International Research Centre MagTop, Institute of Physics, Polish Academy of Sciences, Aleja Lotnikow 32/46, PL-02668 Warsaw, Poland}
\affiliation{Department of Applied Physics, Aalto University, 00076 Aalto, Espoo, Finland}
\affiliation{Computational Physics Laboratory, Physics Unit, Faculty of Engineering and Natural Sciences, Tampere University, FI-33014 Tampere, Finland}

\date{\today}

\maketitle

\section{Analytical solution of the scattering states}

\begin{figure}[ttb]
\vspace{-0.2cm}
\includegraphics[width=0.98\linewidth]{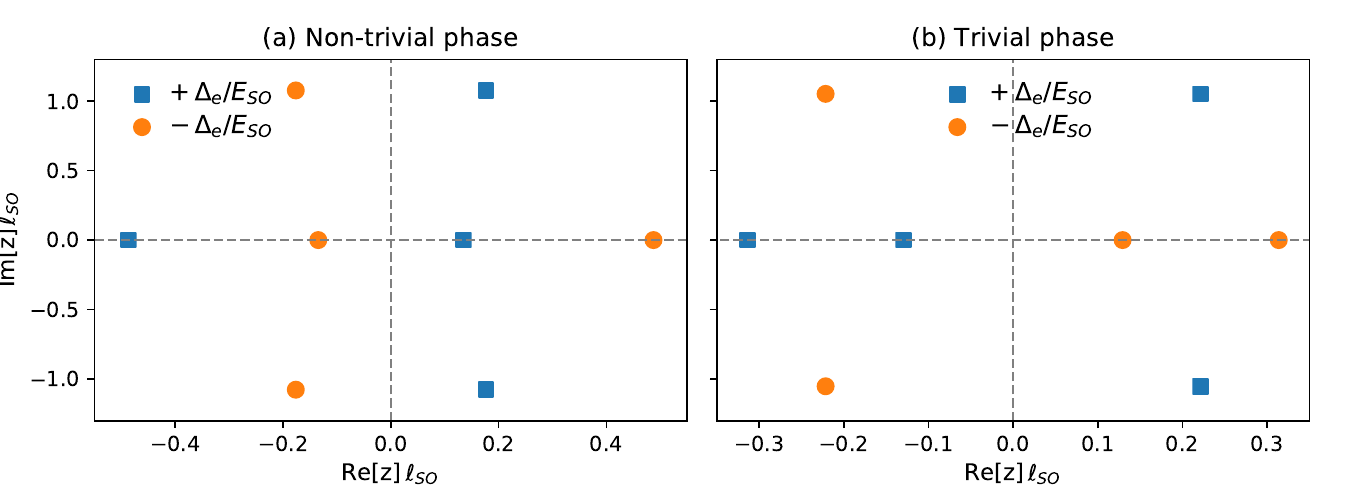}
\vspace{-0.4cm}
\caption{Roots to the characteristic equations (\ref{characteristic-eq}) in the topologically nontrivial and trivial gapped phases. Squares and circles denote the roots of the fourth order polynomial with the sign of the first order coefficient positive or negative, respectively. The values of the parameters used to illustrate the distribution of the roots are: $\Delta_0=E_{SO}$, $\theta=2\pi/5$, $\mu=0$ and (a) $m_0=3E_{SO}/2$ (b) $m_0=E_{SO}/2$.}
\label{roots}
\end{figure}

We consider the scattering states in the lead [$x<0$ in Fig.~1(a) in the main text] at energy $E=0$ and $\omega \to 0$ 
\begin{eqnarray}
    &\Psi_{\rm rot}^{\alpha, L}(x) = \left(\begin{array}{c}\chi_{_{\alpha}} \\ 0 \end{array}\right)e^{ixk_{in}^{\alpha}} + \sum_{\beta}\Bigl[
    \tilde{r}_{ee}^{\beta\alpha}\left(\begin{array}{c}\chi_{_{\beta}}\\ 0 \end{array}\right)e^{-ixk_o^{\beta}}
        +\tilde{r}_{he}^{\beta\alpha}\left(\begin{array}{c} 0 \\ \chi_{_{\overline{\beta}}} \end{array}\right)e^{ixk_o^{\beta}}
        \Bigr], 
\label{psi_lead}        
\end{eqnarray}
composed of an incoming electron with spin-$z$ eigenvalue $\alpha=\pm 1$ in eigenstate $\chi_\alpha=(1+\alpha, 1-\alpha)^T/2$ and  $k^{\alpha} \ell_{SO}=\bigl(\alpha  \!+\! \sqrt{1\!+\! \mu_N/E_{SO} \!}\,\bigr)\big/2$, and four outgoing states with $k_o^{e,\beta} \ell_{SO}=\bigl(-\beta \!+\! \sqrt{1\!+\! \mu_N/E_{SO} \!}\,\bigr)\big/2$. To calculate the reflection coefficients $\tilde{r}_{ee}^{\beta \alpha}$ and $\tilde{r}_{he}^{\beta \alpha}$  we also need to calculate the evanescent states in the ferromagnetic-superconducting nanowire [$x>0$ in Fig.~1(a) in the main text]. The decay lengths of the evanescent states are obtained from the roots of the characteristic equations with $\Re[z_i]>0$
\begin{equation}
\bigg(4z^2 \ell_{SO}^2 +\frac{\mu}{E_{SO}}\bigg)^2 +\bigg(4z \ell_{SO} \pm \frac{\Delta_e}{E_{SO}}\bigg)^2 -\frac{m_{0}^2 \sin^2\theta}{E_{SO}^2}=0, 
\label{characteristic-eq} 
\end{equation}
where $\Delta_e = \sqrt{\Delta_0^2 -m_0^2 \cos^2\theta}$. In Fig.~\ref{roots} we show that the distribution of roots of Eqs.~(\ref{characteristic-eq}) changes depending on whether the system is the topologically trivial or nontrivial gapped phase. The evanescent states in the nontrivial phase stem from triplet ($z_{1,2,3}$) and singlet ($z_4$) roots of the two characteristic equations (\ref{characteristic-eq}) with $\pm$-signs, respectively, whereas  in the trivial phase they stem from two pair of doublets, one doublet for each sign. The wavefunction in both topologically gapped phases is thus represented by 

\begin{eqnarray}
    \label{TopoSol}
    \Psi^{\alpha,R}_{\rm rot}(x) &=& \sum_{j=1}^4 A^\alpha_j\left(
    \begin{array}{c}
    \psi_j \\
    s(j) i e^{ is(j) \lambda}\sigma_z\psi_j
    \end{array}
    \right)e^{-z_jx}, 
    \label{SC_wf}
\end{eqnarray}
\begin{equation*}
\psi_{j}\!=\!\frac{-z_j}{\ell_{SO}}\biggl(4iz_j^2 \ell_{SO}^2+4z_j \ell_{SO}+ \frac{i\mu-s(j)\Delta_e  }{E_{SO}},\frac{i m_0 \sin \theta}{E_{SO}} \biggr)^T,
\end{equation*}
where $e^{i\lambda}\!=\! (\Delta_e \!+\! i m_0\cos\theta)/\Delta_0$, and
\begin{equation}
s(j)=\left\{
\begin{array}{ll}
(-1)^{1+\delta_{j,4}}, & \mathrm{nontrivial\, phase} \\
(-1)^{1+\delta_{j,3}+\delta_{j,4}}, & \mathrm{trivial\, phase}
\end{array}
\right..
\end{equation}
The reflection coefficients are calculated from the continuity equations, $\Psi_{\rm rot}^{\alpha, L}(0)=\Psi_{\rm rot}^{\alpha, R}(0)$ and $\partial_x\Psi_{\rm rot}^{\alpha, L}(0)=\partial_x\Psi_{\rm rot}^{\alpha, R}(0)$. Multiplication of the continuity equations by the $4\times 4$ block diagonal matrix $M=\mathrm{diag}(\mathbb{1},-ie^{i\lambda}\sigma_z )$ facilitates the elimination of some spinors so that in the nontrivial case we obtain

%
\begin{eqnarray}
\label{NtrivrefeLeq}
    \sum_{\beta}\Bigl[ \tilde{r}_{ee}^{\beta\alpha}\chi_{\beta} \!-\!ie^{i\lambda}\tilde{r}_{he}^{\beta\alpha}\sigma_z\chi_{\overline{\beta}}\Bigr] &=& \! A_4\psi_4\Lambda \!-\! \chi_\alpha,  \\
    \sum_{\beta}\Bigl[ i\frac{k_o^{\beta}}{z_4}\tilde{r}_{ee}^{\beta\alpha}\chi_{\beta} \!-\!\frac{k_o^{\beta}}{z_4}e^{i\lambda}\tilde{r}_{he}^{\beta\alpha}\sigma_z\chi_{\overline{\beta}}\Bigr] &=&
     \!A_4\psi_4\Lambda \!+\!i\frac{k_{in}^{\alpha}}{z_4}\chi_{\alpha},
\end{eqnarray}
%
while in the trivial we get
\begin{eqnarray}
\label{trivrefLeq}
    \sum_{\beta}\Bigl[ \tilde{r}_{ee}^{\beta\alpha}\chi_{\beta} \!-\!ie^{i\lambda}\tilde{r}_{he}^{\beta\alpha}\sigma_z\chi_{\overline{\beta}}\Bigr] &=& 
    \! A_3\psi_3\Lambda \!+\! A_4\psi_4\Lambda \!-\! \chi_\alpha,  \\
    \sum_{\beta}\Bigl[ i\frac{k_o^{\beta}}{z_4}\tilde{r}_{ee}^{\beta\alpha}\chi_{\beta} \!-\!\frac{k_o^{\beta}}{z_4}e^{i\lambda}\tilde{r}_{he}^{\beta\alpha}\sigma_z\chi_{\overline{\beta}}\Bigr] &=&
     \!A_3\psi_3\Lambda \!+\! A_4\psi_4\Lambda \!+\!i\frac{k_{in}^{\alpha}}{z_4}\chi_{\alpha}\,.
\end{eqnarray}
The different structure of these equations is the reason why conductance and the spin pumping are correlated (uncorrelated) in the topologically nontrivial (trivial) phase.

We now focus on the solutions in the nontrivial phase. By utilizing $k_{in}^{\alpha}=k_o^{\overline{\alpha}}$ in the solutions to Eq.~(\ref{NtrivrefeLeq}), we obtain  
\begin{eqnarray}
\tilde{r}_{ee}^{\beta\beta} &=& -e^{i \beta \gamma} \bigl(k_o^{\overline{\beta}} \ell_{SO} - i z_4 \ell_{SO} \bigr)\tilde{A}_4^{\beta}, \nonumber\\
\tilde{r}_{ee}^{\overline{\beta}\beta} &=& -i\beta  \bigl(k_o^{\beta} \ell_{SO} -i z_4 \ell_{SO} \bigr)\tilde{A}_4^{\beta}, \nonumber\\
\tilde{r}_{he}^{\overline{\beta}\beta} &=& -i\beta e^{-i\lambda}\Bigl\{
e^{i\beta \gamma} \bigl(k_o^{\beta} \ell_{SO} + i z_4 \ell_{SO} \bigr)\tilde{A}_4^{\beta} +1\Bigr\},\nonumber\\
\tilde{r}_{he}^{\beta\beta} &=& - e^{-i\lambda} \bigl(k_o^{\overline{\beta}} \ell_{SO} + i z_4 \ell_{SO} \bigr)\tilde{A}_4^{\beta}, \label{reflection-coeff}
\end{eqnarray}
where
\begin{equation*}
\tilde{A}_4^\uparrow=A_4^{\uparrow} \frac{ m_0 \sin\theta}{E_{SO}}\frac{\Lambda z_4}{k_o^\uparrow+k_0^{\downarrow}}, \ \tilde{A}_4^\downarrow=i e^{i \gamma} A_4^{\downarrow} \frac{ m_0 \sin\theta}{E_{SO}}\frac{\Lambda z_4}{k_o^\uparrow+k_0^{\downarrow}},
\end{equation*}
$\Lambda = 1+e^{2i\lambda}$ and  
\begin{equation*}
e^{i \gamma}=\frac{4z_4 \ell_{SO} -\Delta_e/E_{SO}  +i(4z_4^2 \ell_{SO}^2+\mu/E_{SO})}{m_0 \sin \theta/E_{SO}}.    
\end{equation*} 
Because the scattering matrix has to be  unitary  we get
\begin{equation*}
\tilde{A}_4^{\beta}=-e^{-i \beta \gamma}\frac{k_o^{\beta} \ell_{SO}-iz_4 \ell_{SO}}{(k_o^{\uparrow})^2 \ell_{SO}^2  + (k_o^{\downarrow})^2 \ell_{SO}^2 + 2z_4^2 \ell_{SO}^2}.
\end{equation*}
Using these formulas we straightforwardly obtain the expressions quoted in the main text.

\section{Reflection coefficients in the  Mahaux-Weidenm\"{u}ller approach and comparison to the numerical results}

\begin{figure}[h]
\vspace{-0.2cm}
\includegraphics[width=0.99\linewidth]{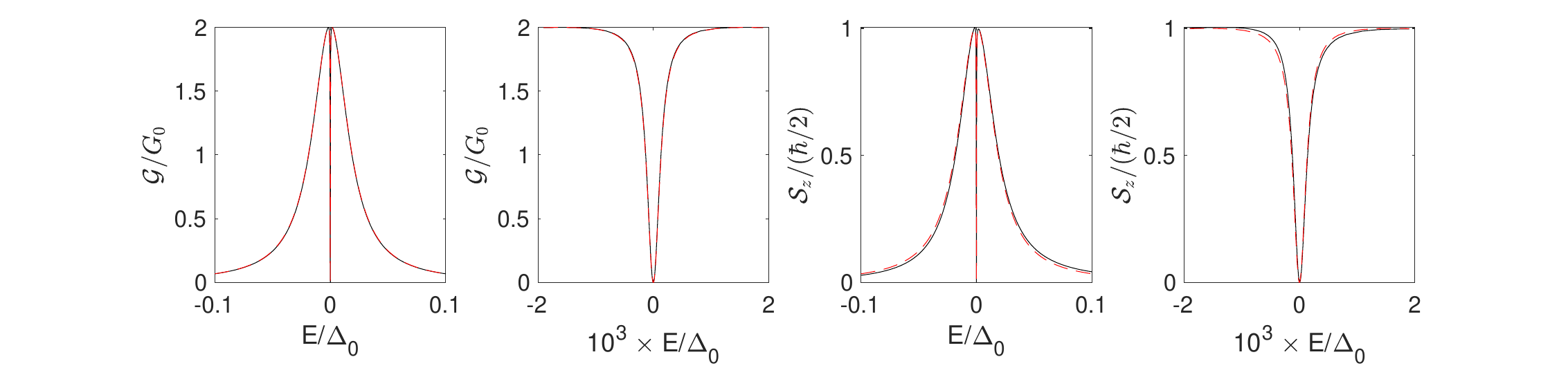}
\vspace{-0.4cm}
\caption{Spectral density of conductance ${\cal G}(E)$ and pumped spin ${\cal S}_z(E)$ as a function of $E$ for $\Delta=E_{SO}$, $m_0=2E_{SO}$ and $L=27 \ell_{SO}$ (solid black lines). In finite wires  both quantities typically have a sharp dip at low energies due to the coupling of MZMs. The magnifications to low energies are shown to make these dips more clearly visible. The dashed red lines show the results obtained using the analytic formula (\ref{supp:usual}) with $\Gamma = 0.0188 \Delta_0$ and
$E_M = 0.0016\Delta_0$.}
\label{suppCondSL27}
\end{figure}

\begin{figure}[h]
\vspace{-0.4cm}
\includegraphics[width=0.99\linewidth]{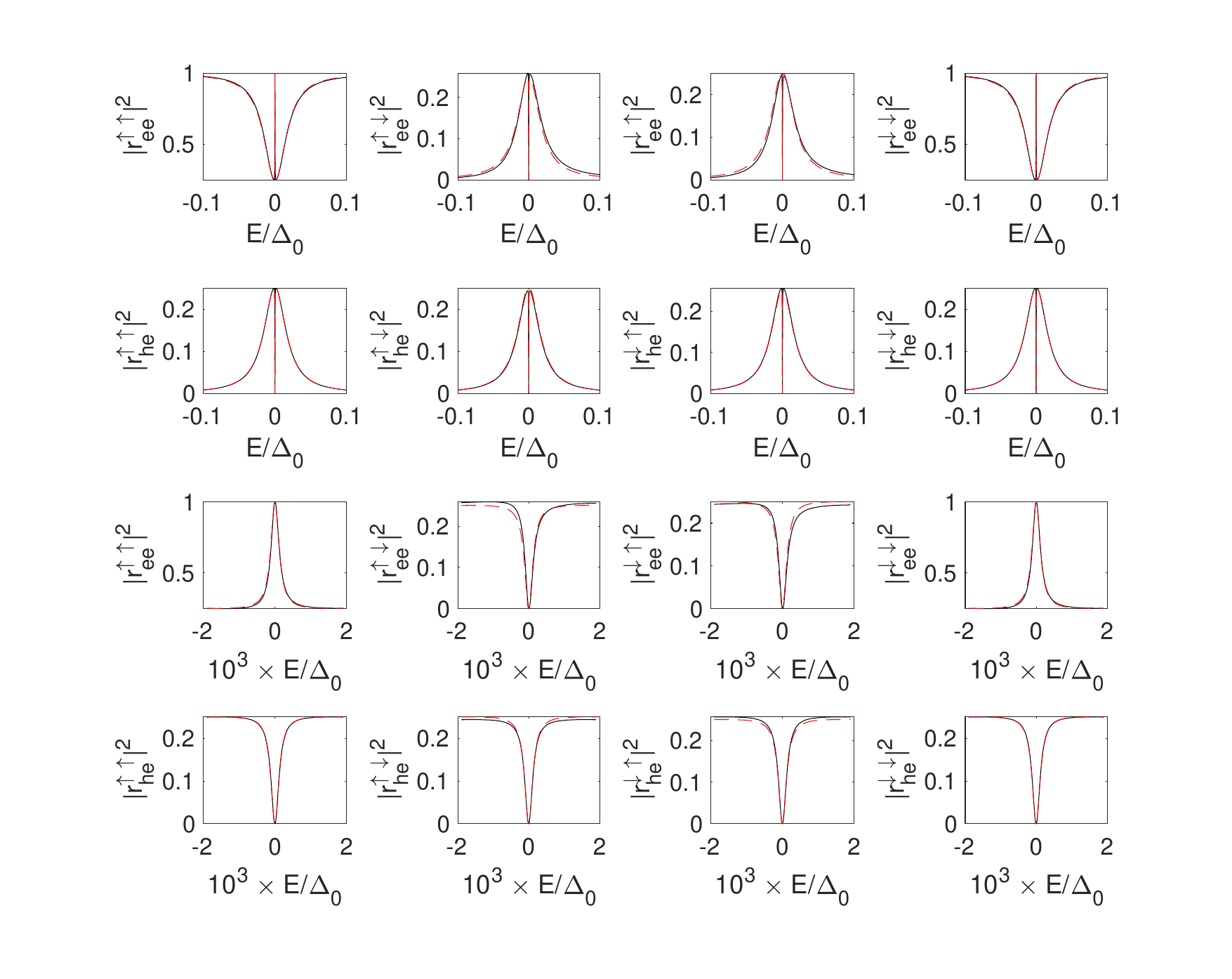}
\vspace{-1.3cm}
\caption{Reflection coefficients as a function of $E$ for $\Delta=E_{SO}$, $m_0=2E_{SO}$ and $L=27 \ell_{SO}$ (solid black lines). The dashed red lines show the results obtained using the analytic formulas  (\ref{ree-supp}) and (\ref{rhe-supp}) with $\Gamma^L_\uparrow = \Gamma^L_\downarrow = \Gamma/2$, $\Gamma = 0.0188 \Delta_0$ and
$E_M = 0.0016\Delta_0$.}
\label{supprefL27}
\end{figure}

As described in the main text the Mahaux-Weidenm\"{u}ller formula for the scattering matrix is 
\begin{equation}
S=1-2 \pi i W^\dag (E-H_M+i \pi W W^\dag )^{-1} W, \label{Mahaux-supp}
\end{equation}
where
\begin{equation}
H_M=\begin{pmatrix} 0 & i E_M \\ -i E_M & 0 \end{pmatrix}, \ W=\begin{pmatrix} w_{\uparrow}^L & w_{\downarrow}^L & w_{\uparrow}^{L*} & w_{\downarrow}^{L*} \\
                  w_{\uparrow}^R & w_{\downarrow}^R & w_{\uparrow}^{R*} & w_{\uparrow}^{R*} \end{pmatrix}, \  0< w_{L\uparrow}, w_{L\downarrow} \in \mathbb{R}.
                  \label{couplings-supp}
\end{equation}
By assuming that $w_{R\sigma} \to 0$ we obtain
\begin{equation}
|r_{ee}^{\uparrow\uparrow}|^2= \bigg|1 + i \frac{E  \Gamma^L_\uparrow}{Z} \bigg|^2, \ |r_{ee}^{\uparrow\downarrow}|^2=|r_{ee}^{\downarrow \uparrow}|^2= \frac{E^2 \Gamma^L_\uparrow \Gamma^L_\downarrow}{|Z|^2}, \   |r_{ee}^{\downarrow\downarrow}|^2= \bigg|1 + i \frac{E  \Gamma^L_\downarrow}{Z} \bigg|^2 \label{ree-supp}
\end{equation}
and
\begin{equation}
|r_{he}^{\uparrow\uparrow}|^2= \frac{(E \Gamma^L_\uparrow)^2}{|Z|^2}, \ |r_{he}^{\uparrow\downarrow}|^2=|r_{he}^{\downarrow \uparrow}|^2= \frac{E^2 \Gamma^L_\uparrow \Gamma^L_\downarrow}{|Z|^2}, \   |r_{he}^{\downarrow\downarrow}|^2= \frac{(E \Gamma^L_\downarrow)^2}{|Z|^2}, \label{rhe-supp}
\end{equation}
where
\begin{equation}
\Gamma^L_\sigma=2 \pi (w^L_\sigma)^2, \ Z=E_M^2 - E^2 - i E \Gamma, \ \Gamma=\Gamma^L_\uparrow +\Gamma^L_\downarrow.
\end{equation}
Using ${\cal G}(E) = 2 G_0  (|r_{he}^{\uparrow\uparrow}|^2+|r_{he}^{\uparrow\downarrow}|^2+|r_{he}^{\downarrow\uparrow}|^2+|r_{he}^{\downarrow\downarrow}|^2 )$ and ${\cal S}_z(E)  = 
\hbar\bigl(|r_{ee}^{\downarrow\uparrow}|^2 + |r_{ee}^{\uparrow\downarrow}|^2 +|r_{he}^{\downarrow\downarrow}|^2 + |r_{he}^{\uparrow\uparrow}|^2 \bigr)/2$ we get
${\cal G}(E)/(2 G_0)={\cal S}_z(E)/(\hbar/2)={\cal F}(E)$, where 
\begin{equation}
{\cal F}(E)= \frac{E^2/\Gamma^2}{E^2/\Gamma^2+(E_M^2-E^2)^2/\Gamma^4}. \label{supp:usual}
\end{equation}
We find that these expressions with $\Gamma^L_\uparrow = \Gamma^L_\downarrow = \Gamma/2$ describe well our numerical results for most of the values of $L$ (see Figs.~\ref{suppCondSL27} and \ref{supprefL27}). It follows from Eq.~(\ref{supp:usual}) that ${\cal G}(E)$ and ${\cal S}_z(E)$ have sharp dips at low energies due to the coupling of MZMs. The width of these dips in energy scales as $\sim E_M^2/\Gamma$ and $E_M$ decreases exponentially with $L$. Therefore, these dips do not affect the quantization of the conductance and spin pumping in long wires in the case of experimentally relevant temperatures and frequencies. 


\begin{figure}[h]
\vspace{-0.2cm}
\includegraphics[width=0.99\linewidth]{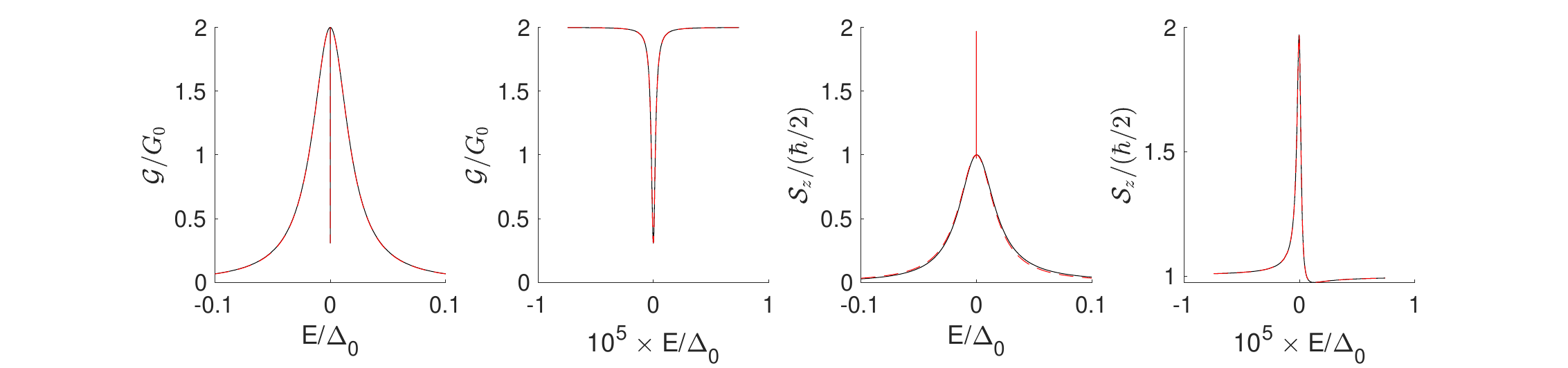}
\vspace{-0.4cm}
\caption{${\cal G}(E)$ and ${\cal S}_z(E)$  for $\Delta=E_{SO}$, $m_0=2E_{SO}$ and $L=26.5 \ell_{SO}$ (solid black lines). For this value of $L$ the hybridization of the MZMs is very small, and although ${\cal G}(E)$ still shows a dip at small energies, ${\cal S}_z(E)$ now has a very sharp peak.  The magnifications to low energies are shown to make the dip and the peak more clearly visible. The dashed red lines show the results obtained using the analytic formulas  (\ref{rhe-full-supp}) and (\ref{ree-full-supp}) with $\Gamma = 0.0186 \Delta_0$, 
$E_M = 1.032 \cdot 10^{-5} \Delta_0$,
$\gamma = 1.9973 \cdot 10^{-7} \Delta_0$,  $x = 1.2866$ and
$y = 1.8586$.}
\label{suppCondSL265}
\end{figure}
\begin{figure}[h]
\vspace{-0.4cm}
\includegraphics[width=0.99\linewidth]{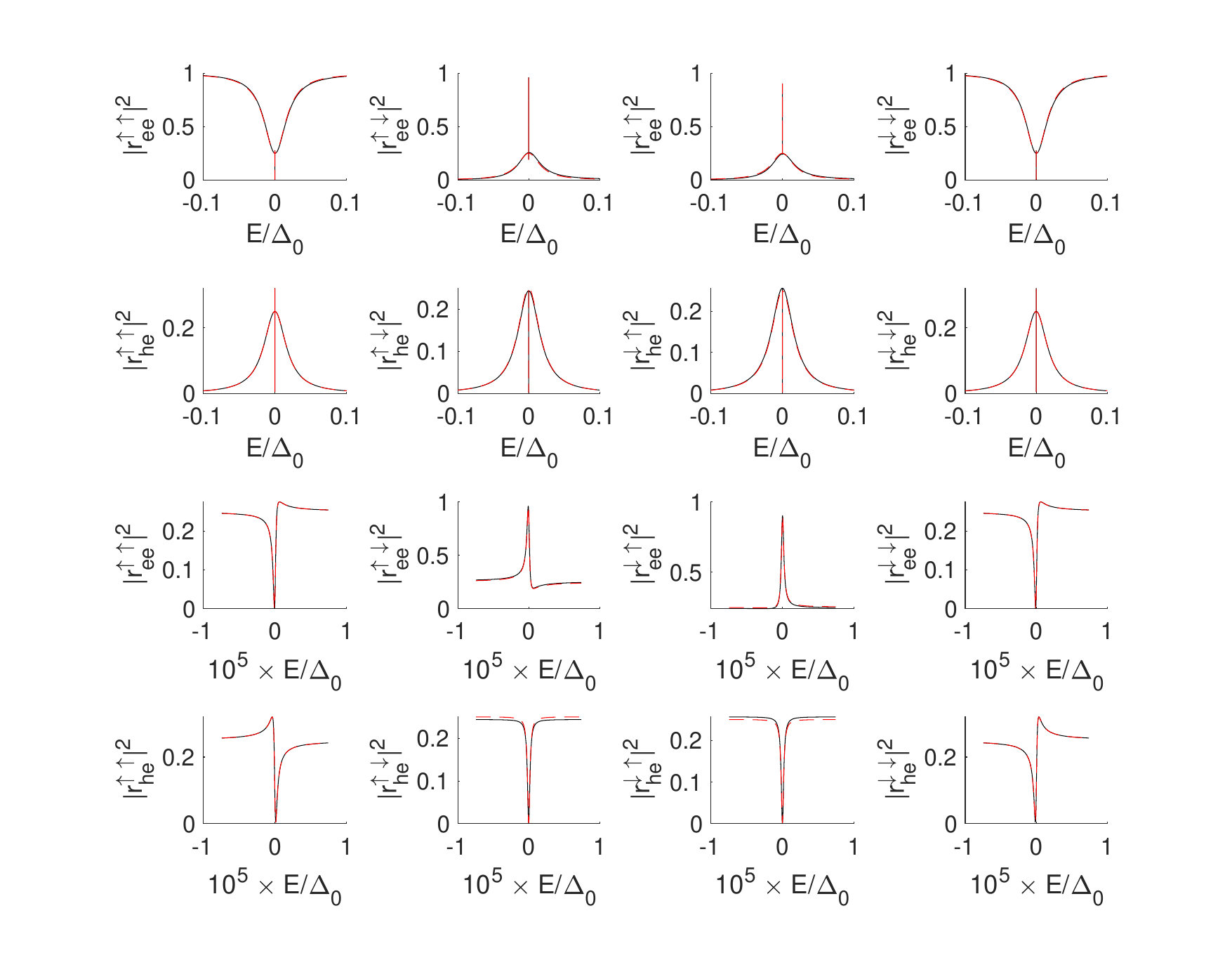}
\vspace{-1.3cm}
\caption{Reflection coefficients as a function of $E$ for $\Delta=E_{SO}$, $m_0=2E_{SO}$ and $L=26.5 \ell_{SO}$ (solid black lines). The dashed red lines show the results obtained using the analytic formulas  (\ref{rhe-full-supp}) and (\ref{ree-full-supp}) with $\Gamma = 0.0186 \Delta_0$, 
$E_M = 1.032 \cdot 10^{-5} \Delta_0$,
$\gamma = 1.9973 \cdot 10^{-7} \Delta_0$,  $x = 1.2866$ and
$y = 1.8586$.}
\label{supprefL265}
\end{figure}

If $E_M \approx 0$ the couplings $w_{R\sigma}$ of the right MZM to the lead can also become important at low energies, and they can turn the sharp dip typically appearing in ${\cal S}_z(E)$ (see Fig.~\ref{suppCondSL27}) into a sharp peak (see Fig.~\ref{suppCondSL265}). The full expression for the scattering matrix obtained from Eqs.~(\ref{Mahaux-supp}) and (\ref{couplings-supp}) is quite complicated. Therefore, for simplicity we approximate the coefficients as
\begin{equation}
w_{\uparrow}^L =  w_{\downarrow}^L =  \sqrt{\frac{\Gamma}{4\pi}}, \  w_{\uparrow}^R= \sqrt{\frac{\gamma}{4\pi}} e^{i x}, \  w_{\downarrow}^R = \sqrt{\frac{\gamma}{4\pi}} e^{i y}.
\end{equation}
This way we get
\begin{eqnarray}
|r_{he}^{\uparrow\uparrow}|^2 &=& \frac{\big|i E \left(\Gamma +\gamma  e^{2 i x}\right)+\gamma  \Gamma  e^{i x} [\cos (y)-\cos (x)]\big|^2}{4 |Z|^2}, \ |r_{he}^{\downarrow\downarrow}|^2= \frac{\big|i E \left(\Gamma +\gamma  e^{2 i y}\right)+\gamma  \Gamma  e^{i y} [\cos (x)-\cos (y)]\big|^2}{4 |Z|^2}, \nonumber \\ |r_{he}^{\uparrow\downarrow}|^2 &=& \frac{\big|i E \left(\Gamma +\gamma  e^{i (x+y)}\right)+E_M \sqrt{\gamma  \Gamma } \left(e^{i x}-e^{i y}\right)-2 \gamma  \Gamma  e^{\frac{1}{2} i (x+y)} \sin ^2\left(\frac{x-y}{2}\right) \cos \left(\frac{x+y}{2}\right)\big|^2}{4 |Z|^2}, \nonumber \\ 
|r_{he}^{\downarrow\uparrow}|^2&=&\frac{\big|i E \left(\Gamma +\gamma  e^{i (x+y)}\right)+E_M \sqrt{\gamma  \Gamma } \left(e^{i y}-e^{i x}\right)-2 \gamma  \Gamma  e^{\frac{1}{2} i (x+y)} \sin ^2\left(\frac{x-y}{2}\right) \cos \left(\frac{x+y}{2}\right)\big|^2}{4 |Z|^2} \label{rhe-full-supp}
 \end{eqnarray}
and
\begin{eqnarray}
|r_{ee}^{\uparrow\uparrow}|^2&=&\bigg|1+\frac{i E (\gamma +\Gamma )-2 i E_M \sqrt{\gamma  \Gamma } \sin (x)+\gamma  \Gamma  ( \cos^2(x)+\cos (x)\cos (y)-2)}{2 Z} \bigg|^2, \nonumber \\
|r_{ee}^{\downarrow\downarrow}|^2&=&\bigg|1+\frac{i E (\gamma +\Gamma )-2 i E_M \sqrt{\gamma  \Gamma } \sin (y)+\gamma  \Gamma  ( \cos^2(y)+\cos (x)\cos (y)-2)}{2 Z} \bigg|^2, \nonumber \\
|r_{ee}^{\uparrow\downarrow}|^2 &=& \frac{\big|i E \left(\Gamma +\gamma  e^{i (y-x)}\right) +E_M \sqrt{\gamma  \Gamma } \left(e^{-i x}-e^{i y}\right)-2 \gamma  \Gamma  e^{\frac{1}{2} i (y-x)} \sin ^2\left(\frac{x+y}{2}\right) \cos \left(\frac{x-y}{2}\right)\big|^2}{4 |Z|^2}, \nonumber \\
|r_{ee}^{\downarrow\uparrow}|^2 &=& \frac{\big|i E \left(\Gamma +\gamma  e^{i (x-y)}\right) +E_M \sqrt{\gamma  \Gamma } \left(e^{-i y}-e^{i x}\right)-2 \gamma  \Gamma  e^{\frac{1}{2} i (x-y)} \sin ^2\left(\frac{x+y}{2}\right) \cos \left(\frac{x-y}{2}\right)\big|^2}{4 |Z|^2}, \label{ree-full-supp}
\end{eqnarray}
where
\begin{equation}
Z=E_M^2-(E+i \gamma) (E+i \Gamma )-\gamma  \Gamma  [\cos (x)+\cos (y)]^2/4.
\end{equation}
These expressions allow to accurately describe our numerical results for the reflection coefficients also in a situation where ${\cal S}_z(E)$ has a sharp peak (see Figs.~\ref{suppCondSL265} and \ref{supprefL265}). Therefore, we can attribute the appearance of the low-energy peak in the ${\cal S}_z(E)$ to the effects arising due to the coupling of the lead to both MZMs. Furthermore, we can easily see that no such peak can appear if the scattering phases satisfy $x=y=0$. Therefore, we can conclude that the peak originates from interference effects between the scattering paths.

\section{Dependence of the transport features with the barrier thickness}

\begin{figure}[h]
\includegraphics[width=0.99\linewidth]{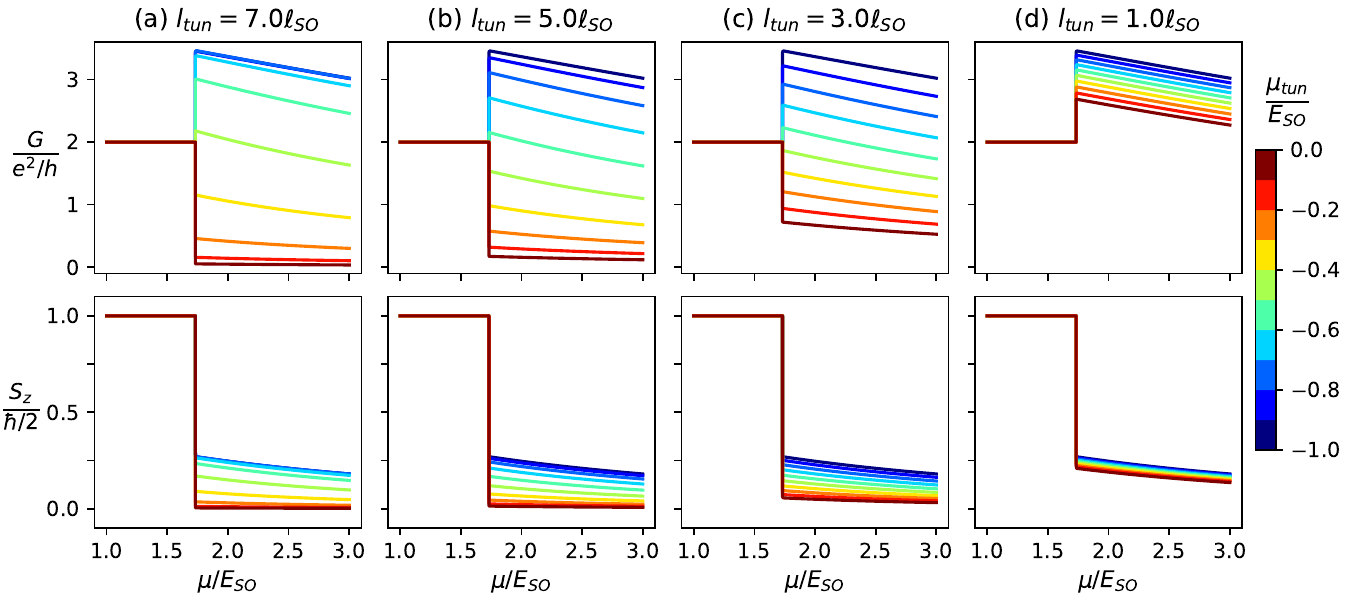}
\vspace{-0.4cm}
\caption{(a)-(d) The dependence of the conductance and spin pumping on $l_{tun}$, $\mu$ and $\mu_{tun}$ in the limit $\omega$, $k_BT\rightarrow 0$. $G$ and $S_z$ are robustly quantized in the topologically nontrivial phase ($\mu < \sqrt{3}E_{SO}$). The model parameters are $\Delta_0=E_{SO}$, $m_0=2E_{SO}$, $\mu_N=0$ and $\theta=2\pi/5$. }
\label{barrthick}
\end{figure}

In the main text we demonstrated a one-to-one correspondence between the conductance and the spin pumping in the nontrivial phase for a barrier with width $l_{tun}=4\ell_{SO}$. In Fig.~\ref{barrthick} we demonstrate that similar results are obtained for other values of $l_{tun}$. The conductance and spin pumping in the nontrivial phase ($\mu<\sqrt{3}E_{SO}$) are robustly quantized for all value of $l_{tun}$. On the other hand, in the trivial phase ($\mu>\sqrt{3}E_{SO}$) the conductance and spin pumping change as a function of $l_{tun}$ and their values are unrelated.           

\section{Interference effects in the presence of strong disorder}

\begin{figure}[h]
\includegraphics[width=0.99\linewidth]{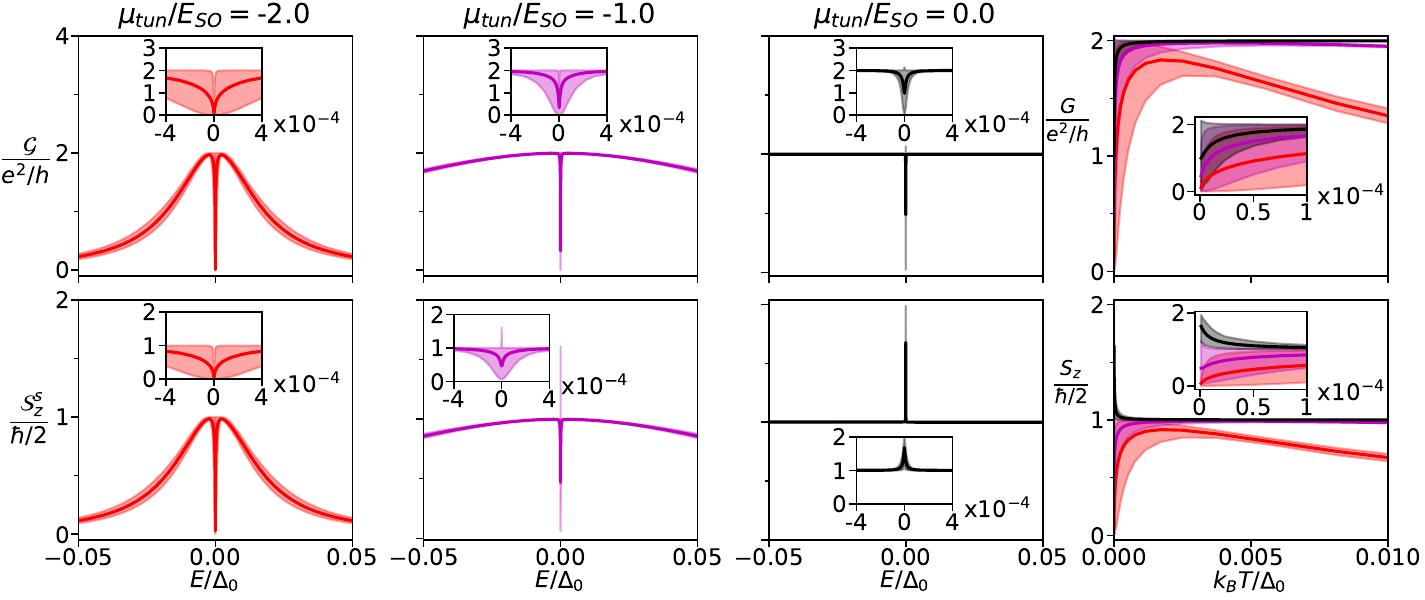}
\vspace{-0.4cm}
\caption{Spectral densities ${\cal G}(E)$ and  ${\cal S}^s_z(E)$ as a function of $E$ for $\mu_{tun}/E_{SO}=-2, -1, 0$. The interference effects lead to sharp dips and peaks in ${\cal G}(E)$ and ${\cal S}^s_z(E)$ at low energies resulting in large sample-to-sample fluctuations in $G$ and $S_z$ at low temperatures (right column). The model parameters are $\Delta_0 = E_{SO} $, $\mu=0$, $m_0 = 2 E_{SO}$, $\theta=\pi/2$ and $L=26.5 \ell_{SO}$. The disorder potential is $V(x)\tau_z$, where the $V(x)$ at each lattice site $x$ are uncorrelated uniformly distributed random numbers between $[-8E_{SO}, 8E_{SO}]$ and we have used lattice constant $d=\ell_{SO}/100$. The error bars denote the $10$th and $90$th percentile values. }
\label{suppCondSL27disorder}
\end{figure}

As discussed in the main text the overall shapes of ${\cal G}(E)$ and ${\cal S}^s_z(E)$ in finite-length wires are similar as in the case of $L \to \infty$, but the hybridization of the MZMs and the interference effects lead to  a very sharp dip or a peak at very low energies. In ${\cal G}(E)$ we find practically always a dip at low energies and the disorder just leads to variation of the width of the dip (see Fig.~\ref{suppCondSL27disorder} top row). On the other hand, in ${\cal S}^s_z(E)$ both peaks and dips are possible depending on the disorder configurations (see Fig.~\ref{suppCondSL27disorder} bottom row).  The conductance $G$ and spin pumping $S_z$ are not affected by these low-energy features at temperatures larger than the typical widths of the sharp peaks and dips, but we find large sample-to-sample fluctuations of $G$ and $S_z$ at very small temperatures (see Fig.~\ref{suppCondSL27disorder} right column).

\section{Feedback on the magnetization dynamics}

The dynamics of the magnet can be inferred from the Landau-Lifshitz-Gilbert equation. For the monodomain dynamics considered in the main text, we can write \cite{TserkovnyakRMP2005}:
\begin{align}
    \dot{\boldsymbol n}(t)=-\gamma[{\boldsymbol n}(t)\times{\boldsymbol B}_{\rm tot}(t)]+\tilde{\alpha}[{\boldsymbol n}(t)\times\dot{\boldsymbol n}(t)]-\frac{\gamma}{M_sV}{\boldsymbol \tau}_{el}(t)\,,
\end{align}
 where ${\bs n}(t)=[\sin\theta\cos\phi(t),\sin\theta\sin\phi(t),\cos\theta]$ is the magnetization direction, $\gamma>0$ is the gyromagnetic factor, $M_s$ the saturation magnetization, $V$ is the volume of the magnet,  $\boldsymbol{B}_{\rm tot}(t)$ is the total (time-dependent) magnetic field,  ${\boldsymbol\tau}_{el}(t)$ is the electronic torque, and $\tilde{\alpha}$ is the Gilbert damping parameter. The interaction Hamiltonian between the magnetization ${\bs n}(t)$ and the electronic spin ${\bs\sigma}$ is
\begin{equation}
    H_{el-m}(t)=m_0{\bs n}(t)\cdot{\bs \sigma}\,\theta(x)\,,
\end{equation}
with the (isotropic) exchange coupling strength $m_0$. The resulting electronic torque becomes
\begin{align}
{\boldsymbol \tau}_{el}(t)=-\frac{i}{2}\langle[{\bs \sigma}, H_{el-m}(t)]\rangle=m_0{\boldsymbol n}(t)\times\langle{\boldsymbol\sigma}\rangle\,,
\end{align}
where $\langle\dots\rangle$ means expectation value over the stationary electronic state. This torque in turn separates into a reactive ($\tau_{el}^{r}$) and dissipative ($\tau_{el}^{d}$) contribution, respectively. The first acts as to modify the resonance frequency of the ferromagnet, while the latter alters its resonance linewidth. The electronic Hamiltonian commutes with $\sigma_z$, which means that in the stationary regime the spin along $z$ created in the topological superconductor by the precession needs to be compensated by the spin flow into the leads (hereby assumed to be perfect spin sinks, i.e. no spin accumulation occurs at the boundary). In a frame rotating with the magnetization that means:
\begin{equation}
\langle I_s\rangle=-\frac{\hbar}{2}\langle\dot{\sigma}_z\rangle=-m_0\langle\sigma_y\rangle\sin{\theta}\,.
\end{equation}
However, $m_0\langle\sigma_y\rangle\equiv\tau_{el}^{d}$ is nothing but the dissipative contribution to the torque, and therefore, in the stationary state
\begin{align}
    \tau_{el}^{d}=-\frac{\langle I_s\rangle}{\sin{\theta}}=-\frac{ \omega}{2\pi\sin{\theta}}S_z\,,
\end{align}
which applies to both the topological and trivial regimes. Consequently the spin current acts as to modify the bare Gilbert damping with the amount:
\begin{equation}
    \Delta\tilde{\alpha}=\frac{\gamma}{2\pi M_sV\sin^2\theta}S_z\,,
\end{equation}
which exhibits the same behaviour as $S_z$. Therefore, it can be used to detect the quantization of the pumped spin current described it this work. 

For the sake of completeness, in the following, we also demonstrate the above relations from microscopics for the semi-infinite system in the absence of a potential barrier. Assuming uniform precession around the $z$ axis, the wave-function on the topological side in the rotating frame is given by Eq.~(\ref{SC_wf}). Then, the associated spin expectation value stemming from energies within the topological gap,  $|E|<\Delta_{gap}$, can be written as:
\begin{align}
    \langle\sigma_\alpha\rangle&=\sum_{j,j'=1,4}\sum_\sigma\int \frac{dE}{2\pi}\rho_\sigma(E)\bar{A}_j^{\sigma}A_{j'}^{\sigma}\phi_j^\dagger(E)\sigma_\alpha\phi_{j'}(E)\int_0^\infty dxe^{-(\bar{z}_j(E)+z_{j'}(E))x}f_\sigma(E)\nonumber\\
    &=\sum_{j,j'=1,4}\int \frac{dE}{2\pi}\rho_\sigma(E)\,\phi_j^\dagger(E)\sigma_\alpha\phi_{j'}(E)\frac{[\bar{A}_j^{\uparrow}A_{j'}^{\uparrow}f_\uparrow(E)+\bar{A}_j^{\downarrow}A_{j'}^{\downarrow}f_\downarrow(E)]}{\bar{z}_j(E)+z_{j'}(E)}\,,
\end{align}
where $\alpha=x,y,z$, $\rho_\sigma(E)\equiv \rho(E)$, and $f_\sigma(E)\equiv f_0(E+\sigma\hbar\omega/2)$ are the density of states and the distribution function in the rotating frame for the spin species $\sigma$, respectively. As demonstrated in the main text, in the absence of a barrier all the functions are weakly dependent on the energy $E$ and $\omega$ for  $|E|\ll\Delta_{gap}$, besides the distribution functions $f_\sigma(E)$. Then, restricting ourselves to the leading order in $\omega$, we can write:
\begin{align}
    \langle\sigma_\alpha\rangle&=\langle\sigma_\alpha\rangle_0+\frac{\hbar\omega}{2}\langle\sigma_\alpha\rangle_\omega+\dots\,,\\
    \langle\sigma_\alpha\rangle_0&=\sum_{j,j'=1,4}\int \frac{dE}{2\pi}\rho_\sigma(E)\,\phi_j^\dagger(E)\sigma_\alpha\phi_{j'}(E)\frac{(\bar{A}_j^{\uparrow}A_{j'}^{\uparrow}+\bar{A}_j^{\downarrow}A_{j'}^{\downarrow})}{\bar{z}_j(E)+z_{j'}(E)}f_0(E)\,,\\
    \langle\sigma_\alpha\rangle_\omega&=\frac{1}{2\pi}\sum_{j,j'=1,4}\rho_\sigma(0)\phi_j^\dagger(0)\sigma_\alpha\phi_{j'}(0)\frac{(\bar{A}_j^{\uparrow}A_{j'}^{\uparrow}-\bar{A}_j^{\downarrow}A_{j'}^{\downarrow})}{\bar{z}_j(0)+z_{j'}(0)}\,,
\end{align}
being the sum of the equilibrium contribution and leading order in $\omega$, respectively. Here  we have used $f_0(E+\sigma\omega/2)=f_0(E)+(\sigma\omega/2)\partial_Ef_0(E)+\dots$, such that at $T=0$ we have $\partial_Ef_0(E)=\delta(E)$. The dissipative torque is determined by $\langle\sigma_y\rangle_\omega$, while the $\langle \sigma_\alpha\rangle_0$ terms act as to modify the free energy of the insulating ferromagnet. The boundary conditions $\Psi_{\rm rot}^{\alpha, L}(0)=\Psi_{\rm rot}^{\alpha, R}(0)$ and $\partial_x\Psi_{\rm rot}^{\alpha, L}(0)=\partial_x\Psi_{\rm rot}^{\alpha, R}(0)$ allow us to obtain all the  coefficients $A_j^\alpha$ for a given set of  nanowire  parameters. We find that in the topological regime $2\pi m_0\langle\sigma_y\rangle_\omega\sin\theta=-1$, in agreement with the torques analysis.   

\section{Quasi-Majorana modes in the topologically trivial phase}\label{qMajoSec}

 As discussed in the main text a smooth tunnel barrier can induce two spatially separated MZMs at the lead-nanowire interface \cite{inh_Brouwer, QD_Liu, QD_Moore_one, QD_Moore_two, inh_workfunc, qMajo_Vuik}, and  in certain cases these quasi-MZMs are so weakly coupled to each other that they can mimic all properties of the MZMs. 
In this section we study the conductance and spin pumping in the presence of quasi-MZMs. For this purpose we consider a  Hamiltonian motivated by Ref.~\cite{qMajo_Vuik}
\begin{equation}
    \mathcal{H}(t)\!=\!\Bigl[\frac{p^2}{2m}- \alpha_R\,p\,\sigma_z-\mu(x) + V(x)\Bigr]\tau_z  \!+\! \bs{m}(x,t)\cdot\bs{\sigma} \!+\! \Delta(x)\tau_x,
    \label{BdG_smoothV}
\end{equation}
where $\bs{m}(x,t)$ and $\Delta(x)$ have the same form as in the main text, but $\mu(x)=\mu_N\Theta(-x) + \mu\Theta(x)$ and $V(x)=V_1\Theta(x)e^{-(x-x_0)^2/2\sigma_1^2}$. To illustrate the emergence of the quasi-MZMs over a wider range of parameters we consider the following values of model parameters: $\Delta_0=20 E_{SO}$, $m_0=2\Delta_0$, $\theta=0.5\pi$, $\mu_{N}=40E_{SO}$, $\sigma_1=4 \ell_{SO}$, $x_0=3\ell_{SO}$, and $k_BT=0.02E_{SO}$. 

\begin{figure}[htb]
\vspace{-0.2cm}
\includegraphics[width=0.94\linewidth]{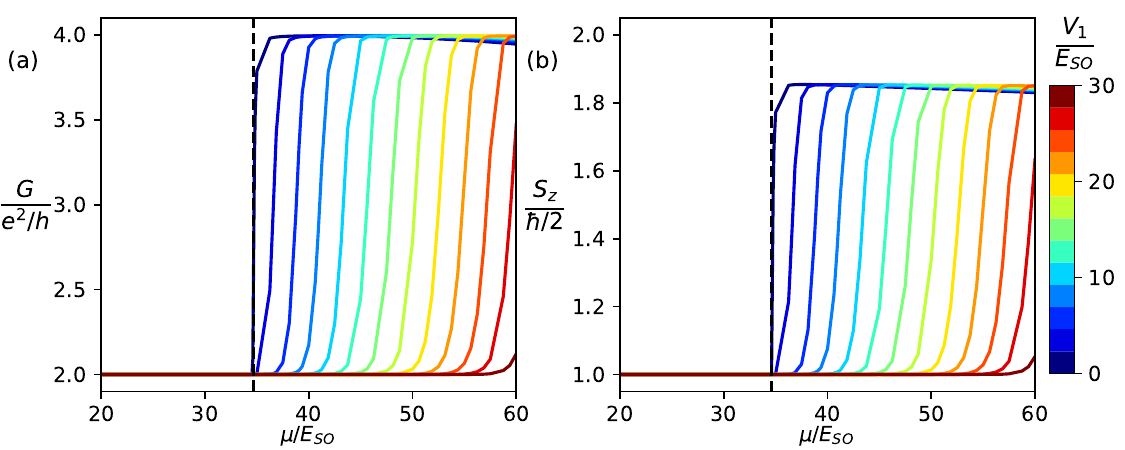}
\vspace{-0.4cm}
\caption{(a) Conductance and (b) spin pumping as a function of $\mu$ in the presence of a smooth tunneling barrier, see the details of the potential in the text. The transition line separating the topologically nontrivial and trivial phases occurring at $\mu = \sqrt{m_0^2 - \Delta_0^2}$ is illustrated with a dashed line. In the presence of a strong smooth tunnel barrier $V_1$ the conductance and spin pumping are quantized also in the trivial wires due to the appearance of quasi-MZMs. However, the quantizatized signatures are  robust upon lowering of the strength of the tunnel barrier $V_1$ only in the non-trivial phase. The model parameters are $k_BT=0.02E_{SO}$,  $\Delta_0=20 E_{SO}$, $m_0=2\Delta_0$, $\theta=0.5\pi$, $\mu_{N}=40E_{SO}$, $\sigma_1=4 \ell_{SO}$, $x_0=3\ell_{SO}$.}
\label{qMajotransfigs}
\end{figure}

Fig.~\ref{qMajotransfigs} shows the transport features as a function of $\mu$ for different barrier heights $V_1$. In the topologically nontrivial phase ($\mu < \sqrt{m_0^2 - \Delta_0^2}\,$) the conductance and spin pumping are quantized independently of strength of the tunnel barrier $V_1$. On the other hand, in the topologically trivial phase ($\mu > \sqrt{m_0^2 - \Delta_0^2}\,$) the quantization can be present for suitably chosen values of $\mu$ and $V_1$ due to the presence of the quasi-MZMs. However, in the case of trivial wires the quantization of $G$ and $S_z$, and their one-to-one the correspondence, is always lost in the strong coupling regime ($V_1 \approx 0$). Therefore the topological MZMs can be distinguished from quasi-MZMs by demonstrating the robustness of the quantized signatures upon lowering of the tunnel barrier.

\section{Fractional entropy change as another distinguishing feature of MZMs}

MZMs can also give rise a fractational change of entropy $\Delta S= \frac{1}{2}k_B \log{2}$ \cite{entro_smirnov_one,entro_Sela}, in sharp contrast with the entropy of fully electronic systems that show integer plateaus \cite{entro_expone,entro_exptwo,entro_expthree}. Entropy measurements have thus gained interest to detect MZMs and other exotic states \cite{entro_Kondo,entro_setup,entro_smirnov_two,entro_smirnov_three}.  In this section, we consider two frameworks to calculate the fractional entropy change related to the MZMs. In the first framework the system is composed by a large and discrete metallic reservoir that is coupled via a tunneling barrier to a superconducting nanowire harboring MZMs. In the second framework the large and discrete reservoir is replaced by a semi-infinite lead having a continuum density of states. 

\subsection{Discrete reservoir}

The Hamiltonian in this case is the same as in Eq.~(\ref{BdG_smoothV}) but with $\Delta(x)=\Delta_0\Theta(L_S-x)\Theta(x)$, $m(x)=m_0\Theta(L_S-x)\Theta(x)$, $V(x)=V_2\Theta(x+\sigma_2)\Theta(-x)$, and $\mu(x)=\mu_N\Theta(x+L_N)\Theta(-x) + \mu\Theta(L_S-x)\Theta(x)$. The entropy is obtained from $S(V_2) = -\frac{dF}{dT}$, where the free energy of a discrete system is $F=-k_BT\sum_{E_i>0}\mathrm{log}\bigl(1 + e^{-E_i/k_BT}\bigr)$, i.e.
\begin{equation}
S(V_2) = k_B\sum_{E_i>0}\mathrm{log}\bigl(1 + e^{-E_i/k_BT}\bigr) + \frac{1}{T} \sum_{E_i>0}\frac{E_ie^{-E_i/k_BT}}{1 + e^{-E_i/k_BT}}.
\label{entropdiscr}
\end{equation}
Specifically, we are interested about the  entropy change when a MZM hybridizes with the states in the metallic reservoir. This can be achieved by tuning the tunnel barrier $V_2$ between the metallic reservoir and the nanowire from a large potential $\tilde{V}$ (nanowire completely decoupled from the metallic reservoir) to a smaller value $V_2$, so that the 
the entropy change is given by $\Delta S_{V_2} = S(V_2) - S(\tilde{V})$ \cite{entro_Sela}. In our simulations we take $\tilde{V} \approx 10^{4}E_{SO}$, and consider a sufficiently long nanowire ($L_S = 80\ell_{SO}$) with a large superconducting proximity-induced gap $\Delta_0=20 E_{SO}$ so that the hybridization of the MZMs localized at the two ends of the wire is negligible. The rest of parameters are  $m_0=2\Delta_O$, $L_N = 1100\ell_{SO}$, $\theta/\pi = 0.5$, $\mu_N = 20 E_{SO}$, $\sigma_2=0.4\ell_{SO}$, and $V_2=30E_{SO}$.   

\begin{figure}[ttb]
\vspace{-0.2cm}
\includegraphics[width=0.7\linewidth]{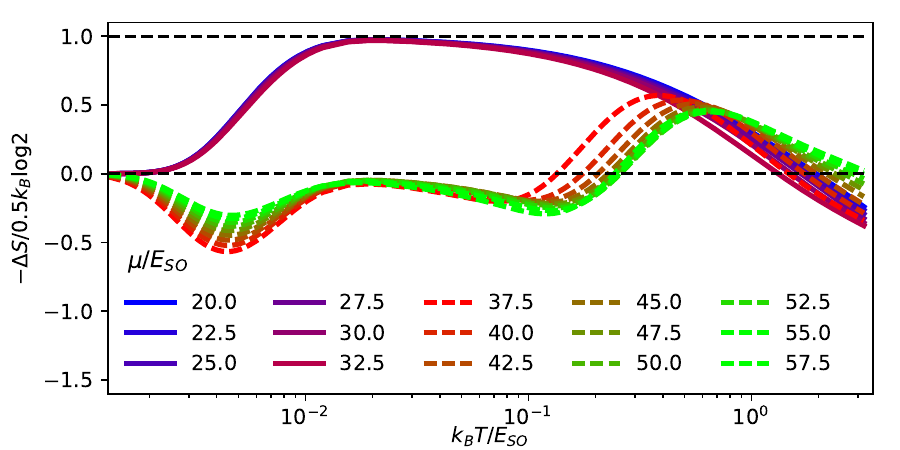}
\vspace{-0.4cm}
\caption{$\Delta S_{V_2}$ as a function of temperature $k_BT$. In the topologically nontrivial phase ($\mu < \sqrt{m_0^2 - \Delta_0^2}\approx 34.6 E_{SO}$) there is a robust plateau of quantized fractional value $\Delta S_{V_2}=-\frac{1}{2}k_B \log{2}$ when  $\delta < k_BT < \Gamma$, where  $\delta$ is the level spacing in the metallic reservoir and $\Gamma$ is the MZM linewidth caused by the coupling of the MZM to the metallic reservoir. In the trivial phase the plateau with fractional value is lost. The parameters used in the calculation are: $\Delta_0=20 E_{SO}$, $m_0 = 2 \Delta_0$, $V_2=30 E_{SO}$, $L_N = 1100\ell_{SO}$, $L_{S}=80\ell_{SO}$, $\theta/\pi = 0.5$, $\mu_N=20 E_{SO}$, and $\sigma_2=0.4\ell_{SO}$. }
\label{Ent_sharp_barr}
\end{figure}

The entropy change $\Delta S_{V_2}$ as a function of temperature is shown in Fig.~\ref{Ent_sharp_barr}, where the ferromagnetic-superconducting nanowire is tuned between nontrivial and trivial phases with the help of chemical potential $\mu$. In the nontrivial phase the entropy shows a robust plateau quantized to $\Delta S_{V_2}=-\frac{1}{2}k_B \log{2}$ for temperatures $\delta < k_BT < \Gamma$, where  $\delta$ is the level spacing in the metallic reservoir and $\Gamma$ is the MZM linewidth caused by the coupling of the MZM to the metallic reservoir. On the other hand, in the trivial phase $\Delta S_{V_2}$ takes small values at these temperatures.  

\subsection{Continuum reservoir}

The Hamiltonian in this case is the same as in Eq.~(\ref{BdG_smoothV}) but with $\Delta(x)=\Delta_0\Theta(x)$, $m(x)=m_0\Theta(x)$, $V(x)=V_2\Theta(x+\sigma_2)\Theta(-x)$, and $\mu(x)=\mu_N\Theta(-x) + \mu\Theta(x)$. In this case we calculate the entropy change according to the formula $\Delta S_{V_2} = - d F_{MZM}/dT - k_B\log{2}$ \cite{entro_Sela}, where $F_{MZM}=-k_BT\int_{-\infty}^{\infty}dE\rho(E)\mathrm{log}\bigl(1 + e^{-|E|/k_BT}\bigr)$ is the free energy of the MZMs with the density of states $\rho(E)=\frac{1}{2}\delta(E) + \frac{1}{2\pi}\frac{\Gamma}{\Gamma^2 + E^2}$. Straightforward calculations yields
\begin{equation}
    \Delta S_{V_2} = \frac{k_B\Gamma}{\pi}\int_{0}^{\infty}dE \frac{\mathrm{log}\bigl(\cosh{E/2k_BT}\bigr)}{\Gamma^2 +E^2}
-\frac{\Gamma}{2\pi T}\int_{0}^{\infty}dE \frac{E\tanh{E/2k_BT}}{\Gamma^2 +E^2}.
\end{equation}

\begin{figure}[ttb]
\vspace{-0.2cm}
\includegraphics[width=0.98\linewidth]{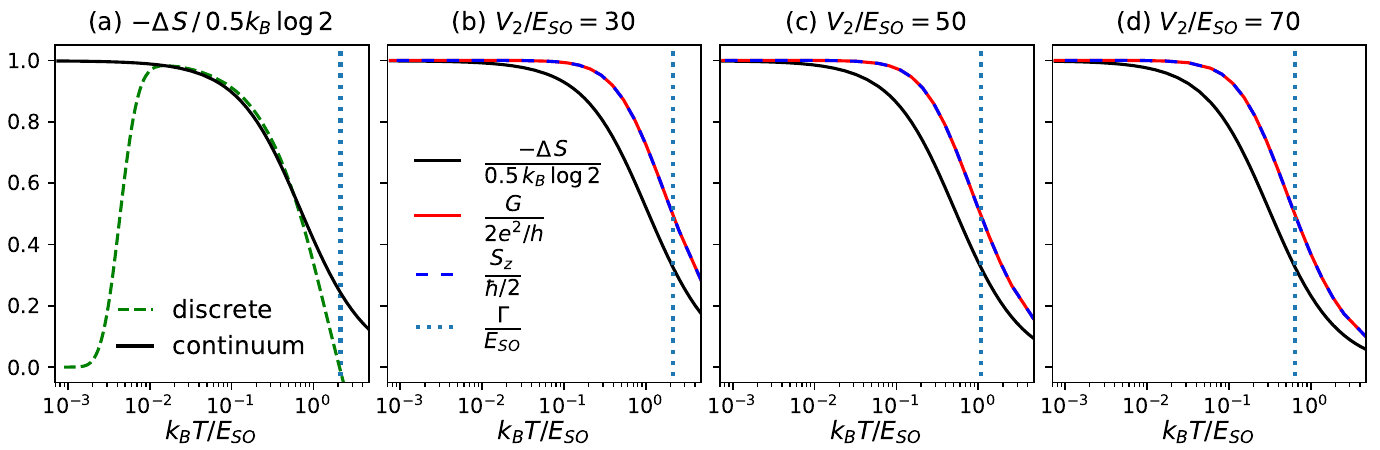}
\vspace{-0.4cm}
\caption{(a) Comparison of the $\Delta S_{V_2}$  calculated using the two different approaches for $V_2=30 E_{SO}$ (see the details in the text). (b)-(d) Comparison of $G$, $S_z$, $\Delta S_{V_2}$ for different barrier heights $V_2$ and temperatures $k_B T$. The dashed vertical lines show the MZM linewidths $\Gamma$ obtained from the zero-bias conductance peak. The parameters used in the calculation are: $\Delta_0=20 E_{SO}$, $m_0 = 2 \Delta_0$, $\theta/\pi = 0.5$, $\mu_N=20 E_{SO}$, $\mu=0$, and $\sigma_2=0.4\ell_{SO}$. For the discrete system $L_N = 1100\ell_{SO}$ and $L_{S}=80\ell_{SO}$. }
\label{Entro_compa}
\end{figure}

In Fig.~\ref{Entro_compa}(a) we compare $\Delta S_{V_2}$ calculated for the discrete and continuous systems. We observe that the two calculations agree well in an interval where $\delta < k_BT < \Gamma$, when $\Gamma$ is extracted from the width of the zero-bias conductance peak.  In Figs.~\ref{Entro_compa}(b)-(d) we compare the $k_BT$ and $V_2$ dependencies of the conductance, spin pumping and entropy change in the topologically nontrivial phase. At low temperatures all three quantities are quantized, but the conductance and spin pumping plateaus extend to slightly higher temperatures than the quantized  $\Delta S_{V_2}$ plateau. 

In the trivial phase $G$, $S_z$ and $\Delta S_{V_2}$ are not quantized and their values are unrelated, see Figs.~\ref{barrthick} and \ref{Ent_sharp_barr}.                

\section{Fractional entropy change related to the quasi-Majorana modes}

\begin{figure}[ttb]
\vspace{-0.2cm}
\includegraphics[width=0.7\linewidth]{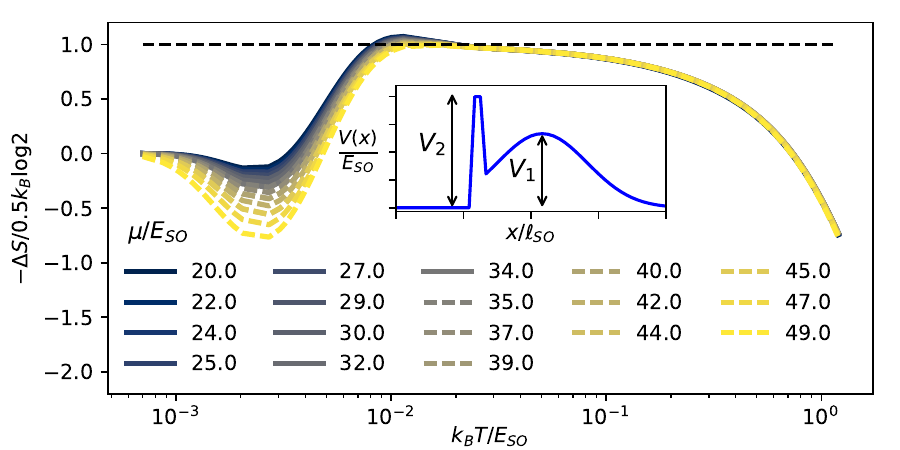}
\vspace{-0.4cm}
\caption{$\Delta S_{V_2}$ as a function of $k_BT$ for different values of $\mu$ in the case of smooth confining potential that facilitates the emergence of quasi-MZMs in the trivial phase. There exists a plateau of fractional value $\Delta S_{V_2}=-\frac{1}{2}k_B \log{2}$ in topologically nontrivial ($\mu < \sqrt{m_0^2 - \Delta_0^2}\approx 34.6 E_{SO}$) and trivial ($\mu > \sqrt{m_0^2 - \Delta_0^2}$)  phases, respectively. Here $V_2=30E_{SO}$ and barrier height $\tilde{V}=80E_{SO}$ is used to decouple the nanowire from the metallic reservoir. The parameters used in the calculation are: $\Delta_0=20 E_{SO}$, $m_0 = 2 \Delta_0$, $V_1=30E_{SO}$, $\sigma_1=4 \ell_{SO}$, $x_0=3\ell_{SO}$, $\sigma_2=2\ell_{SO}$, $L_N = 920\ell_{SO}$, $L_{S}=80\ell_{SO}$, $\theta/\pi = 0.5$, and $\mu_N=40 E_{SO}$. }
\label{Entropy_Majo_qMajo}
\end{figure}

In this section we show that also quasi-MZMs can give rise to a fractional entropy change. Here, the entropy is computed according to the discrete framework with Hamiltonian (\ref{BdG_smoothV}), where $\Delta(x)=\Delta_0\Theta(L_S-x)\Theta(x)$,  $m(x)=m_0\Theta(L_S-x)\Theta(x)$, $V(x)=V_1\Theta(x)e^{-(x-x_0)^2/2\sigma_1^2} + V_2\Theta(x+\sigma_2)\Theta(-x)$, and $\mu(x)=\mu_N\Theta(x+L_N)\Theta(-x) + \mu\Theta(L_S-x)\Theta(x)$. $\Delta S_{V_2}$ as a function of temperature is shown in Fig.~\ref{Entropy_Majo_qMajo}, where $\mu$ is used to tune the system between non-trivial and trivial phases. In this case fractional plateau $\Delta S_{V_2}=-\frac{1}{2}k_B \log{2}$ appears in both phases because of the formation of quasi-MZMs. Thus, $G$, $S_z$ and $\Delta S_{V_2}$ behave in a similar way also in the presence of quasi-MZMs.

\bibliography{amainrefbiblio}